\documentclass[fleqn,usenatbib]{mnras}
\usepackage{newtxtext,newtxmath}

\usepackage[T1]{fontenc}
\usepackage{ae,aecompl}
\usepackage{graphicx}	
\usepackage{amsmath}	
\usepackage{mathtools}
\usepackage{cases}
\usepackage{cuted}
\usepackage{bigstrut}
\usepackage{subfig}
\usepackage{paralist}
\usepackage{scrextend}
\usepackage{lscape}
\usepackage{wrapfig}
\usepackage[T1]{fontenc}
\usepackage{aecompl}
\usepackage{adjustbox} 
\usepackage{underscore}
\usepackage{float}
\usepackage{amsfonts}
\usepackage{color}
\usepackage{booktabs}
\usepackage[flushleft]{threeparttable}
\usepackage[english]{babel}
\usepackage{lastpage}
\usepackage{stmaryrd}
\usepackage[compact]{titlesec}
\usepackage{xcolor}
\titlespacing{\section}{0pt}{2ex}{1ex}
\titlespacing{\subsection}{0pt}{2ex}{1ex}
\titlespacing{\subsubsection}{0pt}{2ex}{1ex}

\newcommand{\civ}{\mbox{C\,{\sc iv}}}
\newcommand{\ciii}{\mbox{C\,{\sc iii}}}
\newcommand{\cii}{\mbox{C\,{\sc ii}}}
\newcommand{\nii}{\mbox{N\,{\sc ii}}}
\newcommand{\niii}{\mbox{N\,{\sc iii}}}
\newcommand{\feii}{\mbox{Fe\,{\sc ii}}}

\newcommand{\caii}{\mbox{Ca\,{\sc ii}}}
\newcommand{\siiv}{\mbox{Si\,{\sc iv}}}
\newcommand{\siiii}{\mbox{Si\,{\sc iii}}}
\newcommand{\siii}{\mbox{Si\,{\sc ii}}}
\newcommand{\siv}{\mbox{S\,{\sc iv}}}

\newcommand{\oi}{\mbox{O\,{\sc i}}}
\newcommand{\oii}{\mbox{O\,{\sc ii}}}
\newcommand{\oiii}{\mbox{O\,{\sc iii}}}

\newcommand{\ovi}{\mbox{O\,{\sc vi}}}
\newcommand{\nv}{\mbox{N\,{\sc v}}}

\newcommand{\ari}{\mbox{Ar\,{\sc i}}}

\newcommand{\CLOUDY}{\textsc{cloudy}}
\newcommand{\mgii}{\ifmmode {\rm Mg}{\textsc{ii}} \else Mg\,{\sc ii}\fi}
\newcommand{\heii}{\ifmmode {\rm He}{\textsc{ii}} \else He\,{\sc ii}\fi}

\newcommand{\angstrom}{\text{\normalfont\AA}}
\newcommand{\sqcm}{cm$^{-2}$}  
\newcommand{\hi}{\mbox{H\,{\sc i}}}
\newcommand{\mgi}{\ifmmode {\rm Mg}{\textsc{i}} \else Mg\,{\sc i}\fi}

\newcommand{\fuse}{\it FUSE}

\newcommand{\lya}{Ly$\alpha$}
\newcommand{\lyb}{Ly$\beta$}

\newcommand{\lyd}{Ly$\delta$}

\def\kms{\hbox{km~s$^{-1}$}}
\def\cmsq{\hbox{cm$^{-2}$}}
\def\cc{\hbox{cm$^{-3}$}}
\definecolor{darkgreen}{rgb}{0.0, 0.2, 0.13}
\setcitestyle{notesep={; },round,aysep={},yysep={,}}

\title[Multiphase modeling of QALs]{Cloud-by-cloud, multiphase, Bayesian modeling: Application to four weak, low ionization absorbers}

\author[Sameer et al.]
{Sameer,$^{1}$\thanks{E-mail: sxx15@psu.edu}  
J. C. Charlton,$^{1}$ 
J. M. Norris,$^{2}$
M. Gebhardt,$^{1}$
C. W. Churchill,$^{3}$
G. G. Kacprzak,$^{4,5}$
\newauthor
S. Muzahid,$^{6,7}$
Anand Narayanan,$^{8}$
N. M. Nielsen,$^{4,5}$
Philipp Richter,$^{9}$
and
Bart P. Wakker$^{10}$
\\
$^{1}$Department of Astronomy \& Astrophysics, 525 Davey Lab,
The Pennsylvania State University, University Park, PA 16802, USA\\
$^{2}$Department of Earth and Space Science, Osaka University, Toyonaka, Osaka, 560$-$0043, Japan\\
$^{3}$Department of Astronomy, New Mexico State University, Las Cruces, NM 88003, USA\\
$^{4}$Centre for Astrophysics and Supercomputing, Swinburne University of Technology, Hawthorn, Victoria 3122, Australia\\
$^{5}$ ARC Centre of Excellence for All Sky Astrophysics in 3 Dimensions (ASTRO 3D)\\
$^{6}$IUCAA, Post Bag 04, Ganeshkhind, Pune 411007, India\\
$^{7}$Leibniz-Institut für Astrophysik Potsdam (AIP), An der Sternwarte 16, D-14482 Potsdam, Germany\\
$^{8}$Department of Earth and Space Sciences, Indian Institute of Space Science \& Technology, Thiruvananthapuram 695547, Kerala, INDIA\\
$^{9}$Institut für Physik und Astronomie, Universität Potsdam, Haus 28, Karl-Liebknecht-Str. 24/25, D-14476, Potsdam, Germany\\
$^{10}$Department of Astronomy, University of Wisconsin-Madison, 475 N. Charter Street, Madison, WI 53706, USA
}
\date{Accepted 2020 November 28. Received 2020 November 24; in original form 2020 September 28}

\pubyear{2020}

\begin{document}
\label{firstpage}
\maketitle

\begin{abstract}

\noindent We present a new method aimed at improving the efficiency of component by component ionization modeling of intervening quasar absorption line systems. We carry out cloud-by-cloud, multiphase modeling making use of \textsc{cloudy} and Bayesian methods to extract physical properties from an ensemble of absorption profiles.  Here, as a demonstration of method, we focus on four weak, low ionization absorbers at low redshift, because they are multi-phase but relatively simple to constrain. We place errors on the inferred metallicities and ionization parameters for individual clouds, and show that the values differ from component to component across the absorption profile. Our method requires user input on the number of phases and relies on an optimized transition for each phase, one observed with high resolution and signal-to-noise. The measured Doppler parameter of the optimized transition provides a constraint on the Doppler parameter of {\hi}, thus providing leverage in metallicity measurements even when hydrogen lines are saturated. We present several tests of our methodology, demonstrating that we can recover the input parameters from simulated profiles.  We also consider how our model results are affected by which radiative transitions are covered by observations (for example how many {\hi} transitions) and by uncertainties in the $b$ parameters of optimized transitions. We discuss the successes and limitations of the method, and consider its potential for large statistical studies. This improved methodology will help to establish direct connections between the diverse properties derived from characterizing the absorbers and the multiple physical processes at play in the circumgalactic medium.

\end{abstract}



\renewcommand*{\thefootnote}{\textsuperscript{\arabic{footnote}}}

\section{INTRODUCTION}
\label{sec:Intro}
The circumgalactic medium (CGM) is the tenuous, multiphase medium in the environment of a galaxy. It plays an important role in steering the evolution of galaxies. The massive reservoir of gas in the CGM seeds the formation of stars which evolve and distribute their energy and content back into the galaxy. CGM gas of galaxies can be probed in intricate detail by using metal absorption lines~\citep[e.g.,][]{bergeron1986mg,bergeron1991sample,Lanzetta1992,steidel1992mg,Churchill1996,Churchill2000,Churchill2020,adelberger2005connection,kacprzak2008halo,kacprzak2015azimuthal,Kacprzak2019ApJ,chen2009probing,steidel2010structure,Nielsen20131,Nielsen20132,stocke2013characterizing,werk2013cos,bordoloi2014cos,turner2014metal,johnson2015possible,richter2016,keeney2017characterizing,muzahid2018cos} observed in the spectra of background quasars. Absorption-line spectroscopic studies provide information about the metallicities, densities, temperatures, and kinematic structure of various parcels of gas along the lines of sight. These studies are important because they potentially allow us to learn about the processes that supply inflowing gas to the galaxy, enrich the surroundings with metal-rich outflows, minor and major mergers, and send enriched gas back to the galaxy as recycled accretion.

\smallskip

In this paper, we present a new Bayesian method for extracting the physical properties of the multiphase gaseous medium probed by quasar absorption-line systems.  To demonstrate the method, we apply it to four, well-studied, weak, low ionization, systems, also known as weak {\mgii} absorbers~\citep{narayanan2005survey}. These systems were chosen because they are relatively simple in the number of low ionization components, but still have multi-phase structure, and also because the origins of these tiny pockets of high metallicity, far from galaxy centers, are of particular interest.

\smallskip

Weak \mgii-absorbers are defined by rest frame equivalent width $W_r^{2796}<$0.3~\AA.  Unlike strong {\mgii} absorbers, which are typically found within $\sim 60$~kpc of a luminous galaxy~\citep{Nielsen20132,nielsen2018magiicat},
the weak \mgii~absorbers are found to be associated with lower neutral hydrogen column densities, and are at larger impact parameters within the circumgalactic environment, typically $>40$~kpc \citep[]{Nielsen20132,nielsen2018magiicat}. \citet{churchill1999population} conducted the first systematic survey of weak \mgii~absorbers in the redshift range $0.4 < z < 1.4$ using HIRES/Keck spectra, and found that the number of weak absorbers per unit redshift of $d\mathcal{N}/dz\approx1.74\pm0.10$ is $\approx$4 times higher than that of the Lyman limit systems (LLSs) which have neutral hydrogen column density $N(\hi)>10^{17.2}$~\sqcm, indicating an origin of the weak absorbers in sub-LLSs~\citep{Churchill2000}. A tantalizing aspect of the weak \mgii-absorbers is that, generally, their low ionization phases exhibit near-solar to super-solar metallicities \citep[]{rigby2002population,lynch2007physical,misawa2008supersolar,narayanan2008chemical} in spite of the fact that luminous galaxies are typically 50$-$200~kpc away.  Detailed photoionization models of the intermediate redshift, weak \mgii-absorbers show that they often have two phases, a high-density region/s that is 1-300 pc thick and produces narrow (a few \kms) low-ionization lines, and kiloparsec-scale, lower density region/s that produces broader, high-ionization lines \citep[]{charlton2003high,narayanan2007survey}. By comparing the relative number of \mgii~and \civ~absorbers,
\citet{milutinovic2006nature} concluded that the low ionization absorption must occur in flattened or filamentary structures that occupy more than half the solid angle subtended by structures that produce only \civ~absorption.

\smallskip

Studies of weak absorbers at low-$z$ offer the advantage of ease of detection of host galaxies. At low redshift, $0 < z < 0.3$, \citet{narayanan2005survey} measured $d\mathcal{N}/dz = 1.00\pm0.20$ for weak \mgii~absorbers (with $W_r^{2796}>0.02$~{\AA}) using \siii~and \cii~as proxies in $HST$/COS spectra. This is smaller than expected based on the evolution expected for the intermediate redshift population, given the decreasing extragalactic background radiation. This is because the low redshift population should consist both of 1-300~pc absorbers that gave rise to \mgii~absorption at intermediate redshift, and larger clouds that gave rise only to higher ionization (\civ) absorption at intermediate redshift.  Understanding which of these types of absorbers dominate the population at low redshift is important to understanding the processes that produce these mysterious, high metallicity objects abundant in the CGM.   

\smallskip

More generally, both observational and theoretical studies~\citep[e.g.,][]{stinson2012magicc,ford2013hydrogen,shen2013circumgalactic,suresh2016ovi} of absorbers reveal the existence of cool (10$^4$ K) and warm-hot (10$^{5}$-10$^{6}$ K) phases in the CGM, often all detected along the same sightline to the background quasar. Different ionization states of material trace different phases of gas, which have different densities and temperatures. Cooler, higher-density gas is traced by lower-ionization species (e.g., \mgii, \siii) and shows structure on smaller scales compared to the warmer and/or lower-density phase (e.g., \civ~or \ovi). Length scales of the lower ionization clouds range from $\sim$ 1 pc~\citep{mccourt2012thermal,liang2020model} to a few kpc. 
As mentioned above, strong \mgii~absorption is detected along most sightlines within an impact parameter of 50 kpc from a galaxy \citep{Nielsen20131}, while weaker \mgii~absorption is found at a median impact parameter of 166 kpc \citep{muzahid2018cos}. {\civ} is often seen out to 100 kpc \citep{chen2001origin,bordoloi2014cos}, while \ovi~is seen out to at least 150 kpc \citep{Prochaska2011,tumlinson2011large,stocke2013characterizing,liangchen2014,pointon2019relationship} of star forming galaxies, but not around quiescent galaxies. Gas traced by {\hi} absorption is seen well beyond \citep{liangchen2014,Borthakur2015}. 

\smallskip

Interpretation of the rich absorption line data is aided by identification of galaxies near the quasar sightlines. It is expected that the orientations of the galaxies will correlate with absorption properties~\citep{Bordoloi2011,bouche2012,kacprzak2012tracing,kacprzak2015azimuthal,Nielsen2015}. In the simplest view, galactic winds due to episodes of star formation are expected to produce high metallicities in the outflowing gas, and should be seen primarily along the minor axis~\citep{muzahid2015extreme,Nielsen2015,pointon2019relationship,Schroetter2019,Peroux2020}. Inflowing pristine gas from the intergalactic medium would be expected to flow in along the major axis. Another important observational result is a bimodality in the metallicity of partial Lyman limit systems, with the high metallicity peak at $\log Z$ = $-0.4$ and the low metallicity peak at $\log Z$ = $-1.7$~\citep{lehner2013bimodal}. It has been hypothesized that this bimodality could be due to the contrast between outflowing and infalling material. 


\smallskip

A coherent and logical picture is beginning to emerge from the data \citep{tumlinson2017circumgalactic}, however, there are some significant problems plaguing interpretations, and the root cause of these problems could be the methods used to derive the properties of the gas from the absorption profiles, particularly the metallicities. Often the total column density (integrating over all components at all velocities) of a given metal line transition is compared to the total column density of hydrogen, in the context of photoionization models, to derive the metallicity. This overlooks the importance of phase structure, with more than one phase contributing to absorption at a single velocity i.e. some transitions could have contributions from two different phases which can lead to an inaccurate estimate of ionization parameter, and thus a systematic error in deriving a single metallicity value. In other words, the average parameters inferred for a system may ``wash out'' contributions from regions of gas that have metallicities and densities that differ significantly from the means, and the possibility of conditions that significantly change with velocity because of changes in physical conditions along the line of sight. Simulations~\citep{Churchill2015,peeples2019figuring} indicate absorption arises in lots of places along the sightline, motivating why multiple metallicities are expected.

\smallskip

In a recent study, \citet{zahedy2019characterizing} have carried out a systematic analysis of 16 intermediate-redshift ($z$=0.21-0.55) luminous red galaxies (LRGs), finding metal-poor absorbing components with \hbox{$<$ 1/10} solar metallicity in half of the LRG haloes in tandem with solar and super-solar metallicity gas in the same halo, suggestive of poor chemical mixing and complex multiphase structure in these haloes. This highlights the importance of resolving multiple components of an absorption system using high resolution absorption spectra, which otherwise is missed by ionization studies that make use of only the integrated {\hi} and metal column densities along individual sightlines. 

\smallskip

This work improves the efficiency of component by component modeling that has been successful in recovering the physical conditions for various individual absorbers~\citep[e.g.][]{churchill1999multiple,charlton2000anticipating,ding2003quadruple,charlton2003high,ding2003multiphase,zonak2004absorption,ding2005absorption,masiero2005models,lynch2007physical,misawa2008supersolar,lacki2010z,jones2010bare,muzahid2015extreme,richter2018,rosenwasser2018understanding}. Rather than averaging over components and phases, it is possible to determine how much of the {\hi} is associated with these different phases in order to derive separate metallicities for various clouds. Resolving the individual clouds allows us to break the degeneracy for components on the flat part of the {\lya} curve of growth, even with coverage of just saturated {\hi} lines, and derive metallicity constraints for different parcels of gas along the line of sight. It is important to do so because different processes e.g., outflows~\citep{bouche2012,bordoloi2014cos,Rubin2014,Schroetter2016}, pristine accretion~\citep{Rubin2012,martin2012,Danovich2015}, recycled accretion~\citep{Ford2014}, minor and major mergers~\citep{martin2012,angles2017}, are surely contributing to the same system, and it is expected that conditions will vary significantly along a line of sight which can span hundreds of kpc spatially~\citep{Churchill2015,peeples2019figuring}. This will lead to a more meaningful comparison to galaxy properties. For example, \citet{pointon2019relationship} did not find a difference between the metallicities of absorbers found within an impact parameter of 200~kpc along the major and the minor axes of isolated galaxies.  Based on cosmological hydrodynamic simulations, a larger metallicity is expected along the minor axis due to outflows and a lower metallicity along the major axis due to inflows \citep{Peroux2020}. However, an observational trend could exist, for example, for the minor axis to have some, but not all, high metallicity components, or for the minor axis to have one or more low metallicity components. Such results would be ``washed out'' by deriving an average metallicity for all gas along a line of sight, which clearly often has multiple complex origins.  For some datasets/projects the new analysis could be transformative, however to make it feasible to use for large statistical studies it is important that the analysis is semi-automated and robust.

In this paper, we describe a new method to derive the physical properties of quasar absorption line systems using the photoionization code \CLOUDY~\citep[ver 17.01;][]{cloudy17},
Bayesian methods, and parallel processing to efficiently obtain results.
Though the code requires some human intervention to choose transitions that define the phase structure, it is mostly automated, and provides uncertainties on parameters in the context of the assumed structure.
The paper is organized as follows: in Section~\ref{sec:observations} we describe the spectral observations that are analysed in this work; in Section~\ref{sec:methodology} we describe the methodology used to determine the physical conditions of an absorber; in Section~\ref{sec:Results_Discussion} we present the results and discuss our findings, comparing to previous work, applying tests, and discussing possible limitations. We conclude in Section~\ref{sec:discussion} by discussing the limitations and potential of our cloud-by-cloud, multiphase, Bayesian modeling (CMBM) method. Throughout this work, we assume a cosmology with $H_{0}\,=\,70$~km~s$^{-1}$~Mpc$^{-1}$, $\Omega_{\rm M}\,=\,0.3$, and $\Omega_{\Lambda}\,=\,0.7.$  Abundances of heavy elements are given in the notation $\rm [X/Y] = \log (X/Y) - \log (X/Y)_{\odot}$ with solar relative abundances taken from \citet{grevesse2011chemical}. All the distances given are in physical units. All the logarithmic values are presented in base-10.

\section{OBSERVATIONS}
\label{sec:observations}

\begin{table*}
\renewcommand\thetable{1}
\centering
	\caption{List of the four modeled weak {\mgii} analogs from the COS-Weak Sample}
	\begin{tabular}{ccccccccc}
		\hline\hline
		QSO. & RA  &  Dec & $z_{qso}^{a}$ & $z_{abs}^{b}$ & Instrument & PID$^{c}$ & Observed  & Observation \\
		& (J2000) & (J2000) & & & & & Wavelength (\angstrom) & Date \\
		\hline
		 HE0153-4520 & 28.80500 & $-45.10333$  & 0.451 & 0.22596 & COS & 11541 & 1135-1794 & 2002-12-19\\
		 &  &  &   & & UVES & 293.A-5038(A) & 3045-6650 & 2014-10-26\\\hline		 
		 PG1116+215 &    169.78583 &     21.32167 &    0.176 & 0.13849 &  {\fuse} & P101 & 979-1189 & 2001-04-22\\
		  &     &      &     & &  COS & 12038 & 1136-1794 & 2011-10-25\\
		  &     &      &     & &  STIS (E140M) & O5E702010 & 1155-1602 & 2000-06-30\\	  
		  &     &      &     & &  STIS (E230M) & O5A302030 & 2010-2817 & 2000-05-15\\
        &     &      &     & &  HIRES & U152Hb & 3104-5901 & 2006-06-02\\\hline
PHL1811 &    328.75623 &     -9.37361 &    0.190 &  0.07776, 0.08094 &  {\fuse} & P108 & 979-1189& 2003-06-02 \\        
        & & & & &  COS & 12038 & 1135-1794 & 2010-10-29\\
        & & & & & STIS (G230MB) & O6CT46010 & 2758-2910 & 2001-10-22\\
        & & & & & HIRES & U149Hr & 3101-5897 & 2007-09-17\\\hline
 
        	\end{tabular}
\begin{flushleft} \small
Notes-- $^{a}$ QSO redshift from NED; $^{b}$ Absorber redshift; \hbox{$^{c}$ proposal ID.}
\end{flushleft}        	
	\label{tab:sample}
\end{table*}

We apply our methods to four low redshift absorption line systems from the COS-Weak sample~\citep{muzahid2018cos}. The background quasars in this study have UV spectra from the Cosmic Origins Spectrograph (COS) instrument on the Hubble Space Telescope (\textit{$HST$}) and optical spectra from High Resolution Echelle Spectrometer (HIRES) installed on the Keck telescope or Ultraviolet and Visual Echelle Spectrograph (UVES) installed on the Very Large Telescope (VLT). The four absorption systems that we consider have spectroscopic redshifts $z < 0.30$.  These weak, low-ionization systems are relatively simple ones compared to some stronger low-ionization absorbers, but they are expected to have multiple gas phases and provide good test cases for our methodology.  Table~\ref{tab:sample} presents the details of the observations of the four systems studied in this work.

\subsection{COS, {\fuse} and STIS Spectra}\label{sec:uvobs}

The UV spectra in the COS-Weak sample are taken from the archive of \textit{$HST$}/COS observations using the G130M and G160M gratings spanning observed wavelength ranges 1135-1457~{\AA} and 1399-1794~{\AA}, respectively. The spectra have an average resolving power of $R \approx 20,000$ and cover a range of ions including the {\hi} Lyman series, {\cii}, {\ciii}, {\civ}, {\nv}, {\ovi}, {\siii}, {\siiii} and {\siiv}. The \textit{$HST$}/COS spectra were obtained from the archive already reduced using the {\sc calcos} pipeline software~\citep{massa2013cos}, with individual exposures aligned and co-added, weighted by their exposure times. The spectra are rebinned to the Nyquist sampling rate. {\fuse} spectra for three absorbers in our sample were available. The observations were obtained already processed with the \textsc{calfuse} pipeline (ver. 3.2.3). Zero point offset correction was carried out on this pre-processed data using the approach described in \citet{wakker2006fuse}. The spectra have an average resolving power of $R \approx 15,000$. STIS spectra for three absorbers in our sample were available. We make use of COS data when STIS and COS both have coverage of a transition of interest because the COS spectra have significantly higher signal-to-noise, $S/N$, ratio compared to the STIS data. When a transition is not covered by COS, we make use of STIS observations. 

\smallskip

We perform continuum normalization of the data by fitting a cubic spline to the spectrum. We estimated the uncertainties in the continuum fits using ``flux randomization'' Monte Carlo simulations (e.g. \citealt{peterson1998optical}), varying the flux in each pixel of the spectrum by a random Gaussian deviate based on the spectral uncertainty. The pixel-error weighted average and standard deviation of 1000 iterations was adopted as the flux and uncertainty of the continuum fit, respectively.

\subsection{Ground-based Optical Quasar Spectra}
When available, we use higher resolution optical spectra to complement the UV spectra because transitions including {\mgi}, {\mgii}, {\feii}, and {\caii} are especially useful in revealing the presence of multiple components for absorption systems at redshifts of $z_{abs} \gtrsim 0.1$. The spectra have an average resolving power of $R\approx 45,000$. The HIRES spectra were reduced using the Mauna Kea Echelle Extraction (MAKEE) package. The UVES spectra were reduced using the European Southern Observatory (ESO) pipeline \citep{dekker2000optical} and the UVES Post-Pipeline Echelle Reduction (UVES POPLER) code~\citep{murphy2016,murphy2019}. The program numbers and dates of these observations are also listed in Table~\ref{tab:sample}.

\section{METHODOLOGY}
\label{sec:methodology}

We model the observed absorption system using the photoionization code {\CLOUDY}, and synthesize the expected absorption profiles and compare them to the observed profiles in order to infer the physical conditions of the absorbing gas. What makes this method different from other researchers' methods is that we model the entire absorption system across all transitions by splitting it into its constituent clouds. 

\smallskip

We begin by choosing a ``trustworthy'' low ionization species (the ``optimized ion'') for a Voigt profile (VP) fit, meaning one with lines that are unsaturated, and observed at a high resolution, and with a large \textbf{$S/N$}. The starting point is a fit of Voigt profiles to all transitions within this single ionization species (each parcel of gas described by a single VP will be called a ``cloud'').  For this optimized transition we determine the parameters of column density ($N$), Doppler parameter ($b$), and redshift ($z$) for each cloud using a Monte Carlo technique employed using PyMultiNest~\citep{buchner2014x}. PyMultiNest is a \textsc{python} implementation of MultiNest~\citep{feroz2009multinest}, which is a robust nested sampling method that draws a new, uniformly random point with higher likelihood through an ellipsoidal rejection sampling scheme~\citep{shaw2007efficient}. It is efficient at sampling from a multimodal posterior distribution. We model Voigt profiles using an analytic approximation~\citep{garcia2006voigt} for the Voigt-Hjerting function as implemented in the VoigtFit package~\citep{krogager2018voigtfit}. We obtain VP fits to the absorption profiles for the optimized ion, allowing for more than one component. We fit the profiles of the optimized ion with a VP model corresponding to the highest log-evidence and the least model complexity. After we determine the observed values of $\log N$, $b$, and $z$ for each cloud, we use the parameters of that VP fit for the optimized ion as a starting point for models in \CLOUDY.

\smallskip

{\CLOUDY} has many parameters that can be adjusted. The two most important parameters are metallicity relative to solar abundance ($Z$) and ionization parameter ($U$), defined by $U = n_{\gamma} /n_{H}$, where $n_{\gamma}$ is the volume density of ionizing photons. Ionization parameter serves as a direct proxy for the hydrogen number density, but may be more intuitive when thinking of contributions of different clouds to the observed ionization states. We adopt the \citet{haardt2012radiative}[HM12] model for the extragalactic background radiation (EBR) to determine $n_{\gamma}$ at the given redshift. The systematic effects arising from the choice of the EBR are discussed in~\citet{keeney2017characterizing}. We adopt the HM12 model as it best reproduces the low-redshift \hi~column density distribution~\citep{danforth2016HST} for $\log [N(\hi)/\cmsq] >$ 14~\citep{shull2014galaxies}.  We assume a solar abundance pattern~\citep{grevesse2011chemical} for our modeling.  Detailed constraints on abundance pattern are not possible without higher $S/N$ coverage of a larger number of transitions.

\smallskip

It is a time-consuming endeavor to construct {\CLOUDY} models component-by-component and phase-by-phase for individual absorbers because of the limitations imposed by serial computing. However, it is clear that doing so will lead to a significant gain in our ability to correctly infer the presence of metallicity structure in the CGM. To increase efficiencies we have developed robust \textsc{python} based codes that employ distributed parallel processing on the CyberLAMP infrastructure~\footnote{\url{https://wikispaces.psu.edu/display/CyberLAMP}} to obtain models in a shorter time. 

\begin{figure*}
    \centering
    \subfloat{{\includegraphics[scale=0.55]{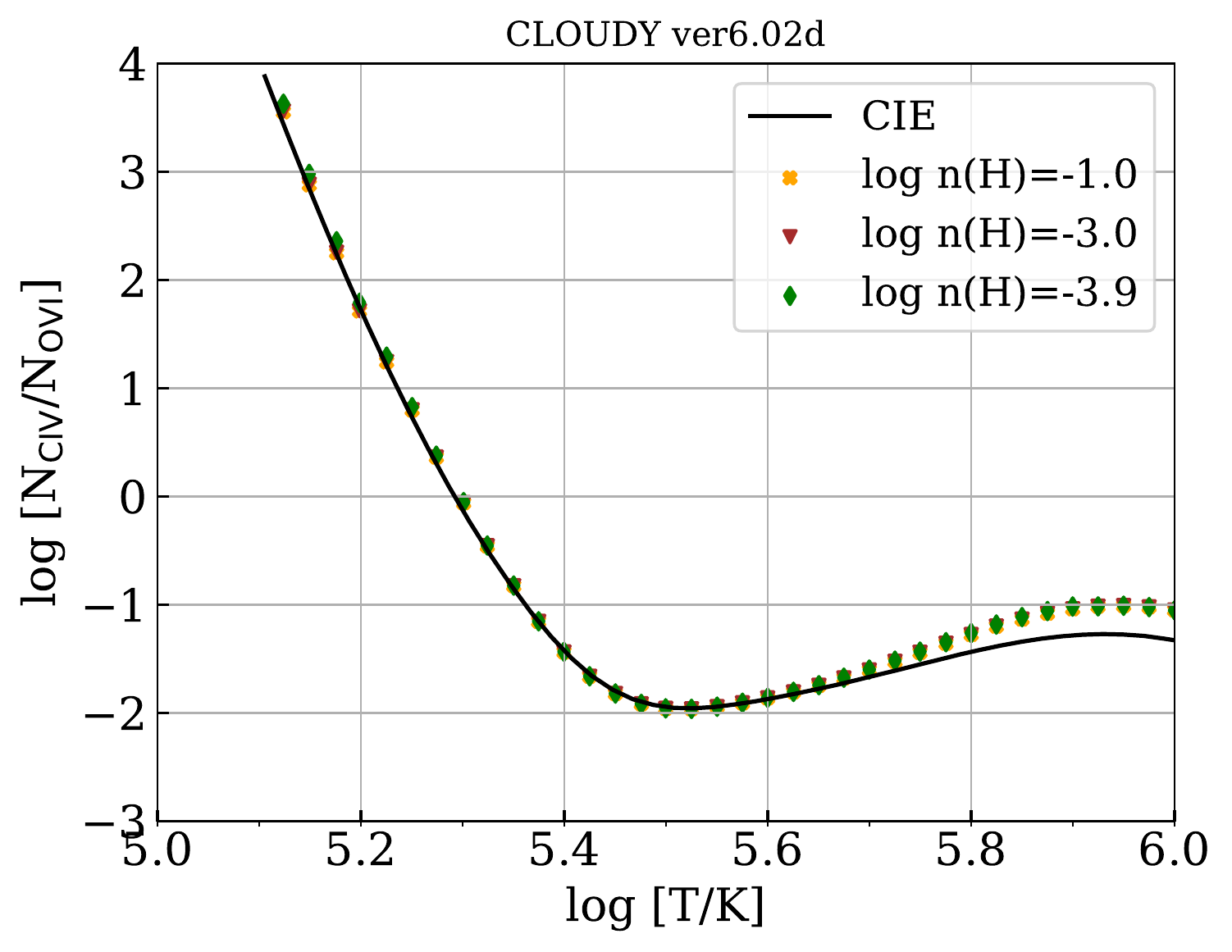} }}%
    \qquad
    \subfloat{{\includegraphics[scale=0.55]{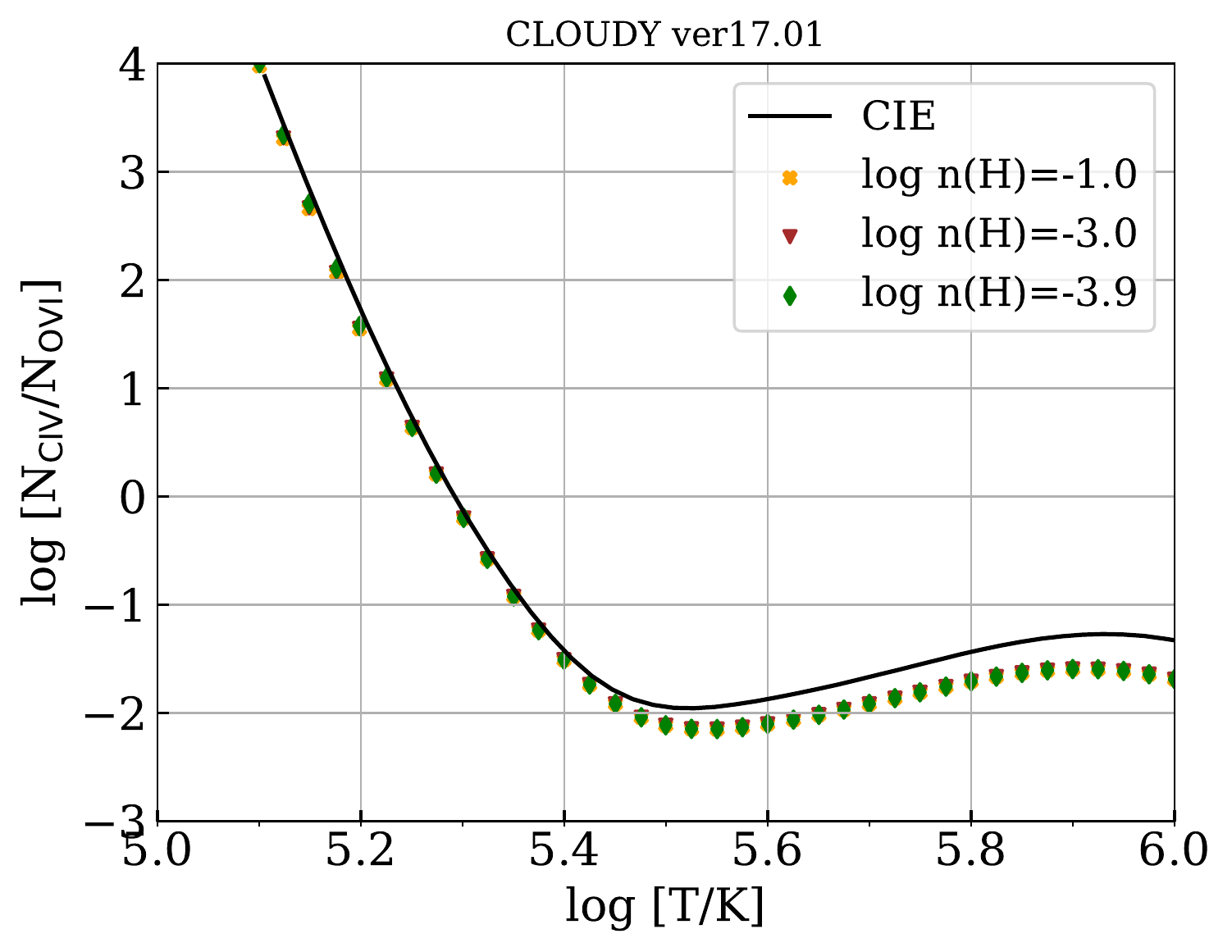} }}%
    \caption{Comparison of CIE model (black; \citet{gnat2007time} and collisional ionization models obtained with {\CLOUDY} ver6.02d (left panel) and ver17.01 (right panel) for a metallicity of $\log Z = -1$ and different hydrogen densities ($\log n(H)$). The disagreement of the CIE model with collisional ionization models obtained with {\CLOUDY} ver17.01 for $\log T > 5.5$ is attributed to the evolving atomic database \citep{chatzikos2015implications}. We use {\CLOUDY} ver17.01 and adopt $\log [n(H)/\cc] = -3.9$ for collisional ionization modeling.}%
    \label{fig:cloudycollisional}%
\end{figure*}

\begin{figure*}
\begin{center}
\includegraphics[width=\linewidth]{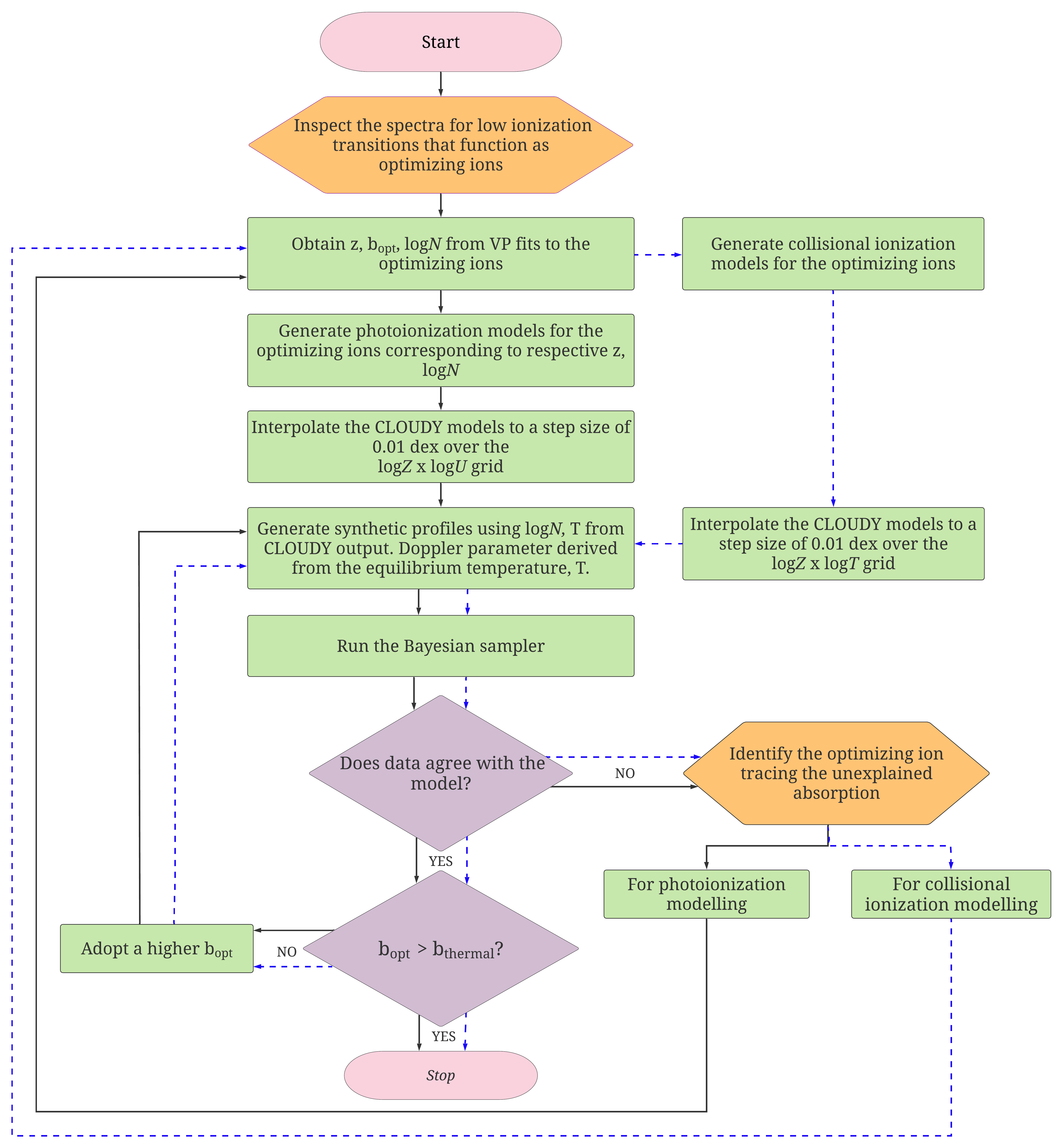}
\caption{A flowchart showing the workflow involved in obtaining the models that best describe the conditions present in an absorption system.}
\label{fig:flowchart}
\end{center}
\end{figure*}

\smallskip

The {\CLOUDY} photoionization equilibrium (PIE) models are generated for the identified optimized ion tracing the low ionization gas phase for a broad range of values over the $\log Z$ $\times$ $\log U$ grid in the range: $\log Z$ $\in$ [-3.0,1.5] and $\log U$ $\in$ [-6.0, 0.0], obtained with a step-size of 0.1. The {\CLOUDY} models are then linearly interpolated onto the $\log Z$ $\times$ $\log U$ grid with a spacing of 0.01 dex. The interpolating function is generated by triangulating the input data with Qhull~\citep{barber1996quickhull}, and on each triangle performing linear barycentric interpolation. The interpolation error is of the order $\sim O(h^{2})$, $h$ being 0.1, and introduces only small errors $\sim$ 0.01 dex in $\log Z$ and $\log U$. The total column density of hydrogen, $N(H)$, is adjusted for each model until the column density of the optimized ion is matched for each point on the $\log Z$ $\times$ $\log U$ grid. We allow for an uncertainty of 0.05 dex in the column density matching when generating the {\CLOUDY} models. The uncertainties arising due to the interpolation error and the error in column density are not accounted for in our error estimation, but are usually negligible in comparison to other sources of error. We quote 2$\sigma$ uncertainties in our results for the best model parameter values, which account only for the photon noise in the comparison between the synthesized model profiles and the observed spectra.

\smallskip

We then determine the parameters that best describe the conditions ($\log Z$, $\log U$) present in each one of the absorbers again using PyMultiNest. The posterior distribution of the parameters, which is proportional to the product of priors and likelihood, is determined by using uniform priors on $\log Z$ and $\log U$, spanning the grid range mentioned above. The log-likelihood function, $P$, is formulated as

\begin{equation}
    P_{i,j} = \frac{data_{i,j}-model_{i,j}}{\sigma_{i,j}}
\end{equation}
\begin{equation}
    P = -\frac{1}{2}\Big[\bigparallel_{i,j}^{} P_{i,j}\Big] \Big[\bigparallel_{i,j}^{} P_{i,j}\Big]^T   
\end{equation}

\noindent where $data_{i,j}$ denotes the array of data pixels corresponding to species $i$, and transition $j$ of the species observed in the spectral data. $\sigma_{i,j}$ denotes the 1$\sigma$ uncertainty array corresponding to data$_{i,j}$. $\bigparallel_{i,j}^{} P_{i,j}$ denotes the concatenated arrays from all the data$_{i,j}$ corresponding to different species (i) and transitions (j) observed in the spectrum. The pixels span a velocity range of 500 {\kms} centered at the redshift of the absorber. We ensure that the pixels which are affected by blending or are too noisy are masked when evaluating the log-likelihood. In some cases, we ``deactivate'' a line transition when data$_{i,j}$ is severely affected by blending/noise during the log-likelihood evaluation. $model_{i,j}$, denotes the flux array synthesized using the column densities obtained from the {\CLOUDY} output for the different components in species $i$, centered at the redshift, $z$, of the optimized ion tracing the phase. The Doppler parameters for these components, $b_{i}$, are determined as 

\begin{equation}
\label{eq:3}
   b_{opt}^{2} = b_{turb}^{2} + \frac{2kT}{m_{opt}}
\end{equation}
\begin{equation}
\label{eq:4}
   b_{i}^{2} = b_{turb}^{2} + \frac{2kT}{m_{i}}
\end{equation}

\noindent where $b_{opt}$ is the Doppler parameter determined from the Voigt profile fit to the optimized ion. $b_{turb}$ is the same for all species in the phase traced by the optimized ion. $b_i$ is then determined from the equilibrium temperature, $T$, given by {\CLOUDY}. Voigt profiles are then synthesized with the parameters $b_i$, $z$, $\log N$ (from {\CLOUDY} output), for each component, and the resulting synthetic profile is convolved with a wavelength dependent instrumental line spread function to obtain $model_{i,j}$. The instrumental line spread function is determined for each transition based on the instrument used for observations as listed in Table~\ref{tab:sample}. When there is a significant uncertainty in $b_{opt}$, it could be possible that the adopted $b_{opt}$ leads to a {\CLOUDY} equilibrium temperature which is too high, and which results in an unphysical value for $b_{turb}$. In such a case, we adopt a higher value for the $b_{opt}$ within the 2$\sigma$ uncertainty limit.

\smallskip

Typically, all of the observed line transitions cannot be reproduced with one phase of gas. It is often the case that a model that fits the low-ionization transitions (e.g., \mgii, \feii) cannot produce the observed amount of high-ionization absorption (e.g., \civ, \ovi), or adequately match the profiles of the full Lyman series. In such cases, we will use more than one optimized ion. The other optimized ions represent the different phases of gas, with distinct regions having different densities along the line-of-sight. For example, we would typically optimize on {\mgii} for the low-ionization gas, and on {\civ} for high-ionization gas. In some intervening absorbers, \ovi, because of its comparatively high ionization potential ($\sim$ 114 eV), arises in gas with \hbox{T $>$ 10$^{5}$ K} where the ionization is governed by collisions between energetic free electrons and metal ions~\citep{jayadev20}. We model the gas phase traced by {\ovi} to be under collisional ionization equilibrium (CIE). For such collisional ionization models, the {\CLOUDY} grids are run over $\log Z$ $\times$ $\log T$ space: $\log Z$ $\in$ [-3.0,1.5] and $\log [T/K]$ $\in$ [5.0, 5.8]. Similar to the PIE modeling, the CIE models are obtained by varying the total column density of hydrogen, $N(H)$, until the column density of {\ovi} is matched for each point on the $\log Z$ $\times$ $\log T$ grid. In Figure~\ref{fig:cloudycollisional}, we show for a metallicity of $\log Z = -1.0$ and different gas densities that {\CLOUDY} can produce accurate results when modeling gas in CIE. The column density ratios of {\civ}/{\ovi} agree well with the CIE model~\citep{gnat2007time} for temperatures in the range of $10^{5}-10^{5.8}$ K, and are independent of the hydrogen number densities. We contrast the modeling results obtained with {\CLOUDY} ver6.02d and ver17.01 to show that the apparent discrepancy with the \citet{gnat2007time} model arises because of the evolving atomic database~\citep{chatzikos2015implications}. The {\CLOUDY} models are generated for an assumed value of $\log n_{H} = -3.9$, based on the Milky Way halo density at 50kpc~\citep{Kaaret2020}. 

To summarize, we start with the assumption that a single, low ionization phase describes an absorption system. If such a simple, one-phase model fails to account for the observed column densities in all the transitions, we incorporate additional phases until the synthesized spectra from {\CLOUDY} best match the observed spectra for all transitions. Generally, we find that a single phase is insufficient to explain the observed abundances of all the transitions even when the transitions coincide in velocity space. If the synthetic spectra fit the data for all transitions, then it is plausible that the physical conditions in the {\CLOUDY} model match those in the gas that produces the absorption. This allows us to determine precise allowed ranges of metallicity, density, and temperature for each individual cloud that create the observed transition lines. A summary of the methodology is given in the form of a flowchart in Figure~\ref{fig:flowchart}.

\section{RESULTS}
\label{sec:Results_Discussion}

\subsection{Application to four quasar absorption line systems}


We discuss the findings from the application of our methodology to four, weak low-ionization absorption line systems, and compare our results to those obtained from earlier studies adopting the HM12 model for the EBR.

\subsection{The $z = 0.07776$ absorber towards the quasar PHL1811}
\label{sec:z = 0.07776}

\begin{table*}
\renewcommand\thetable{2}
\begin{center}
\caption{\bf The gas phases in the $z = 0.07776$ absorber towards PHL1811}
\label{tab:phl18110.07776model}
 
\begin{tabular}{c@{\hspace{0.25\tabcolsep}}c@{\hspace{0.55\tabcolsep}}c@{\hspace{0.55\tabcolsep}}c@{\hspace{0.55\tabcolsep}}c@{\hspace{0.55\tabcolsep}}@{\vline}c@{\hspace{0.55\tabcolsep}}c@{\hspace{0.55\tabcolsep}}c@{\hspace{0.55\tabcolsep}}c@{\hspace{0.55\tabcolsep}}c@{\hspace{0.55\tabcolsep}}c@{\hspace{0.55\tabcolsep}}@{\vline}@{\hspace{0.55\tabcolsep}}c@{\hspace{0.55\tabcolsep}}c@{\hspace{0.55\tabcolsep}}c@{\hspace{0.55\tabcolsep}}} \hline \hline

& &  & & & & & Lyman lines & & & & & {\lya}-only  & \\  
\hline
Optimized & $z$ & $b$ & $b_{used}$ & $\log \frac{N}{\cmsq}$ & $\log Z$ & $\log U$ & $\log \frac{n(H)}{\cc}$ & $\log \frac{N(\hi)}{\cmsq}$ & $\log \frac{L}{kpc} $ & $\log \frac{T}{K} $ & $\log Z$ & $\log U$ & $\log \frac{N(\hi)}{\cmsq}$\\  
ion & & (\kms) & (\kms) & & & & & & & & & &\\
(1) & (2) & (3) & (4) & (5) & (6) & (7) & (8) & (9) & (10) & (11) & (12) & (13) & (14)\\\hline
\textcolor{blue}{\ciii_0} & 0.07766 & $3.2^{+3.7}_{-1.0}$  & 5.0 & $13.49^{+0.90}_{-0.55}$ & $0.85^{+0.11}_{-0.18}$ & $-3.11^{+0.08}_{-0.06}$  & $-3.22^{+0.06}_{-0.08}$ & $13.93^{+0.15}_{-0.13}$ & $-2.19^{+0.19}_{-0.11}$ & $4.02^{+0.11}_{-0.06}$ & $0.26^{+1.10}_{-0.67}$ & $-2.47^{+0.23}_{-0.49}$  & $13.19^{+0.41}_{-0.27}$ \\    

\textcolor{red}{\siiii_0} & 0.07774 & $2.6^{+6.7}_{-2.2}$  & 3.1 & $12.45^{+1.45}_{-0.40}$ & $-0.32^{+0.12}_{-0.12}$ & $-3.68^{+0.07}_{-0.06}$  & $-2.64^{+0.06}_{-0.07}$ & $15.72^{+0.11}_{-0.09}$ & $-1.46^{+0.14}_{-0.09}$ & $4.18^{+0.03}_{-0.03}$ & $-0.65^{+0.27}_{-0.24}$ & $-3.81^{+0.07}_{-0.07}$  & $16.21^{+0.26}_{-0.30}$ \\    

\textcolor{red}{\siiii_1} & 0.07779 & $3.7^{+5.5}_{-3.5}$  & 5.7 & $12.21^{+1.64}_{-0.28}$ & $-0.85^{+0.08}_{-0.07}$ & $-3.04^{+0.03}_{-0.02}$  & $-3.29^{+0.02}_{-0.03}$ & $14.26^{+0.11}_{-0.16}$ & $-0.27^{+0.15}_{-0.17}$ & $4.68^{+0.06}_{-0.05}$ & $-0.62^{+0.19}_{-0.22}$ & $-2.90^{+0.08}_{-0.09}$  & $14.01^{+0.05}_{-0.12}$ \\  

\textcolor{darkgreen}{\ovi_0} & 0.07770 & $47.0^{+21.0}_{-19.0}$ &  47.0 & $13.50^{+0.17}_{-0.21}$ & $-1.16^{+0.34}_{-0.33}$ & -  & $-3.90$ & $13.37^{+0.16}_{-0.29}$ & - &  $5.38^{+0.05}_{-0.02}$ & $-1.30^{+0.25}_{-0.36}$ & - & $13.46^{+0.12}_{-0.17}$ \\

\hline 

\end{tabular} \\

\end{center}

Properties of the different gas phases contributing to the $z = 0.07776$ absorber towards PHL1811 traced by their respective optimized ions. Notes: (1) Optimized ion tracing a phase; (2) Redshift of the component; (3) Doppler parameter of optimized ion; (4) Adopted Doppler parameter for the optimized ion; (5) log column density of optimized ion; (6) log metallicity;  (7) log ionization parameter; (8) log hydrogen number density; (9) log hydrogen column density (10) log thickness in kpc; (11) log temperature in Kelvin. The marginalized posterior values of model parameters are given as the median along with the upper and lower bounds associated with a 95\% credible interval. For the collisionally ionized {\ovi} phase, the quantities $\log U$ and $\log L$ are not determined as they are dependent on the assumed value of $\log n(H)$. The synthetic profiles based on these models are shown in Figure~\ref{fig:sysplotphl18110.07}. The marginalized posterior distributions for the VP fit parameters (columns 2, 3, and 5) of the optimized ions are presented in Figures~\ref{fig:voigtCIII0phl18110.07}, \ref{fig:voigtSiIIIphl18110.07}, and \ref{fig:voigtOVIphl18110.07}. The marginalized posterior distributions for the cloud properties (columns 6, 7, 8, 9, 10, 11) are presented in Figures~\ref{fig:CIII0PHL18110.07}, \ref{fig:SiIII0PHL18110.07}, \ref{fig:SiIII1PHL18110.07}, and \ref{fig:OVI0PHL18110.07}. Columns (12), (13), and (14) describe the marginalized posterior distributions of log metallicity, log ionization parameter, and log hydrogen column density, for a {\lya}-only model described in \S~\ref{subsec:0.07776caveats}.

\end{table*}

\begin{figure*}
\begin{center}
\includegraphics[width=\linewidth]{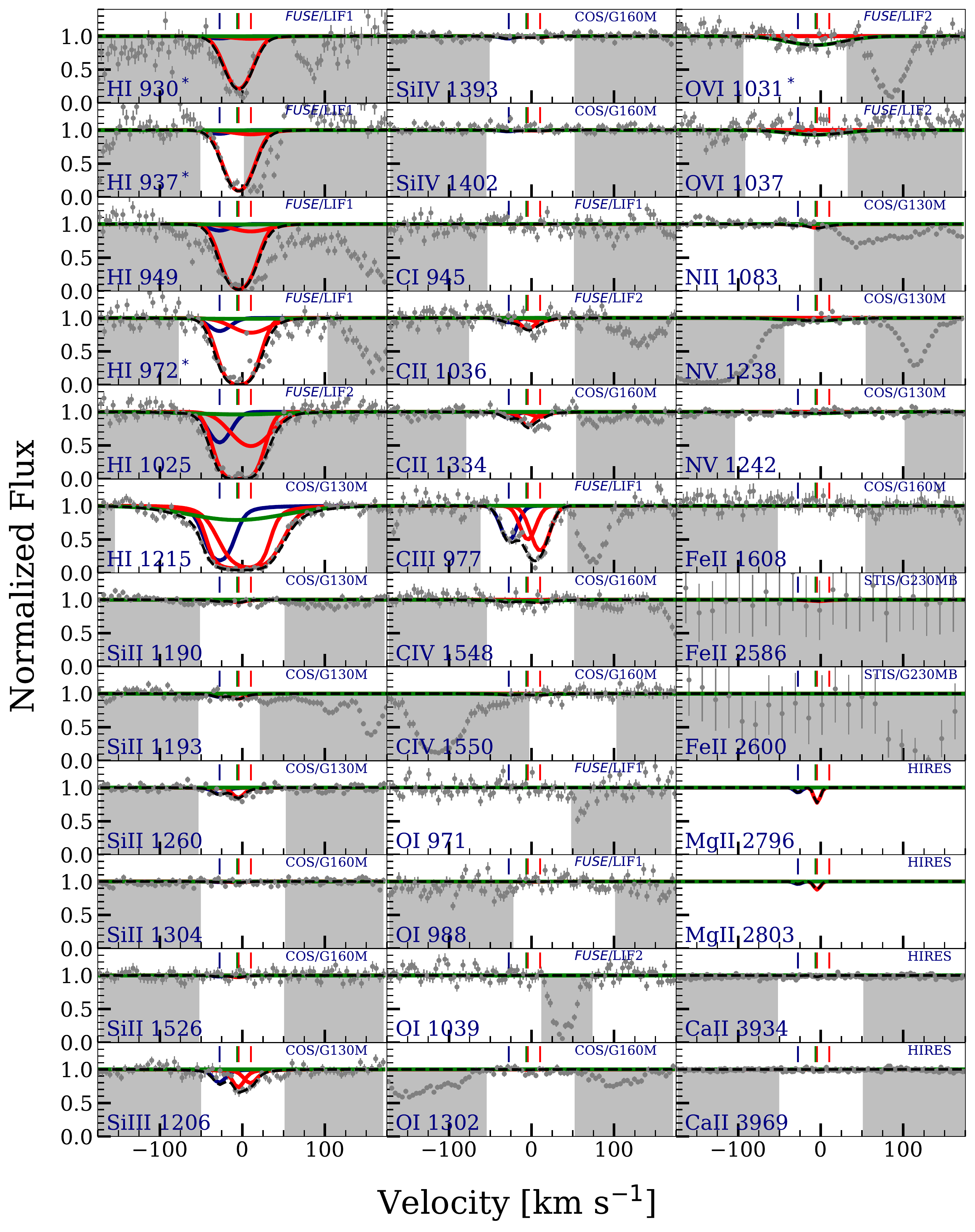}
\caption{\CLOUDY~models for the $z = 0.07776$ absorber towards the quasar PHL1811 obtained using the MLE values. The spectral data are shown in gray with 1$\sigma$~error. The instruments used for observation in different regions of the spectrum are indicated in the panels. The centroids of absorption components as determined from the VP fits to optimized ions are indicated by the vertical tick marks on top of each panel. The low ionization photoionized phase is traced by three clouds - the blueward cloud optimized on {\ciii} (shown as a blue curve), and the blended redward clouds optimized on {\siiii} (shown as a red curve). The high ionization collisionally ionized phase (shown by a green curve) is obtained by optimizing on \ovi. The superposition of these three models is shown by the black dashed curve. The models are obtained using the maximum likelihood parameter values. The region shaded in grey shows the pixels that were not used in the evaluation of the log-likelihood. The lines {\hi} $\lambda$930, {\hi} $\lambda$937, {\hi} $\lambda$972, and {\ovi} $\lambda$1031 are corrected for blending as described in \S~\ref{sec:z = 0.07776} and earmarked with an asterisk.}
\label{fig:sysplotphl18110.07}
\end{center}
\end{figure*}

A system plot of the $z=0.07776$ absorber, with the constraining transitions shown, is given in Figure~\ref{fig:sysplotphl18110.07}. This system has weak absorption detected in {\siii} and \cii. Though {\mgii} is not covered in the STIS/G230MB spectrum, it is expected to be weak as well. The {\feii}, covered in the COS/G160M and STIS/G230MB spectra, is not detected. The intermediate ionization transitions, {\siiii} and \ciii, are detected, however, {\siiv} and {\civ} are not detected. The {\ovi} is detected in the $\lambda$1031 transition, but not in the $\lambda$1037 transition. \ovi~$\lambda$1031 is contaminated at $-35$ {\kms} by Galactic \feii~$\lambda$1112. We account for the contribution of \feii~$\lambda$1112 using other Galactic {\feii} lines from COS data, and display the corrected \ovi~$\lambda$1031 profile. The {\lya} transition was covered by $HST$/COS observations, however, all constraining higher order Lyman series lines plotted (down to \hi~$\lambda$930) come from {\fuse} observations. The lines \hi~$\lambda$937, and \hi~$\lambda$930 are affected by contamination due to the Galactic molecular hydrogen lines. \hi~$\lambda$937 is affected by H$_{2}$ W 8$-$0 Q(2)$\lambda$1010.94 at 59 {\kms}, and \hi~$\lambda$930 by H$_{2}$ L 8$-$0 P(1)$\lambda$1003 at 26 \kms. We divide out the contribution due to these blends using several of the Galactic hydrogen molecular lines present in the {\fuse} observations. The lines H$_{2}$ L 6$-$0 R(0)$\lambda$~1024.37 and H$_{2}$ L 6$-$1 R(1)$\lambda$~1024.99 affect the continuum placement of \hi~$\lambda$949 on its redward side. We omit \hi~$\lambda$949 as a constraining ion. The \hi~$\lambda$972 is blended at \hbox{10 {\kms}} with Galactic \ari~$\lambda$1048; we divide out its contribution to the \hi~$\lambda$972 profile using the \ari~$\lambda$1066 line. The {\lyb} is strongly contaminated by Galactic \siiii~$\lambda$1205~\citep{savage2014properties}, and though the contamination is at the center of the saturated profile, {\lyb} is still not used as one of the constraining transitions. 

\smallskip

We begin by optimizing on the column density of {\ciii} for the blueward cloud and {\siiii} for the redward cloud. We find that the VP fit to {\ciii} yields a large uncertainty on its $b$-parameter ($3.2^{+3.7}_{-1.0}$ \kms). The best-fit {\CLOUDY} model for the {\ciii} cloud corresponds to $\log Z = -1.36$, $\log U = -3.00$, and $\log T = 4.87$. However, this temperature is too high to be consistent with the best VP fit $b$-parameter for {\ciii}. We therefore tried a 2$\sigma$ higher $b$-parameter (b(\ciii) = 7.00 \kms) for the {\ciii} cloud and found the temperature ($\log T = 4.42$) to be considerably lower because of the resulting higher metallicity ($\log Z = 0.85$). We find a similar high value of metallicity for a $b$-parameter only 1$\sigma$ higher than the best fit value ($b$(\ciii) = 5.00 \kms). In Table~\ref{tab:phl18110.07776model} we present the model with $b$(\ciii) = 5.00 \kms.

\smallskip

For the redward {\siiii} cloud we find that a single component fit results in overproduction of {\lyb} and underproduction of {\ciii} on the redward side of the profiles. An adjustment of the $b$ of the optimized {\siiii} cannot resolve both of these discrepancies. A two component fit is thus preferred for {\siiii}, and the column densities and Doppler parameters for the optimized {\siiii} are listed in Table~\ref{tab:phl18110.07776model}. Again, we found that the best VP fit $b$-parameters for the two {\siiii} components resulted in a model with a temperature too high to be self-consistent. Therefore, we adopt 1$\sigma$ higher values for the $b$-parameters of the two redward components in \siiii. The posterior distributions for $z$, $b$ and $\log N$ for the {\ciii} and two {\siiii} components are presented in Figures~\ref{fig:voigtCIII0phl18110.07} and \ref{fig:voigtSiIIIphl18110.07}. For the redward cloud, the lower ionization {\siii} and {\cii} transitions might have been used as optimized transitions, however the VP fits would be quite uncertain due to their noisier profiles. Also, {\ciii} is affected by line saturation, so {\siiii} is the best choice for the optimized transition for the redward cloud. It is clear that the ionization parameters of the three low ionization clouds must be relatively low, since \siiv~and \civ~are not detected, such that {\ovi} could not be produced in the same phase with the {\ciii} and \siiii. We therefore determine that this absorber is best explained by invoking the presence of a high ionization phase giving rise to the {\ovi} absorption. The VP fit posterior distribution for {\ovi} is presented in Figure~\ref{fig:voigtOVIphl18110.07}.

\smallskip

In Table~\ref{tab:phl18110.07776model} we present the posterior results for the low ionization phase, which are summarized in Figures~\ref{fig:CIII0PHL18110.07}, \ref{fig:SiIII0PHL18110.07}, and \ref{fig:SiIII1PHL18110.07} for the {\ciii} component and the two {\siiii} components, respectively.  Figure~\ref{fig:sysplotphl18110.07} shows \CLOUDY~models, using the maximum likelihood estimate (MLE) values superimposed on the data. These plots show that the metallicities of the two redward components (red curve) are constrained to be $\log Z = -0.32^{+0.11}_{-0.11}$ and $\log Z = -0.85^{+0.06}_{-0.06}$ in order to fit the {\lya} profile, and several of the Lyman series transitions. The metallicity of the weaker, blueward {\ciii} component is constrained to be $\log Z = 0.85^{+0.11}_{-0.18}$, and it fits the blue wing of {\lya} in combination with the middle {\siiii} cloud and the {\ovi} cloud.

\smallskip

Table~\ref{tab:phl18110.07776model} summarizes the constraints placed on the metallicities, ionization parameters, number density of hydrogen, $N(\hi)$ values, thickness, and temperatures for the two low ionization clouds. The blueward cloud has a thickness of $\sim$ 6 pc, and the two redward clouds have thicknesses of $\sim$ 35 pc and $\sim$ 0.5 kpc respectively. The three clouds fall far short of producing a partial Lyman limit break (combined $\log [N(\hi)/\cmsq]\sim 15.75$).
 
 \smallskip
 
This low ionization phase does not produce {\ovi} absorption. While it is possible that the absorption is a broad {\lya} component unrelated to the detected broad {\ovi} absorption, a collisionally ionized phase, such as a Coronal Broad Lyman-alpha Absorber (CBLA; \citealt{richter2020}), optimized on {\ovi}, is also consistent with the data. This CBLA would have a temperature of $\log [T/K] = 5.38$, so as not to produce detectable {\civ} absorption, and to fit the wings of the {\lya} line.  For $b(\ovi) \sim 47$~{\kms}, the wings of {\lya} are fit for a metallicity of $\log Z=-1.16^{+0.34}_{-0.33}$. The posterior results for this high ionization phase are summarized in Figure~\ref{fig:OVI0PHL18110.07}. 

\smallskip

\begin{figure*}
\begin{center}
\includegraphics[width=\linewidth]{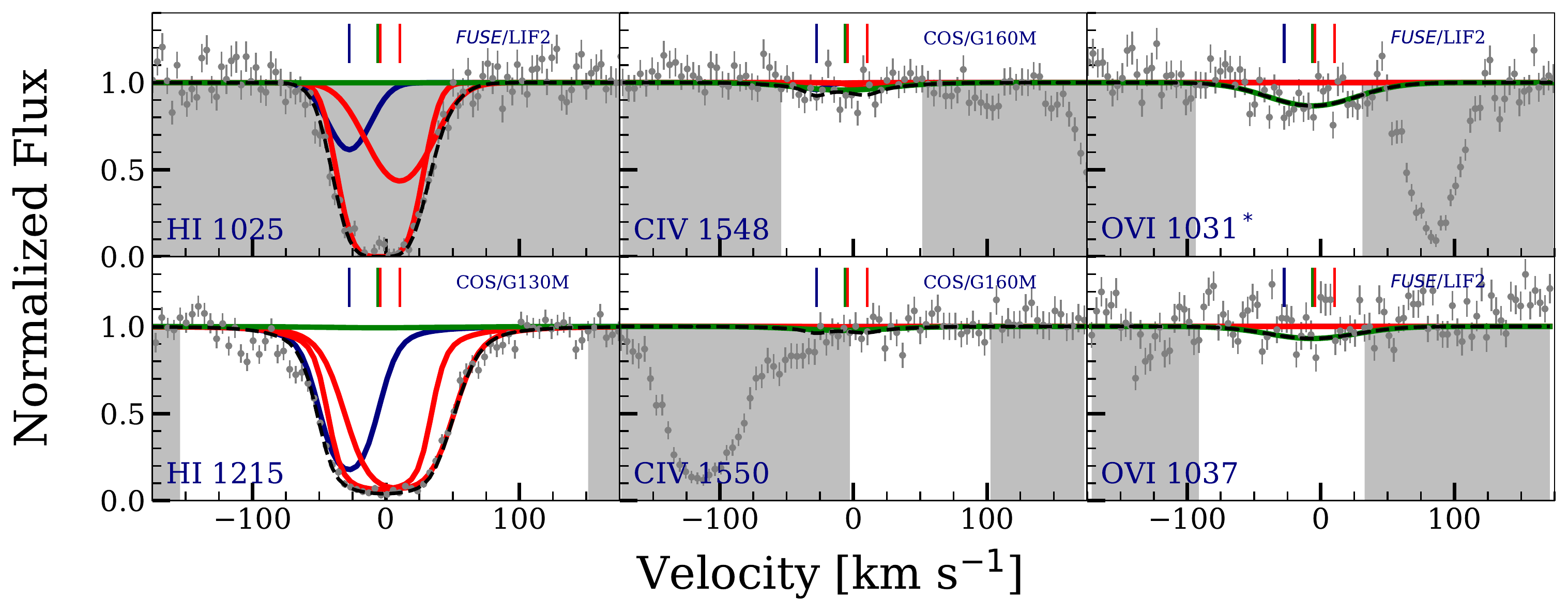}
\caption{{\CLOUDY} photoionization models for some chosen transitions for the $z=0.07776$ absorber towards PHL1811. Symbols, labels, and masked pixels are the same as shown in Figure~\ref{fig:sysplotphl18110.07} but the high phase producing {\ovi} is modeled as a photoionized phase. In such a model the blueward wing of {\lya} is unexplained.}
\label{fig:sysplotphl18110.07phot}
\end{center}
\end{figure*}

We also investigate the alternative of a photoionized phase in {\ovi}. A photoionized model with $\log U = -1.5$ and a supersolar metallicity could fit the {\ovi} profile, and also well explains the observed abundances of other transitions, however the blueward wing of {\lya} is unexplained as shown in Figure~\ref{fig:sysplotphl18110.07phot}. A model with lower ionization parameter could fit the wing of {\lya} but would overproduce the {\civ}. Hence, we favor a collisional ionization model for the {\ovi} phase.

\subsubsection{Comparison to Previous Models}
\label{subsec:0.07776comparisontoprev}
Our modeling reveals three components seen in the low ionization transitions, and required by the observed profiles. The responsible photoionized clouds are found to differ from each other in metallicity, from $\sim$ 1/10th solar metallicity to highly supersolar. There is also a separate, collisionally ionized phase ($\log [T/K] = 5.38$) that produces broad {\ovi} and {\hi} absorption. The total {\hi} column density from our model is $\log [N(\hi)/\cmsq]=15.75$.

\citet{lacki2010z} conducted multiphase modeling for this system, using a wider grid of $\log U$ and $\log Z$, and the HM01~\citep{Haardt2001} EBR. Also, {\civ} appears to be detected in the $HST$/STIS spectrum that they used, which we do not detect in our $HST$/COS coverage. For the blueward cloud in {\ciii}, they determine $\log Z = -1.0$ and $\log U = -1.0$ producing {\civ} and {\ovi}. In contrast to their metallicity, we find $\log Z = 0.85^{+0.11}_{-0.18}$ with a lower ionization parameter of $\log U = -3.11^{+0.08}_{-0.06}$ such that {\civ} is not produced. We find that a low metallicity, $\log Z \sim -1.0$, with our lower ionization parameter, would yield a high temperature phase that violates equation~\ref{eq:3}.

For the redward low ionization absorption, they find a single low metallicity value of $\log Z = -1.0$ and $\log U = -3.0$. In agreement with their result, we determine two separate low metallicities of $\log Z = -0.32^{+0.11}_{-0.11}$ and $\log Z = -0.85^{+0.06}_{-0.06}$ with ionization parameters of $\log U = -3.67^{+0.06}_{-0.05}$ and $\log U = -3.04^{+0.02}_{-0.02}$, respectively.

Other recent models for this system find average values using a single, low ionization component. \citet{keeney2017characterizing} found $\log Z=-0.15$ and $\log U = -3.05$, using $\log [N(\hi)/\cmsq]=15.4$, and using the same HM12 as we have. \citet{muzahid2018cos} measured $\log [N(\hi)/\cmsq]=16.0$ and, using the KS15~\citep{KS15} EBR, found $\log Z=-0.4$ and $\log U=-3.6$. In our models, the centered {\siiii} cloud has a similar metallicity, and produces nearly all of the absorption in higher order Lyman series lines because of its larger {\hi} column density.  The redward {\siiii} cloud is, however, essential to producing the {\lya} absorption in the base of the profile, and is thus required to have a lower metallicity. Our conclusions regarding the presence and properties of a collisionally ionized {\ovi} phase contributing to this system are consistent with those of~\citet{savage2014properties} who determine $\log T = 5.49^{+0.14}_{-0.19}$ and [O/H] = $-1.93^{+1.32}_{-0.11}$.

\subsubsection{Caveats and Tests}
\label{subsec:0.07776caveats}

\begin{figure*}
\begin{center}
\includegraphics[width=\linewidth]{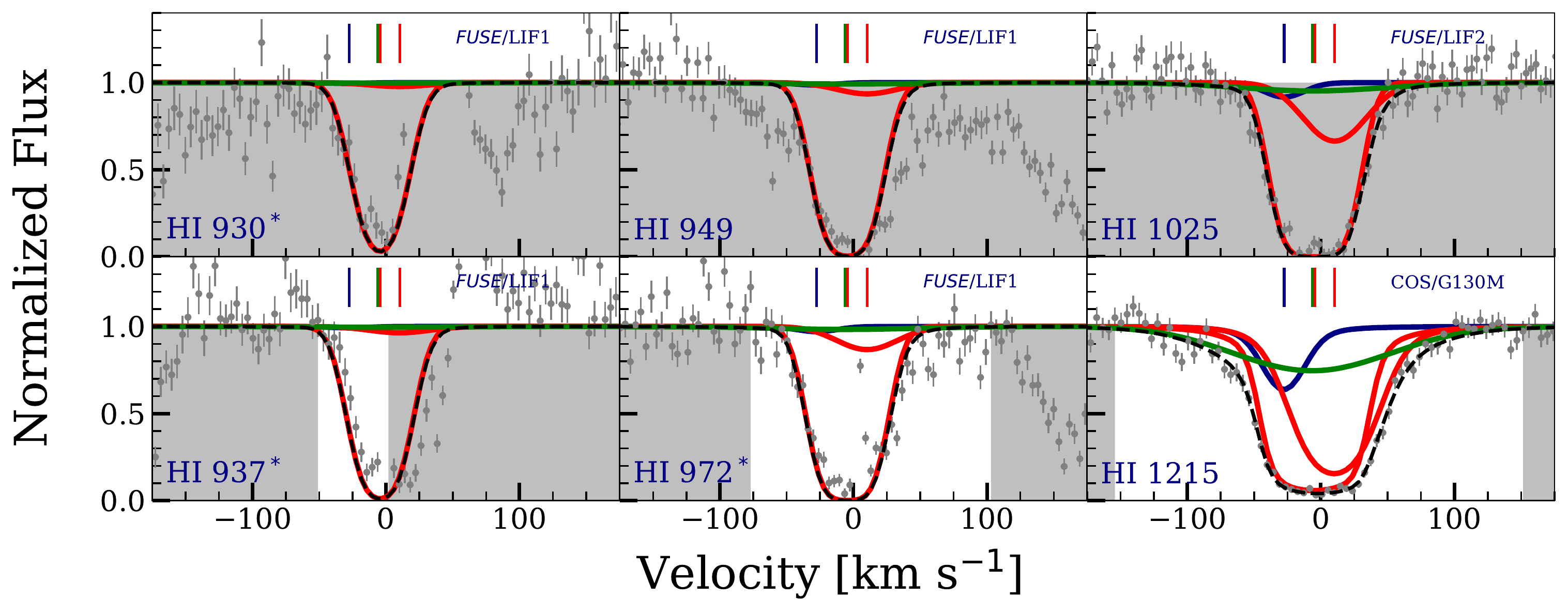}
\caption{{\CLOUDY} models for Lyman lines for the $z = 0.07776$ absorber showing that a ``{\lya} - only'' model very slightly overpredicts the observed higher order Lyman series lines. Symbols, labels, and masked pixels are the same as in Figure~\ref{fig:sysplotphl18110.07}. However, the parameters for this model are consistent with the parameters obtained with the full model.}
\label{fig:Modelsphl18110.07nolyaseries}
\end{center}
\end{figure*}

\begin{figure*}
\begin{center}
\includegraphics[width=\linewidth]{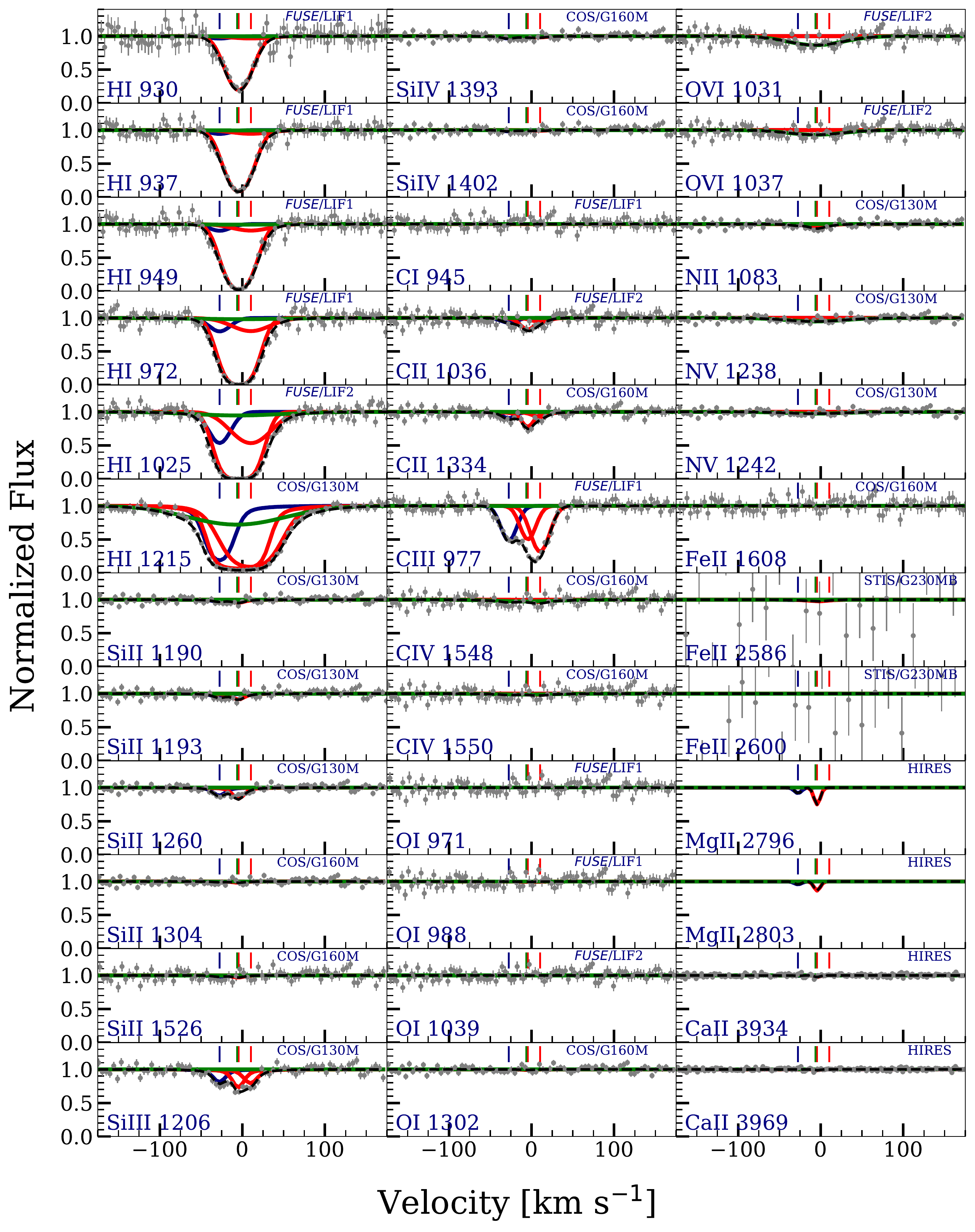}
\caption{{\CLOUDY} models overlaid on a mock spectrum modeled after the $z = 0.07776$ absorber towards PHL1811. Symbols, labels, and curves are the same as in Figure~\ref{fig:sysplotphl18110.07}. We find that our method recovers the parameters with which the spectrum was generated as described in \S~\ref{subsec:0.07776caveats}. The 1$\sigma$ errors on the spectral data correspond to an average error observed in that part of the spectrum.}
\label{fig:simulreal}
\end{center}
\end{figure*}
Though we do not have wavelength coverage of the key {\mgii} transition in any of our spectra for this quasar, we do show the predicted profiles from our favored model in Fig.~\ref{fig:sysplotphl18110.07}. The {\mgii} would most likely classify as a single component, weak {\mgii} absorber with $W_r = 28$ {m\angstrom}. Only the blueward {\siiii} cloud produces detectable {\mgii}, though we do see a slight contribution from the {\ciii} cloud.

The four systems we have chosen to model in this paper, demonstrating our methodology, have richer datasets than the typical low $z$ weak, low ionization absorber~\citep{muzahid2018cos}.  Most importantly, in many cases there is no {\fuse} coverage of the Lyman series lines. To see what the effect would be on constraints, we have run our models considering only the {\lya} line.  Table~\ref{tab:phl18110.07776model} lists the constraints from this experiment.  We find the values for $\log Z$ and $\log U$ to be consistent, for all four clouds, with our model that makes use of all the Lyman series coverage. There is, however, a large uncertainty in the values for the ``{\lya} only'' model, particularly for the blueward {\ciii} cloud.  The ``{\lya} only'' model is shown in Fig.~\ref{fig:Modelsphl18110.07nolyaseries}. The ability to constrain parameters using only {\lya} is important given that many of the systems in the literature only have one to two {\hi} lines, and thus the robustness of the method to limited {\hi} data is very important.

As another test of our methodology, we synthesized artificial spectra modeled after this $z = 0.07776$ absorber. We add Gaussian noise to the modeled flux of the synthetic spectra, and flux uncertainties are based on the average $S/N$ observed in that part of the spectrum. We included one blueward {\ciii} cloud, two blended {\siiii} clouds and a warm/hot {\ovi} cloud, as we found for this absorber. For our experiment, we adopted four different sets of metallicity, $\log Z$, and ionization parameter, $\log U$ values, including a combination similar to the real system shown in Figure~\ref{fig:simulreal}. These parameter choices for our simulations are listed in Table~\ref{tab:simulandrecov} along with the recovered values. In two cases we altered the metallicities in the synthetic spectra, and in one other case we adjusted the ionization parameter of the blueward {\ciii} and the blended redward {\siiii} clouds while keeping the parameters of {\ovi} cloud fixed at the values of the best fit model.  In all cases we were able to recover the correct model parameters, with the errors, that we specified for the synthetic spectra. This provides confidence that our CMBM approach is recovering accurate results from the observed spectra. 

\begin{table}
\caption{\bf Simulated and recovered values \label{tab:simulandrecov}}
\noindent  
\centering
{\begin{tabular}{p{0.8cm}p{0.5cm}p{0.5cm}p{0.5cm}p{1.0cm}p{1.0cm}} \hline \nonumber
& & Simulated  &  & Recovered & \\ 
Test ID & Phase & $\log Z$ & $\log U$ & $\log Z$ & $\log U$\\\hline
A &  \textbf{\textcolor{blue}{\ciii}} & $0.85$ & $-3.13$ & $0.87^{+0.09}_{-0.09}$ & $-3.16^{+0.04}_{-0.04}$ \\
 &  \textbf{\textcolor{red}{\siiii}} & $-0.34$ & $-3.66$ & $-0.31^{+0.05}_{-0.06}$ & $-3.69^{+0.04}_{-0.03}$ \\
 &  \textbf{\textcolor{red}{\siiii}} & $-0.86$ & $-3.04$ & $-0.88^{+0.04}_{-0.05}$ & $-3.03^{+0.01}_{-0.01}$ \\\hline
 
B &  \textbf{\textcolor{blue}{\ciii}} & 0.10 & $-3.13$ & $0.08^{+0.20}_{-0.45}$ & $-3.12^{+0.12}_{-0.10}$ \\
 &  \textbf{\textcolor{red}{\siiii}} & 0.10 & $-3.66$ & $0.13^{+0.05}_{-0.06}$ & $-3.66^{+0.02}_{-0.02}$ \\
 &  \textbf{\textcolor{red}{\siiii}} & 0.10 & $-3.04$ & $0.15^{+0.13}_{-0.19}$ & $-3.03^{+0.11}_{-0.09}$ \\\hline

C &  \textbf{\textcolor{blue}{\ciii}} & $-1.0$ & $-3.20$ & $-0.73^{+0.27}_{-0.27}$ & $-3.31^{+0.11}_{-0.13}$ \\
 &  \textbf{\textcolor{red}{\siiii}} & $0.0$ & $-4.00$ & $0.01^{+0.04}_{-0.04}$ & $-4.01^{+0.02}_{-0.02}$ \\
 &  \textbf{\textcolor{red}{\siiii}} & $0.75$ & $-4.00$ & $0.75^{+0.01}_{-0.01}$ & $-4.00^{+0.01}_{-0.01}$ \\\hline

D &  \textbf{\textcolor{blue}{\ciii}} & $1.00$ & $-3.00$ & $0.94^{+0.22}_{-0.17}$ & $-3.05^{+0.13}_{-0.09}$ \\
 &  \textbf{\textcolor{red}{\siiii}} & $-1.50$ & $-3.68$ & $-1.48^{+0.11}_{-0.20}$ & $-3.69^{+0.12}_{-0.10}$ \\
 &  \textbf{\textcolor{red}{\siiii}} & $1.00$ & $-3.06$ & $1.14^{+0.19}_{-0.24}$ & $-3.03^{+0.15}_{-0.17}$ \\

\hline 
\end{tabular}} \\

The results of testing the algorithms' ability to recover the information of simulated spectra. Test A corresponds to a synthesized spectrum with the parameters determined for the $z=0.07776$ absorber towards PHL1811. The models superimposed on the data are shown in Figure~\ref{fig:simulreal} for Test A. Test B, C, and D correspond to different combinations of metallicity and ionization parameter; in all the cases we are able to recover expected values within the errors.

\end{table}

The final consideration for this absorber is the relatively large uncertainties in the VP fit parameters for the blended optimized {\siiii} clouds and for the {\ciii} cloud.   The best VP fit for the {\siiii} yielded $b = 2.6$~{\kms} for the blueward component and $b=3.7$~{\kms} for the redward component, and for {\ciii} we found $b = 3.2$~{\kms}, as noted in Table~\ref{tab:phl18110.07776model}. The photoionization models based on these best VP fit parameters for the blueward {\ciii}, and the two redward {\siiii} produce high temperatures that are not self-consistent with equation~\ref{eq:3}. We resolved the problem by adopting higher $b$ values within the limits of 2$\sigma$ uncertainty. If we did adopt the best VP fit $b$ parameters, {\CLOUDY} models would find consistent values of $\log Z$ and $\log U$ for the {\siiii} clouds, but would find a drastically lower metallicity (of $\log Z = -1.36$) for the {\ciii} cloud. This indicates that in the absence of higher resolution coverage of an optimized transition, there is some level of systematic uncertainty in the derived model parameters. The possible effects of different $b$ values of blended components should always be considered.

\subsubsection{Galaxy Properties and Physical Interpretation}
\label{subsec:0.07776galax}
The closest galaxy to the sightline, at a coincident redshift, is a quiescent, inclination $i=50\deg$, $L_g = 0.26 L^*$ galaxy at an impact parameter of 237~kpc~\citep{keeney2017characterizing}. It is unlikely that the galaxy itself would produce significant absorption for impact parameters $>$200 kpc~\citep{liangchen2014}, however~\citet{muzahid2018cos} notes that there are 7 galaxies within 1 Mpc and 500 {\kms} of the absorber. This suggests a group environment.

\smallskip

Only one of the three low ionization clouds in this system, the middle one with $\log N({\hi})=15.7$, would produce detectable weak {\mgii} absorption. With a metallicity of $\log Z=-0.32$, and a size of $\sim 30$~pc, it is consistent with the weak, low ionization absorber population at intermediate redshift \cite{rigby2002population,charlton2003high,misawa2008supersolar}.  However, those absorbers often have a lower density, {\civ} phase aligned in velocity with the {\mgii}. The blueward cloud is also very small and has a supersolar metallicity. The redward cloud, however, has a lower metallicity of $\log Z=-0.85$, and a larger line-of-sight thickness of $\sim 500$~pc.
This cloud would be consistent with lines of sight through infalling gas in filamentary structures as seen in numerical simulations~\citep{Hafen2019}.  It is unclear if the three clouds are structurally related, despite their close proximity in velocity space.

\smallskip

Because of the lack of detected {\civ}, this absorber is similar to the well studied $z=0.0053$ Virgo cluster absorber found toward 3C 273~\citep{stocke2004discovery}. Though they are not common~\citep{milutinovic2006nature}, there is reason to believe that the lower amplitude EBR will lead to some systems with less prominent {\civ} absorption \citep{narayanan2005survey}. In this absorber, though {\civ} is not detected, we do see collisionally ionized gas giving rise to {\ovi} absorption, a transition not often covered for the higher redshift systems. A survey of low redshift, weak low ionization absorbers is needed to see if such warm/hot gas is commonly found, perhaps arising in an intragroup medium, such as it is observed in the Local group \citep{bouma19}, or at the interface between cooler clouds and hotter surrounding regions \citep{Oppenheimer2016,Pointon2017}.

\subsection{The $z = 0.08094$ absorber towards the quasar PHL1811}
\label{sec:z = 0.08094}

\begin{table*}
\renewcommand\thetable{4}
\begin{center}
\caption{\bf The gas phases in the $z = 0.08094$ absorber towards PHL1811}
\label{tab:phl18110.08modelpar}

\begin{tabular}{c@{\hspace{0.25\tabcolsep}}c@{\hspace{0.55\tabcolsep}}c@{\hspace{0.55\tabcolsep}}c@{\hspace{0.55\tabcolsep}}c@{\hspace{0.55\tabcolsep}}@{\vline}c@{\hspace{0.55\tabcolsep}}c@{\hspace{0.55\tabcolsep}}c@{\hspace{0.55\tabcolsep}}c@{\hspace{0.55\tabcolsep}}c@{\hspace{0.55\tabcolsep}}c@{\hspace{0.55\tabcolsep}}@{\vline}@{\hspace{0.55\tabcolsep}}c@{\hspace{0.55\tabcolsep}}c@{\hspace{0.55\tabcolsep}}c@{\hspace{0.55\tabcolsep}}} \hline \hline

& &  & & & & & Lyman lines & & & & & {\lya}-only  & \\  
\hline
Optimized & $z$ & $b$ & $b_{used}$ & $\log \frac{N}{\cmsq}$ & $\log Z$ & $\log U$ & $\log \frac{n(H)}{\cc}$ & $\log \frac{N(\hi)}{\cmsq}$ & $\log \frac{L}{kpc} $ & $\log \frac{T}{K} $ & $\log Z$ & $\log U$ & $\log \frac{N(\hi)}{\cmsq}$\\  
ion & & (\kms) & (\kms) & & & & & & & & & &\\
(1) & (2) & (3) & (4) & (5) & (6) & (7) & (8) & (9) & (10) & (11) & (12) & (13) & (14)\\\hline
\textcolor{blue}{\ciii_0} & 0.08057 & $16.6^{+6.0}_{-4.2}$ & 12.0 & $12.88^{+0.08}_{-0.09}$ & $0.21^{+0.07}_{-0.07}$ & $-3.87^{+0.03}_{-0.03}$  & $-2.46^{+0.03}_{-0.03}$ & $15.34^{+0.04}_{-0.03}$ & $-2.35^{+0.06}_{-0.07}$ & $3.96^{+0.04}_{-0.04}$ & $-0.09^{+0.56}_{-0.61}$ & $-3.10^{+0.11}_{-0.59}$ & $14.27^{+0.63}_{-0.19}$\\    

\textcolor{orange}{\civ_0} & 0.08074 & $18.0^{+6.0}_{-4.7}$ & 12.0 & $13.44^{+0.07}_{-0.08}$ & $-2.02^{+0.01}_{-0.03}$ & $-3.03^{+0.01}_{-0.01}$  & $-3.29^{+0.01}_{-0.01}$ & $15.55^{+0.01}_{-0.01}$ & $1.80^{+0.01}_{-0.01}$ &  $4.90^{+0.01}_{-0.01}$ & $-1.04^{+0.04}_{-0.01}$ & $-2.71^{+0.01}_{-0.05}$ & $14.27^{+0.02}_{-0.05}$\\

\textcolor{red}{\civ_1} & 0.08092 & $13.0^{+2.6}_{-2.4}$ & 13.0 & $13.87^{+0.07}_{-0.06}$ & $0.63^{+0.14}_{-0.12}$ & $-2.83^{+0.03}_{-0.03}$  & $-3.49^{+0.03}_{-0.03}$ & $15.28^{+0.14}_{-0.18}$ & $-0.16^{+0.15}_{-0.15}$ &  $4.25^{+0.05}_{-0.05}$ & $0.63^{+0.12}_{-0.12}$ & $-2.86^{+0.03}_{-0.03}$ & $15.34^{+0.13}_{-0.14}$\\

\textcolor{darkgreen}{\siii_0} & 0.08092 & $6.1^{+0.9}_{-0.9}$ & 5.00 & $14.02^{+0.17}_{-0.10}$ & $-0.30^{+0.02}_{-0.02}$ & $-3.81^{+0.04}_{-0.04}$  & $-2.52^{+0.04}_{-0.04}$ & $18.03^{+0.02}_{-0.02}$ & $0.24^{+0.08}_{-0.04}$ &  $4.12^{+0.01}_{-0.01}$ & $-0.29^{+0.01}_{-0.03}$ & $-3.84^{+0.06}_{-0.03}$ & $18.06^{+0.02}_{-0.02}$ \\

\hline 

\end{tabular} 
\end{center}

Properties of the different gas phases present in the $z = 0.08094$ absorber towards PHL1811 traced by their respective optimized ions. Notes: (1) Optimized ion tracing a phase; (2) Redshift of the component; (3) Doppler parameter of optimized ion; (4) Adopted Doppler parameter for optimized ion; (5) log column density of optimized ion; (6) log metallicity;  (7) log ionization parameter; (8) log hydrogen number density; (9) log hydrogen column density (10) log thickness in kpc; (11) log temperature in Kelvin. The marginalized posterior values of model parameters are given as the median along with the upper and lower bounds associated with a 95\% credible interval. The synthetic profiles based on these models are shown in Figure~\ref{fig:Modelsphl18110.08}. The marginalized posterior distributions for the VP fit parameters (columns 2, 3, and 5) of the optimized ions are presented in Figures~\ref{fig:voigtCIII0phl18110.08}, \ref{fig:voigtCIVphl18110.08}, and \ref{fig:voigtSiIIphl18110.08}. The marginalized posterior distributions for the cloud properties (columns 6, 7, 8, 9, 10, and 11) are presented in Figures~\ref{fig:CIII0PHL18110.08}, \ref{fig:CIV0PHL18110.08}, \ref{fig:CIV1PHL18110.08}, and \ref{fig:SiIIPHL18110.08}. Columns (12), (13), and (14) describe the marginalized posterior distributions of log metallicity, log ionization parameter, and log hydrogen column density, for a {\lya}-only model described in \S~\ref{subsec:0.08094caveat}.
\end{table*}

\begin{figure*}
\begin{center}
\includegraphics[width=\linewidth]{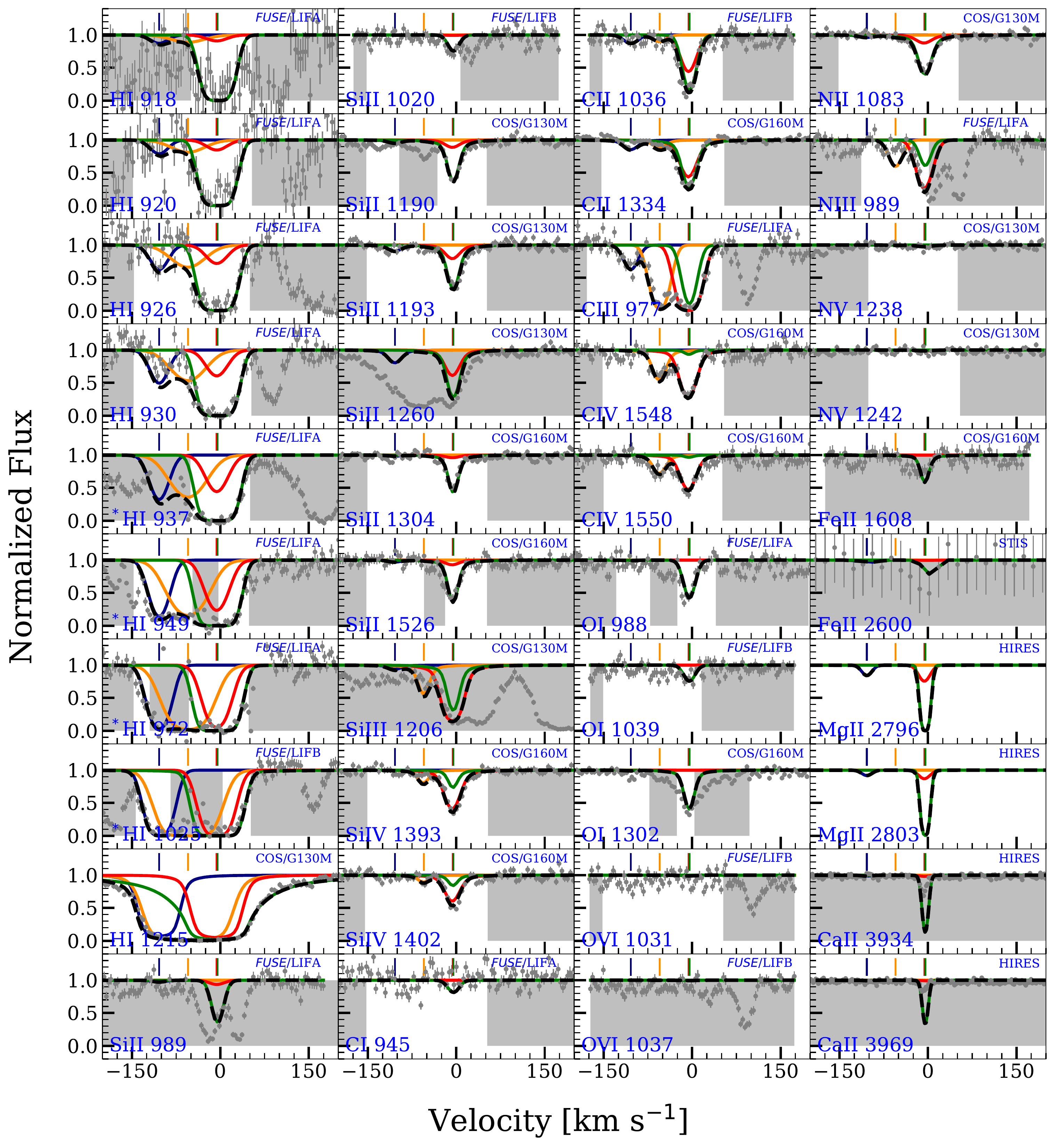}
\caption{\CLOUDY~models for the $z = 0.08094$ absorber towards the quasar PHL1811 obtained using the MLE values. The spectral data are shown in gray along with the 1$\sigma$~error. The instruments used for observation in different regions of the spectrum are indicated in the panels. The centroids of absorption components as determined from the VP fits for optimized ions are indicated by the vertical tick marks on top of each line. The absorption system is modeled using four clouds. A low ionization phase traced by the blueward {\ciii} cloud (shown as blue curve), an intermediate ionization phase traced by {\civ} (shown as yellow curve), and two coincident phases - a high ionization phase traced by {\civ} (shown as red curve) and a low ionization phase traced by {\siii} (shown as green curve). The superposition of these two models are shown by the black dashed curve. The region shaded in grey shows the pixels that were not used in the evaluation of the log-likelihood. The lines {\hi} $\lambda$937, {\hi} $\lambda$949, {\hi} $\lambda$972, and {\hi} $\lambda$1025 are corrected for blending as described in \S~\ref{sec:z = 0.08094} and earmarked with an asterisk.
}
\label{fig:Modelsphl18110.08}
\end{center}
\end{figure*}

Spectral regions covering constraining transitions for the $z=0.08094$ absorber, from {\fuse}, $HST$/COS, $HST$/STIS, and Keck/HIRES, are shown in Fig.~\ref{fig:Modelsphl18110.08}. The low ionization absorption in {\siii} and \cii~shows two components, with the redward being the strongest. The system might classify as a multiple component weak {\mgii} absorber, but the {\mgii} transitions were not covered to confirm that expectation. The \hi~$\lambda$1025, \hi~$\lambda$972, \hi~$\lambda$949, and \hi~$\lambda$937 are affected by blending due to the Galactic molecular hydrogen lines. \hi~$\lambda$1025 is contaminated by H$_{2}$:L 0$-$0 R(1) $\lambda$1108.63 at $-32$ {\kms}, \hi~$\lambda$972 by H$_{2}$:L 4$-$0 P(1) $\lambda$1051.03 at $-78$ {\kms} and by H$_{2}$:L 4$-$0 R(2)$\lambda$1052.49 at 56 {\kms}, \hi~$\lambda$949 by H$_{2}$:L 6$-$0 R(2)$\lambda$1026.53 at $-63$ {\kms}, and \hi~$\lambda$937 by H$_{2}$:L 7$-$0 R(1)$\lambda$1013.44 at $-95$ {\kms}. We divide out the contribution of these lines by making use of several other Galactic hydrogen molecular lines observed in the {\fuse} spectrum. 

\smallskip

This system was previously analyzed by \citep{jenkins2003absorption,jenkins05,richter2016,richter2020}. The strongest low ionization component is also detected in \oi, \nii, \feii, and \caii. There are three components detected in \ciii, and the two redward components in {\civ} as well. In a noisy part of the {\fuse} spectrum, {\ovi} provides only an upper limit. Lyman series lines, from {\lya} down to \hi~$\lambda$920, provide useful constraints on the metallicities of the three components, with the blueward component clearly separated in the higher order lines. The metallicity of the strongest, redward component is best constrained by the redward side of all the Lyman series profiles. A full Lyman break is observed at this redshift in the {\fuse} coverage \citep{jenkins2003absorption}. Table~\ref{tab:phl18110.08modelpar} summarizes the constraints placed on the metallicities, ionization parameters, number densities, neutral hydrogen column densities, temperatures, and thickness for the four clouds we find necessary to explain the multiphase absorption. 

\smallskip

Since {\ciii} is saturated for the strongest component, and several {\siii} lines were covered, the model is optimized on the measured {\siii} column density. This cloud was constrained to have metallicity, $\log Z=-0.30$, and gives rise to the Lyman limit break. It also fits the distinctive ``wing'' on the redward side of the {\lya} profile, and also produces a small amount of absorption consistent with the smaller ``wing'' on the blueward side. For an ionization parameter of $\log U=-3.81$, the observed \oi, \feii, and {\nii} are fully produced, but all the higher ionization transitions are underproduced, calling for a separate, higher-ionization phase at this same velocity. The {\caii} was not included as a model constraint because it is often depleted onto dust \citep{Richter2011}, and we do find it to be overproduced by 0.75 dex by the $\log U=-3.8$ model. The large $N(\hi)$ needed to reproduce this relatively strong profile results in a cloud thickness of 1.7~kpc.

\smallskip

To account for the \civ, \siv, \ciii, \siii, \niii, and \cii, coincident with the strongest {\siii} component, we included an additional cloud, optimized on the {\civ} at this velocity.  This cloud has the highest ionization parameter of $\log U = -2.8$ among the absorbers present in this system, this cloud can produce the observed absorption in all these transitions.  Its metallicity must be relatively high (we obtain $\log Z = 0.6$) in order that it does not affect the fit to the Lyman series lines.  The thickness of this cloud is $0.7$~kpc.

\smallskip

For the middle component, the optimized transition is \civ, since the {\ciii} is saturated, and detections are weak or absent in the low ionization transitions. This cloud has an ionization parameter of $\log U = -3.0$. A metallicity of $\log Z=-2.0$ allows this cloud to fill in the Lyman series profiles in between the contributions from the other clouds, which themselves are well constrained from the blueward and redward sides of the profiles.  Despite its much lower $\log [N(\hi)/\cmsq]$ of 15.5, relative to the strongest cloud, the low metallicity leads to a similar cloud thickness of 63~kpc. This model, however, overpredicts the absorption in \niii, but an abundance pattern variation from solar could resolve this discrepancy.

\smallskip

The third, blueward cloud model is optimized on {\ciii} since it is at most weakly detected in \siii~and \cii. Because {\civ} is not detected, this cloud has a lower ionization parameter of $\log U = -3.9$. A supersolar metallicity ($\log Z=0.2$) is required in order that Lyman series lines are not overproduced. Though the $\log [N(\hi)/\cmsq] = 15.34$ is similar to that for the middle cloud, the thickness is several orders of magnitude smaller, only 5 parsecs. This contrast in thickness is striking, and perhaps surprising, but it seems to be required by the observed profiles.

\smallskip

In Table~\ref{tab:phl18110.08modelpar} we present the posterior results for different phases, which are summarized in Figures~\ref{fig:CIII0PHL18110.08}, \ref{fig:CIV0PHL18110.08}, \ref{fig:CIV1PHL18110.08}, and \ref{fig:SiIIPHL18110.08} for the four clouds.

\subsubsection{Comparison to Other Models}
\label{subsec:0.08094comparison}

It is apparent from our models that the four clouds needed to explain the low ionization absorption profiles have significantly different properties from one another, pointing to different physical origins of the gas. The metallicities of these clouds are found to be $\log Z=-0.3$ for the strongest low ionization component, $\log Z = 0.6$ for the coincident higher ionization component, a very low metallicity of $\log Z=-2.0$ for the middle cloud, and $\log Z=0.2$ for the weak, blueward component. The redward, strong low ionization component also contributes the bulk of the {\hi} absorption and produces a Lyman limit break, as well as giving rise to the wings on the {\lya} profile. \citet{richter2020} has modeled the right, broad wing of the {\lya} absorption as CBLA that traces the collisionally ionized, coronal gas of the CGM host galaxy, however, we find that the wing is explained by the redward {\siii} cloud.

The middle cloud has $\log U = -3.0$, while the redward cloud has $\log U = -3.80$. Another startling difference is perhaps the thicknesses of the clouds, with the lower density, high metallicity cloud that is only 5 pc thick, while the redward spans a few kpc. 

Previous models by \citet{keeney2017characterizing}, \citet{muzahid2018cos}, and \citet{wotta2019cos} which average the components together infer metallicities and ionization parameters consistent (considering differences in EBR models) with those we obtain for the strongest lower ionization component.  However, they miss the richness of information available in the profile about two other regions along the sightline which have vastly different properties and physical origins.

\subsubsection{Caveats and Tests}
\label{subsec:0.08094caveat}
\begin{table*}
\renewcommand\thetable{5}
\begin{center}
\caption{\bf The influence of possible uncertainty in {\CLOUDY} equilibrium temperature}
\label{tab:cloudyeqtemp}

\begin{tabular}{c@{\hspace{1.00\tabcolsep}}@{\vline}c@{\hspace{1.00\tabcolsep}}c@{\hspace{1.00\tabcolsep}}@{\vline}c@{\hspace{1.00\tabcolsep}}c@{\hspace{1.00\tabcolsep}}@{\vline}c@{\hspace{1.00\tabcolsep}}c@{\hspace{1.00\tabcolsep}}} \hline \hline

 & 0.9T  &  & T & & 1.1T \\ \hline
 Phase & $\log Z$ & $\log U$ & $\log Z$ & $\log U$ & $\log Z$ & $\log U$\\\hline
  \textbf{\textcolor{blue}{\ciii\_0}} & $0.16^{+0.07}_{-0.08}$ & $-3.86^{+0.03}_{-0.03}$ & $0.21^{+0.07}_{-0.07}$ & $-3.87^{+0.03}_{-0.03}$ & $0.27^{+0.08}_{-0.07}$ & $-3.87^{+0.03}_{-0.02}$ \\
  \textbf{\textcolor{orange}{\civ\_0}} & $-2.02^{+0.01}_{-0.01}$& $-3.03^{+0.01}_{-0.01}$ & $-2.02^{+0.01}_{-0.01}$& $-3.03^{+0.01}_{-0.01}$ & $-2.02^{+0.01}_{-0.01}$& $-3.03^{+0.01}_{-0.01}$\\
  \textbf{\textcolor{orange}{\civ\_1}} & $0.52^{+0.15}_{-0.12}$ & $-2.82^{+0.04}_{-0.03}$ & $0.63^{+0.14}_{-0.12}$ & $-2.83^{+0.03}_{-0.03}$ & $0.65^{+0.14}_{-0.11}$ & $-2.85^{+0.03}_{-0.03}$\\
 \textbf{\textcolor{darkgreen}{\siii\_0}} & $-0.36^{+0.02}_{-0.02}$ & $-3.69^{+0.03}_{-0.07}$& $-0.30^{+0.02}_{-0.02}$ & $-3.81^{+0.04}_{-0.04}$ & $-0.22^{+0.01}_{-0.07}$ & $-3.95^{+0.11}_{-0.03}$\\\hline

\end{tabular} \\
\end{center}
The results of testing the influence of possible uncertainties in {\CLOUDY} equilibrium temperature on model parameters. We find that a 10\% change in temperature from the value given by {\CLOUDY} still produces qualitatively similar results.

\end{table*}

\begin{figure*}
\begin{center}
\includegraphics[width=\linewidth]{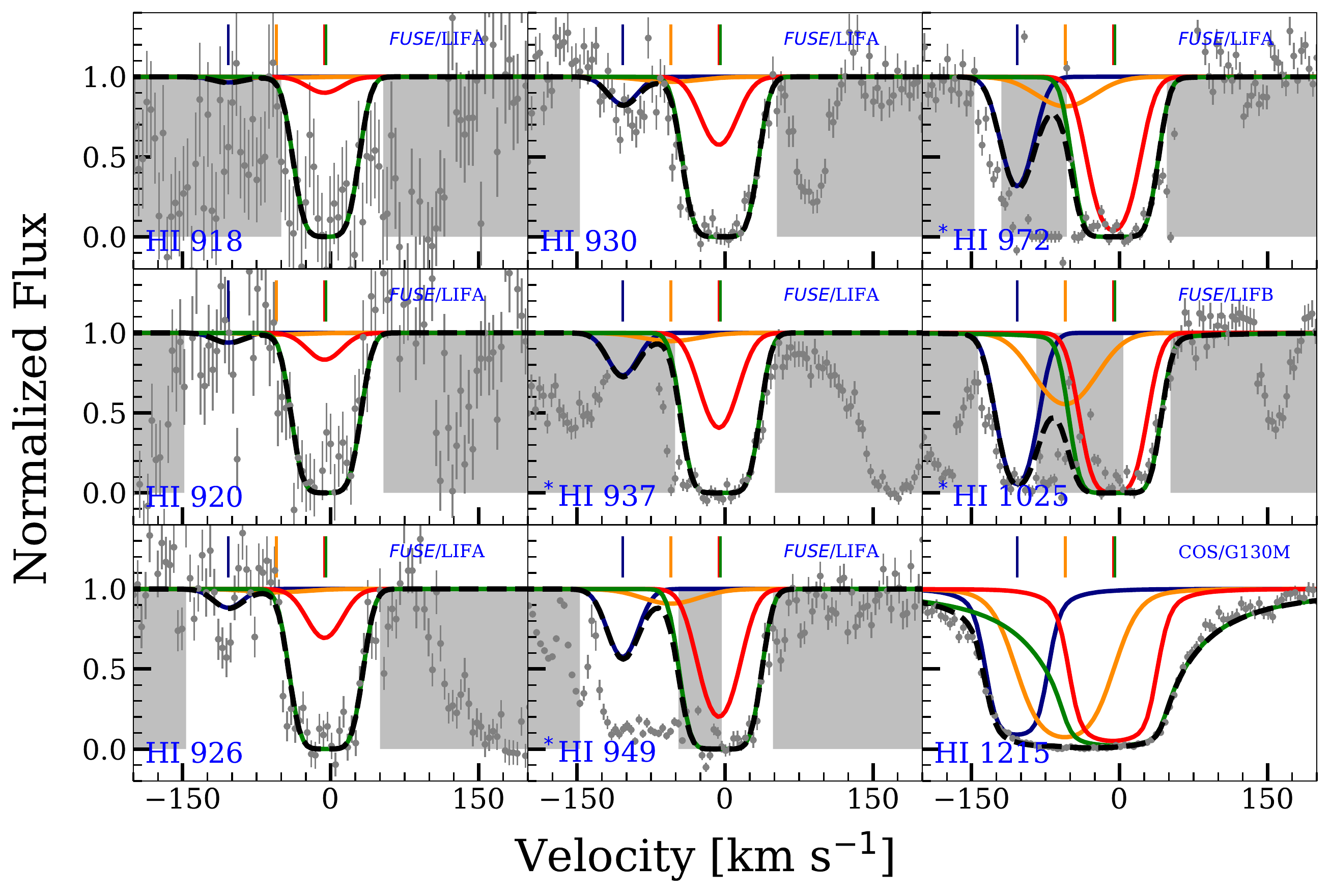}
\caption{{\CLOUDY} models for Lyman lines for the $z = 0.08096$ absorber towards PHL1811 showing that a ``{\lya} only'' model predicts well the properties of all but the middle {\civ} cloud (yellow curve). Symbols, labels, and masked pixels are the same as in Figure~\ref{fig:Modelsphl18110.08}.}
\label{fig:Modelsphl18110.08lyaseries}
\end{center}
\end{figure*}

For this system, we predict the strength of {\mgii}. The equivalent width predicted for {\mgii} is $W_r(2796) = 243$~{m\AA}, dominated by the redward cloud. 

We also conduct a test for this system to determine how much information can be derived using a ``{\lya}~only'' model.   Table~\ref{tab:phl18110.08modelpar} summarizes these constraints for the ``{\lya} only'' model compared to the original one.  Even without access to the Lyman series lines, the ionization parameters and metallicities of all but the middle {\civ} clouds agree well within errors. This disagreement arises due to the fact that the higher order lines provide the information necessary to produce the shape of the profile. The ``{\lya}-only'' model is shown in Fig.~\ref{fig:Modelsphl18110.08lyaseries}.

Central to our ability to discern accurate model parameters, even from saturated {\hi} lines, is the use of the {\CLOUDY} model equilibrium temperature in order to determine the thermal/turbulent contributions to the $b$ parameter, and thus to infer $b(\hi)$ for each component.  We thus investigated for this system the effect of uncertainties in the temperature given by {\CLOUDY}, considering values 10\% smaller and 10\% larger for all four clouds. In order to do this, for each {\CLOUDY} model we took the output temperature and artificially adjusted it before synthesizing model spectra.  When these model spectra, which therefore were produced with different $b$ parameters for non-optimized transitions, were compared to the data, different $\log Z$ and $\log U$ parameters were found to compensate for the different $b$ values. Results are summarized in Table~\ref{tab:cloudyeqtemp}.  This table shows that the effects of this assumed uncertainty \textbf{are} quite small.  As expected, we find that a lower temperature will lead to a larger thermal contribution and a larger $b(\hi)$, and thus a slightly smaller inferred metallicity, while a higher temperature will lead to a slightly larger metallicity.  Results are overall not very sensitive to the equilibrium {\CLOUDY} temperatures.

\subsubsection{Galaxy Properties and Physical Interpretation}
\label{subsec:0.08094galax}
The closest galaxy to the sightline at this redshift is an $L_{g}=0.56L^*$ early type galaxy at an impact parameter of 35 kpc from the sightline~\citep{keeney2017characterizing}.  Although the Sloan survey does not show many other galaxies within 1 Mpc and 500 {\kms} of the line of sight~\citep{muzahid2018cos}, there is one close companion galaxy with $L\sim L^*$ at an impact parameter of 87 kpc~\citep{jenkins05}, which is possibly interacting with the nearer galaxy.

Our inference of four very different gas clouds points toward there being different processes at work along the sightline. The presence of two nearby galaxies suggests that there could be gas associated with each of the two different galaxies, or with debris that resulted from their interaction. There also could be gas associated with the group of galaxies containing the pair and also possibly dwarf galaxies to these L$^{*}$/ sub-L$^{*}$ galaxies, though no other bright galaxies are present besides the two.  

It is very interesting that the smallest of the four clouds is a fairly typical, supersolar, weak, low ionization absorber. These are typically not associated closely with specific galaxies, however in this case there is a galaxy within an impact parameter of 35 kpc.  The lowest metallicity cloud ($\log Z=-2.0$) could be gas infalling into the galaxy pair from a more pristine surrounding environment, consistent with expected values for IGM accretion metallicities from simulations, which yield values between \hbox{$-3 < \log Z < -1.5$} \citep{Hafen2019}. In this case, there is no evidence of hot gas that produces broad \ovi~absorption, consistent with the previous studies of group environments~\citep{Oppenheimer2016,Pointon2017}. Instead, the wings of the {\lya} profile can be fully explained by the low ionization cloud that produces a full Lyman break.

\subsection{The $z = 0.13849$ absorber towards the quasar PG1116+215}
\label{sec:z = 0.13849}

\begin{table*}
\renewcommand\thetable{6}
\begin{center}
\caption{\bf The gas phases in the $z = 0.13849$ absorber towards PG1116+215}
	\label{tab:pg1116modelpar}    
\begin{tabular}{c@{\hspace{0.25\tabcolsep}}c@{\hspace{0.55\tabcolsep}}c@{\hspace{0.55\tabcolsep}}c@{\hspace{0.55\tabcolsep}}c@{\hspace{0.55\tabcolsep}}@{\vline}c@{\hspace{0.55\tabcolsep}}c@{\hspace{0.55\tabcolsep}}c@{\hspace{0.55\tabcolsep}}c@{\hspace{0.55\tabcolsep}}c@{\hspace{0.55\tabcolsep}}c@{\hspace{0.55\tabcolsep}}@{\vline}@{\hspace{0.55\tabcolsep}}c@{\hspace{0.55\tabcolsep}}c@{\hspace{0.55\tabcolsep}}c@{\hspace{0.55\tabcolsep}}} \hline \hline

& &  & & & & & Lyman lines & & & & & {\lya}-only  & \\  
\hline
Optimized & $z$ & $b$ & $b_{used}$ & $\log \frac{N}{\cmsq}$ & $\log Z$ & $\log U$ & $\log \frac{n(H)}{\cc}$ & $\log \frac{N(\hi)}{\cmsq}$ & $\log \frac{L}{kpc} $ & $\log \frac{T}{K} $ & $\log Z$ & $\log U$ & $\log \frac{N(\hi)}{\cmsq}$\\  
ion & & (\kms) & (\kms) & & & & & & & & & &\\
(1) & (2) & (3) & (4) & (5) & (6) & (7) & (8) & (9) & (10) & (11) & (12) & (13) & (14)\\\hline

\textcolor{blue}{\mgii_0} & 0.13846 & $2.5^{+0.4}_{-0.4}$ & 2.9 &  $12.39^{+0.06}_{-0.05}$ & $0.26^{+0.05}_{-0.04}$ & $-3.43^{+0.04}_{-0.05}$  & $-2.79^{+0.05}_{-0.04}$ & $15.70^{+0.04}_{-0.05}$ & $-1.09^{+0.06}_{-0.06}$ & $4.09^{+0.02}_{-0.02}$ & $0.44^{+0.10}_{-0.11}$ & $-3.38^{+0.06}_{-0.05}$ & $15.53^{+0.11}_{-0.11}$\\    

\textcolor{red}{\mgii_1} & 0.13850 & $1.7^{+0.5}_{-0.3}$ & 1.7 & $12.26^{+0.11}_{-0.08}$ & $0.82^{+0.19}_{-0.11}$ & $-4.14^{+0.19}_{-0.09}$  & $-2.08^{+0.09}_{-0.19}$ & $15.33^{+0.09}_{-0.20}$ & $-3.44^{+0.23}_{-0.14}$ &  $2.99^{+0.21}_{-0.21}$ &$1.28^{+0.34}_{-0.57}$ & $-3.86^{+0.25}_{-0.41}$ & $14.86^{+0.60}_{-0.28}$ \\

\textcolor{orange}{\siiv_0} & 0.13847 & $5.4^{+6.7}_{-2.9}$ & 6.0 & $12.88^{+0.33}_{-0.12}$ & $-1.90^{+0.07}_{-0.07}$ & $-3.07^{+0.04}_{-0.03}$  & $-3.15^{+0.04}_{-0.04}$ & $15.97^{+0.02}_{-0.02}$ & $1.76^{+0.11}_{-0.09}$ &  $4.80^{+0.01}_{-0.01}$ & $-1.65^{+0.10}_{-0.11}$ & $-3.08^{+0.03}_{-0.02}$ & $15.83^{+0.07}_{-0.08}$ \\

\textcolor{darkgreen}{\ovi_0} & 0.13848 & $34.7^{+3.2}_{-3.2}$ & 34.7 & $13.82^{+0.03}_{-0.03}$ & $-1.06^{+0.05}_{-0.09}$ & - & $-3.90$  & $13.65^{+0.05}_{-0.05}$ & - &  $5.38^{+0.01}_{-0.01}$ & $-1.14^{+0.06}_{-0.07}$ & - & $13.73^{+0.05}_{-0.05}$\\

\hline 

\end{tabular} 
\end{center}

Properties of the different gas phases present in $z = 0.13849$ absorber towards PG1116+215 traced by their respective optimized ions. Notes: (1) Optimized ion tracing a phase; (2) Redshift of the component; (3) Doppler parameter of optimized ion (4) Adopted Doppler parameter for the optimized ion (5) log column density of optimized ion; (6) Metallicity;  (7) log ionization parameter; (8) log hydrogen number density; (9) log hydrogen column density (10) log thickness in kpc; (11) log temperature in Kelvin. The marginalized posterior values are given as the median along with the upper and lower bounds associated with 95\% credible interval are given. For the collisionally ionized {\ovi} phase, the quantities $\log U$ and $\log L$ are not determined as they are dependent on the assumed value of $\log n(H)$. The synthetic profiles based on these models are shown in Figure~\ref{fig:sysplotPG1116}. The marginalized posterior distributions for the VP fit parameters (columns 2, 3, and 5) of the optimized ions are presented in Figures~\ref{fig:voigtMgIIpg1116}, \ref{fig:voigtSiIVpg1116}, and \ref{fig:voigtOVIpg1116}. The marginalized posterior distributions for the cloud properties (columns 6, 7, 8, 9, 10, and 11) are presented in Figures~\ref{fig:mgii0pg1116}, \ref{fig:mgii1pg1116}, \ref{fig:siivpg1116}, and \ref{fig:ovipg1116}. Columns (12), (13), and (14) describe the marginalized posterior distributions of log metallicity, log ionization parameter, and log hydrogen column density, for a {\lya}-only model described in \S~\ref{subsec:0.13849caveats}.
  
\end{table*}

\begin{figure*}
\begin{center}
\includegraphics[width=\linewidth]{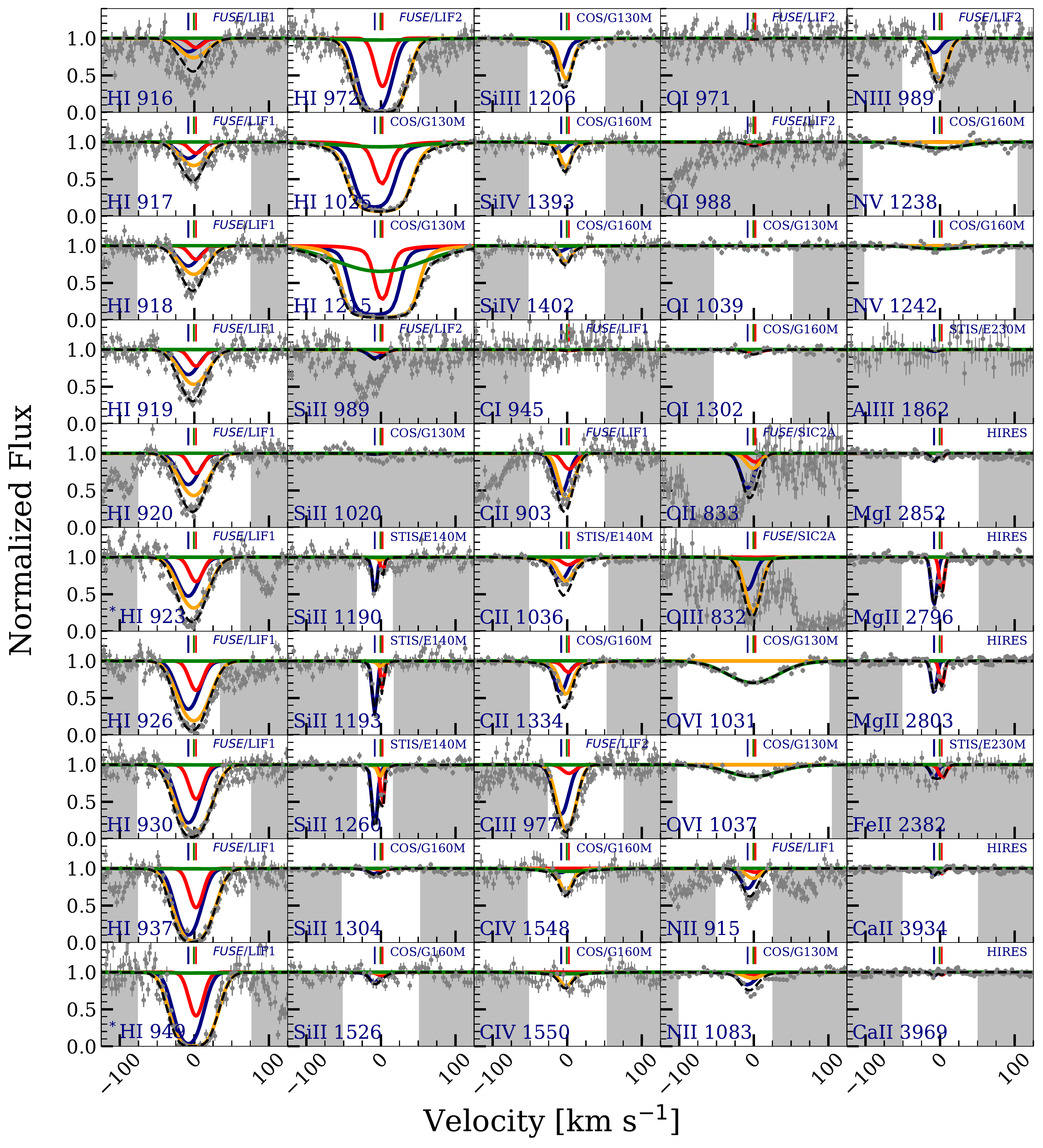}
\caption{{\CLOUDY} models for the $z = 0.13849$ absorber towards PG1116+215 obtained using the MLE values. The spectral data are shown in gray along with the 1$\sigma$~error. The instruments used for observation in different regions of the spectrum are indicated in the panels. The centroids of absorption components as determined from the VP fits for optimized ions are indicated by the vertical tick marks on top of each line. The low ionization photoionized phase consisting of two clouds are shown by the blue and red curves, and is optimized on \mgii, the intermediate ionization photoionized phase shown by the yellow curve is optimized on \siiv, and the collisionally ionized phase shown by the green model is optimized on \ovi. The superposition of all the three models is shown by the black dashed curve. The region shaded in grey shows the pixels that were not used in the evaluation of the log-likelihood. The lines {\hi} $\lambda$923 and {\hi} $\lambda$949 are corrected for blending as described in \S~\ref{sec:z = 0.13849} and earmarked with an asterisk.}
\label{fig:sysplotPG1116}
\end{center}
\end{figure*}

\citet{keeney2017characterizing} have modeled this absorber using $HST$/COS, $HST$/STIS, and {\fuse} spectra and determine a metallicity of $\log Z = -0.19^{+0.07}_{-0.08}$ for a single, low ionization cloud in this $z=0.13849$ absorber. However, the higher resolution, Keck/HIRES data reveals the presence of two narrow components in {\mgii} ($b(\mgii) = 2.5$ and $1.7$~\kms) with $W_r(2796)$ = \hbox{95 $\pm$ 5  m\angstrom}. It is important to note that relatively broad (8$-$20 \kms) low ionization components may often resolve into narrower components (\hbox{1$-$7 \kms}), and the $b$ parameters of these components can substantially affect the constraints on the metallicity in our modeling method. Several of the lines from the $z=0.13849$ system are affected by blends with other systems along the sightline.  The \hi~$\lambda$949 and \hi~$\lambda$923 are affected by the Galactic molecular hydrogen lines. The \hi~$\lambda$949 is contaminated at \hbox{$-62$ {\kms}} by H$_{2}$ L 2–0 P(2) $\lambda$1081.265, and \hi~$\lambda$923 at $-32$ {\kms} by H$_{2}$ L 4–0 P(1)$\lambda$1051.03. We divide out the contribution of these lines using the several other Galactic hydrogen lines observed in the {\fuse} spectrum. The \hi~$\lambda$972 line is contaminated by the {\lyd} of $z=0.16612$ absorber at $57$ {\kms}, and the \hi~$\lambda$926 line is contaminated by {\lyb} of the $z=0.02829$ absorber at $66$ {\kms}. We mask their influence in the log-likelihood evaluation.

The low ionization \mgii~clouds in this absorber are also detected in \nii, \siii, \cii, and \mgi~and marginally in \caii~and \oi. The {\oii} is contaminated by H$_{2}$ W 3$-$0 R(3)$\lambda$948.42, but the profile is still shown for completeness in Figure~\ref{fig:sysplotPG1116}. The region of the $HST$/STIS spectrum covering \feii~$\lambda$~2382 is too noisy to provide a useful constraint. The intermediate ionization states, {\siiii}, {\oiii}~and \ciii~are detected (though {\oiii} is contaminated by H$_{2}$ L 14$-$0 R(2)$\lambda$948.47), and though it is not as strong as the \mgii, \civ~is also detected. These transitions are clearly offset by $\sim 5$ {\kms} redward of the low ionization transitions. A broad, \ovi~cloud is clearly detected in both doublet members, in the high $S/N$ $HST$/COS spectrum.  The \lya~and \lyb~were covered in the high $S/N$ $HST$/COS spectrum as well, but the higher order Lyman series lines, down to $\lambda$918, were covered by {\fuse}. It is critical to note that the \hi~lines appear to be centered on the \siiv~and \civ~absorption rather than on the low ionization absorption. These velocity offsets suggest that two phases are needed to fit this system.

\smallskip

Our modeling yields constraints presented in Table~\ref{tab:pg1116modelpar}, and model curves that are superimposed on the data in Figure~\ref{fig:sysplotPG1116}. The posterior distributions for the parameters are summarized in Figures~\ref{fig:mgii0pg1116}, \ref{fig:mgii1pg1116}, \ref{fig:siivpg1116}, and \ref{fig:ovipg1116} for the different phases. The $b$ parameter ($b(\mgii)$=2.9 {\kms}) adopted for the blueward {\mgii} cloud is $2\sigma$ higher than the best VP fit value, because the lower value yielded a {\CLOUDY} model temperature too high to be consistent with the line width. The two, narrow, low ionization clouds have high metallicities ($\log Z=0.26$ and $0.82)$, low ionization parameters ($\log U = -3.43$ and $-4.14$), low temperatures (10$^4$ K and 10$^{3.0}$ K), and small thickness (100 pc and $<$1 pc).  With these low ionization parameters, only a small fraction of the intermediate ionization absorption can arise in the same phase as the low ionization absorption.  It is also impossible to produce the {\lya} and other \hi~lines with such narrow low ionization clouds, particularly the redward cloud which is best fit with $b(\mgii)$ of only 1.7 \kms.

\smallskip

A slightly broader cloud optimized on the \siiv~column density ($b(\siiv) = 6$ \kms) was added to the model and a maximum likelihood solution is determined combining with the two low ionization clouds, and an additional broad {\ovi} cloud to be discussed below.  This \siiv~cloud produces the majority of the \siiii~absorption and all of the observed \civ~absorption.  It also produces the majority of the \lya~absorption and a substantial fraction of the Lyman series lines as well. The properties of this \siiv~cloud are $\log Z=-1.90$, $\log U = -3.07$, and cloud thickness of $60$~kpc, and it contributes $\log [N(\hi)/\cmsq] = 16.0$ to the model.

\smallskip

In Figure~\ref{fig:sysplotPG1116}, the model, including the three low ionization clouds, slightly overproduces the observed {\cii} $\lambda$ 1036, but it is apparently consistent with the {\cii} $\lambda$ 903 and $\lambda$ 1334 lines, so we assume there may be a small defect in the {\cii} $\lambda$ 1036 data. The results are not dependent on which of the {\cii} transitions are used as constraints.

\smallskip

The broad \ovi~cloud ($b(\ovi) \sim 35$ \kms) is consistent with collisional ionization with $\log [T/K] = 5.4$, and it can produce the wings on the \lya~line that would otherwise be unexplained. The metallicity to provide that fit to \lya~is $\log Z = -1.06$. 

\smallskip

\begin{figure*}
\begin{center}
\includegraphics[width=\linewidth]{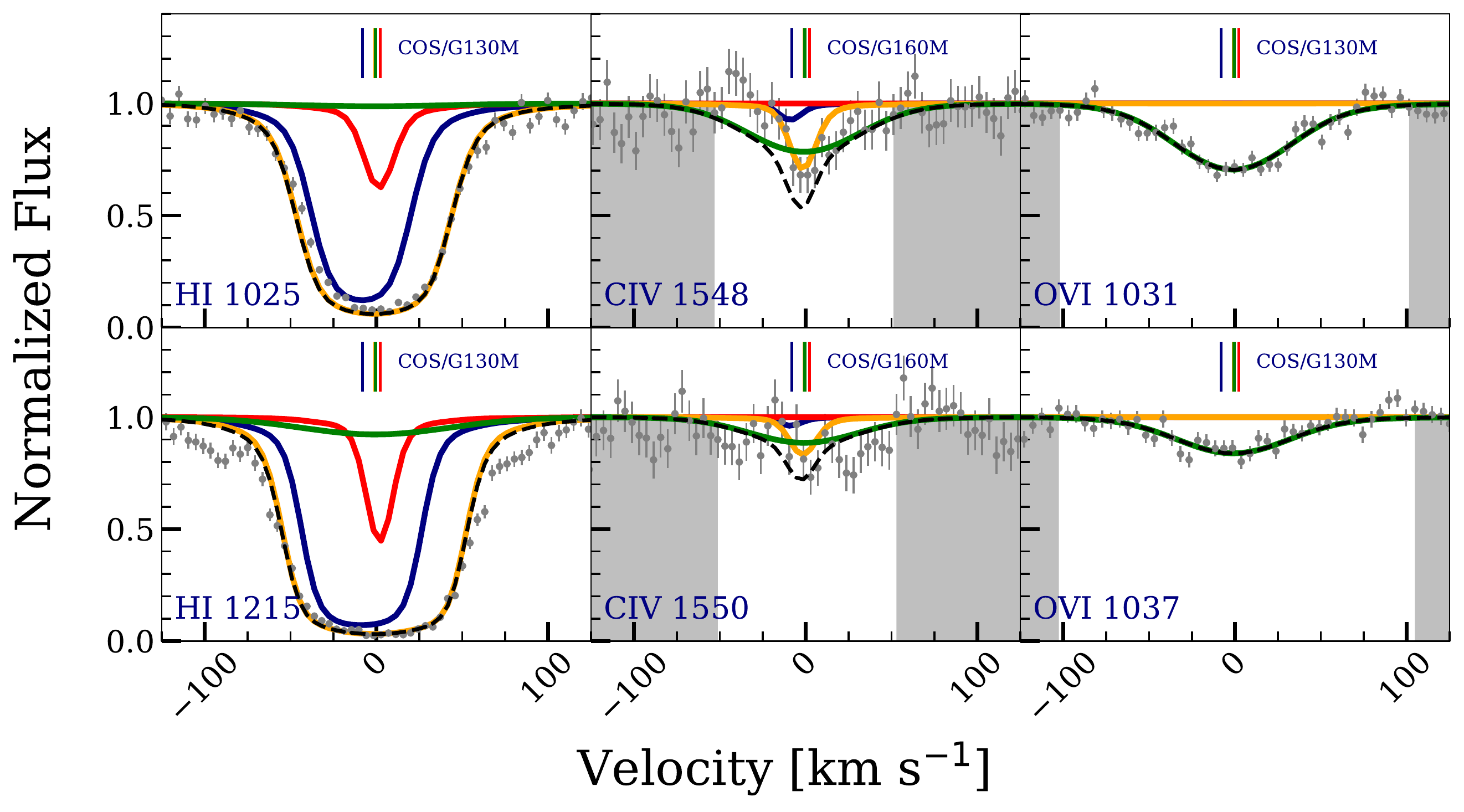}
\caption{Photoionization only models for the $z = 0.13849$ absorber towards PG1116+215. Clearly, the photoionized model for {\ovi} is discrepant as it overproduces {\civ} and the wings of the {\lya} are unexplained. Symbols, labels, and masked pixels are the same as in Figure~\ref{fig:sysplotPG1116}.}
\label{fig:ModelsPG1116OVIPHOT}
\end{center}
\end{figure*}

We also test if {\ovi} could arise in a photoionized phase. A photoionized model with $\log U = -1.6$ and a supersolar metallicity could fit the \ovi~profile, without contributing to the {\hi} however it also produces a broad \civ~profile which is clearly inconsistent with the data, as shown in Figure~\ref{fig:ModelsPG1116OVIPHOT}.

\begin{figure*}
\begin{center}
\includegraphics[width=\linewidth]{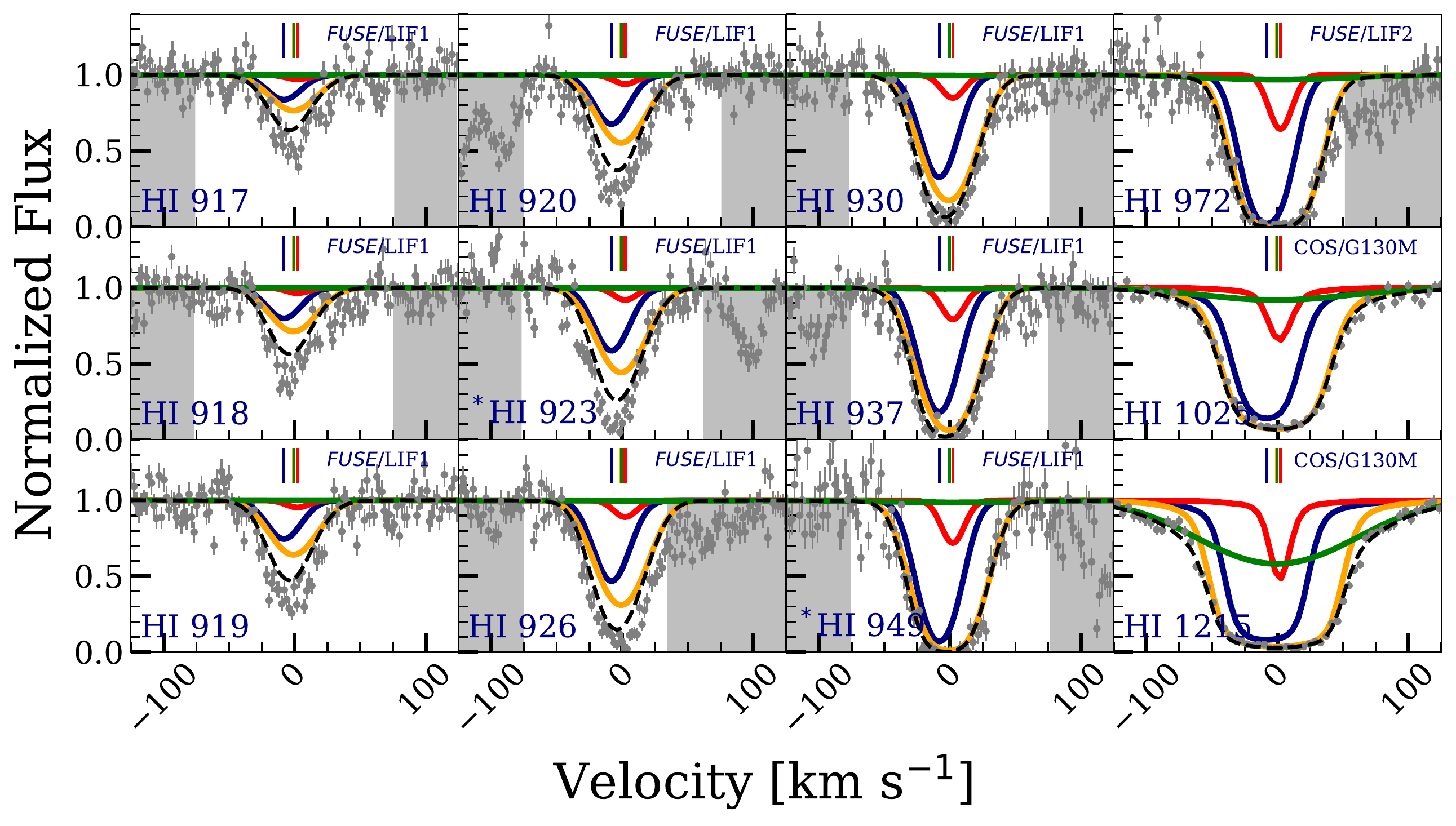}
\caption{{\CLOUDY} models for Lyman lines for the $z = 0.13849$ absorber towards PG1116+215 showing that a ``\lya - only'' model noticeably underproduces several of the higher order Lyman lines. Symbols, labels, and masked pixels are the same as in Figure~\ref{fig:sysplotPG1116}.}
\label{fig:ModelsPG1116newONLYAseries}
\end{center}
\end{figure*}

\subsubsection{Comparison to Previous Work}
\label{subsec:0.13849comparison}
This system might have been fit with just one low ionization component if it were not for higher resolution and high $S/N$ coverage of \mgii, which clearly shows two narrow components. These clouds are typical of weak \mgii~absorbers at higher redshifts ($z=0.4$-$1.0$), with low ionization parameters $\log U=-3.4$ and $-4.1$, super solar metallicities ($\log Z=0.3$ and $\log Z=0.8$), and small thickness (\hbox{100 pc} and \hbox{$<$ 1 pc}).
Our modeling indicates the presence of two additional phases in order to produce the observed high ionization and \hi~absorption, which cannot come from the low ionization clouds.  An intermediate ionization phase, with $\log U = -3.1$ and with low metallicity ($\log Z=-1.9$) produces \siiii, \siiv, and \civ~absorption and is responsible for the majority of \hi~absorption as well, having $\log [N(\hi)/\cmsq] = 16.0$.  This intermediate ionization gas phase is much more extended, with a thickness of $60$ kpc.
Finally, broad \ovi~absorption and wings on the \lya~profile constrain the properties of a higher ionization/hotter phase.  We find it to be consistent with collisional ionization at $\log [T/K] = 5.4$ in line with the {\lya} fit by \citet{Stocke2014} ($b$=86 $\pm$ 11 \kms). \citet{richter2020} successfully fitted an even broader component to the combined STIS and COS data set ($b$=150 \kms). He identified the extended wings of the {\lya} profile as part of a CBLA that traces the collsionally ionized, hot coronal gas of the host galaxy of this CGM absorber, in line with the predicted CBLA properties from his semi-analytic model (\citealt{richter2020}; Sect. 5.5). In contrast, \citet{savage2014properties} inferred a lower temperature of $\log [T/K] = 4.72$, and a contribution from photoionization from their analysis. However, we find that {\ovi} arising in a photoionized phase gives rise to a broad {\civ} which is inconsistent with the data, as shown in Figure~\ref{fig:ModelsPG1116OVIPHOT}.

\smallskip

Previous investigators \citep[e.g.][]{keeney2017characterizing,muzahid2018cos,pointon2019relationship,wotta2019cos} found metallicities ($\log Z\sim -0.3$) intermediate between those of our low metallicity, intermediate ionization cloud, and our supersolar metallicity, low ionization clouds. The ionization parameters of these earlier models tend to match that of our intermediate ionization phase, with $\log U = -3.1$, because they are trying to produce all the absorption in a single phase.  We see a significant variation in ionization parameter/density across the profiles. Again the possible origin of the absorber through a combination of low metallicity inflowing gas, and high metallicity outflowing gas, may be smeared out through modeling that does not consider multiple components and phases.  

\subsubsection{Caveats and Tests}
\label{subsec:0.13849caveats}
For this system, a ``{\lya} only'' model provides constraints that qualitatively match those that we obtained with the full model. These constraints, with errors, are compared in Table~\ref{tab:pg1116modelpar}.  The best ``{\lya} only'' model slightly under predicts several of the higher order {\hi} lines because of the slightly lower metallicity for the {\siiv} phase. The ``{\lya} only'' model is shown in Figure~\ref{fig:ModelsPG1116newONLYAseries}. 

\subsubsection{Galaxy Properties}
\label{subsec:0.13849galax}
There are three galaxies within an impact parameter of 1~Mpc at the redshift of this absorber~\citep{muzahid2018cos}, with the nearest being an $L=1.2L^*$ galaxy at an impact parameter of 139 kpc~\citep[e.g.][]{keeney2017characterizing}. As with the previous systems, this absorber appears to be found in a small group of galaxies.

The low ionization clouds are again typical of the supersolar metallicity, parsec-scale, weak \mgii~absorbers that are pervasive at intermediate redshifts \cite{rigby2002population,narayanan2008chemical}.  The \siiv~cloud is more characteristic of a sightline through the larger, intragroup medium, since it is 60 kpc in extent along the sightline, and has a low metallicity.  It would presumably be gas falling into the group~\citep{Hafen2019}, because of its low metallicity of $\log Z=-1.9$. The warm/hot, collisionally ionized \ovi~phase also has a relatively low metallicity of $\log Z=-1.06$.  Perhaps this gas is more related to the nearest galaxy, or perhaps to an unidentified dwarf in the group, or to the low-metallicity coronal gas of the absorber host galaxy.

\smallskip

\subsection{The $z = 0.22596$ absorber towards the quasar HE0153-4520}
\label{sec:z = 0.22596}

\begin{figure*}
\begin{center}
\includegraphics[width=\linewidth]{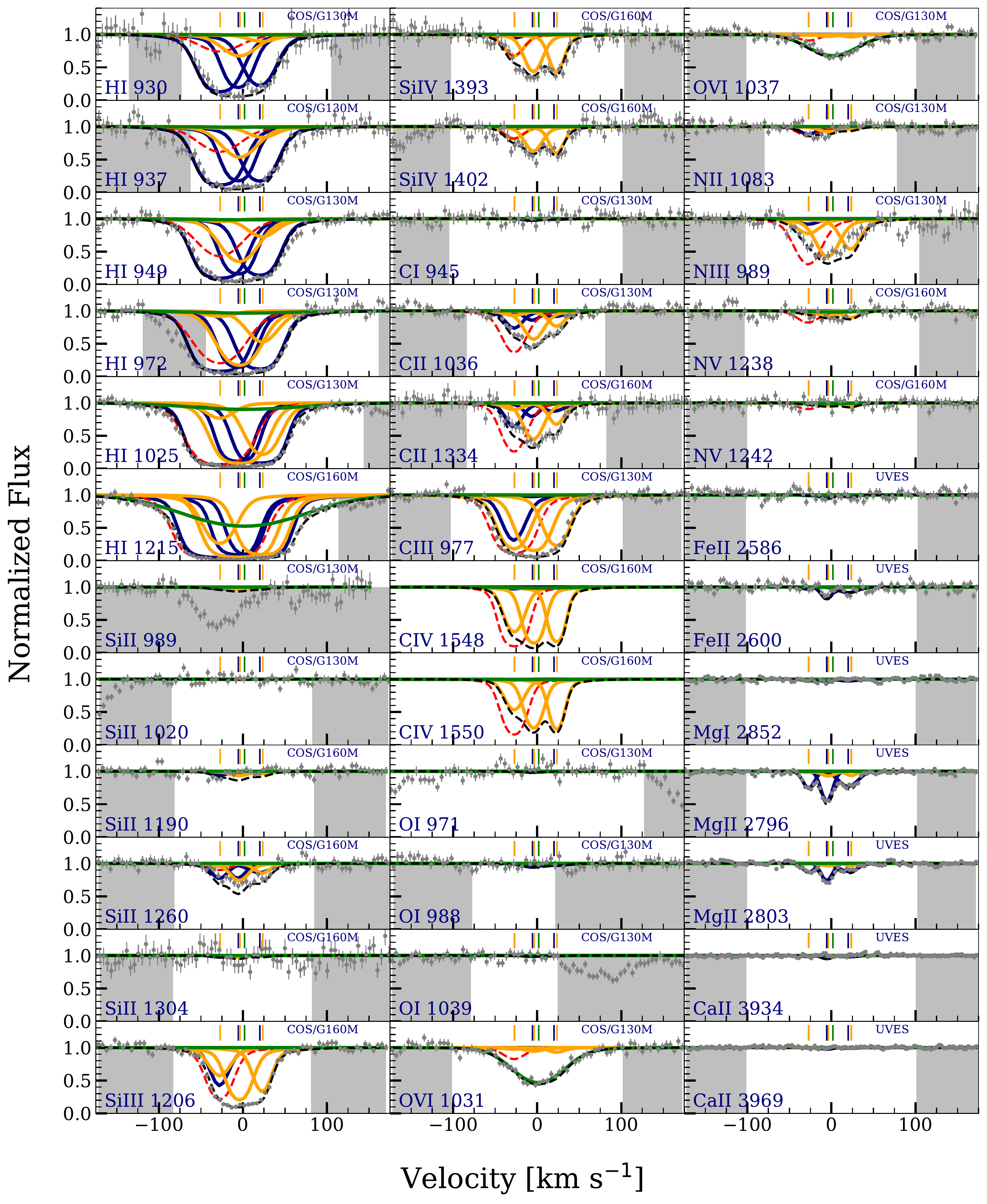}
\caption{{\CLOUDY} models for the $z = 0.22596$ absorber towards HE0153-4520 obtained using the MLE values. The spectral data is shown in gray along with the 1$\sigma$~error. The instruments used for observation in different regions of the spectrum are indicated in the panels. The centroids of absorption components as determined from the VP fits for optimized ions are indicated by the vertical tick marks on top of each line. The low ionization photoionized phase shown by the blue curves is obtained by optimized on the three \mgii~clouds, the intermediate ionization photoionized phase shown by the yellow curve is obtained by optimized on the three \siiv~clouds, and the collisionally ionized phase shown by the green curve is obtained by optimized on \ovi. We also show a model (red dashed curve) with solar metallicity in the intermediate ionization \siiv\_0 cloud which clearly exceeds the absorption seen in {\lya}, \cii, \niii, and \nv. The superposition of all the three models is shown by the black dashed curve. The region shaded in grey shows the pixels that were not used in the evaluation of the log-likelihood.}
\label{fig:ModelsHE0153sysplot}
\end{center}
\end{figure*}

This three component, weak \mgii~absorber (with $W_r(2796) = 184 \pm 6 $ m\angstrom) also shows absorption in the low ionization \siii, \cii, \nii, and {\feii} transitions.  Because the \mgii~is observed with the highest resolution with VLT/UVES, it is used as the optimized transition in our modeling. The detection of \feii, in particular, requires a low ionization parameter/high density for the two blueward, low ionization clouds.  Because of this, the detected \siiii, \ciii, \niii, and \siiv~cannot be produced in the same phase. Our model thus also includes three intermediate ionization clouds, optimized on the \siiv~doublet. Finally, a strong detection of broad \ovi~absorption and the corresponding wings in the \lya~profile call for a high ionization, collisionally ionized gas phase for the absorber, possibly related to a CBLA that traces the hot halo of the host galaxy.

\smallskip

Thus, our optimized \CLOUDY~models are run incorporating seven clouds in three phases.
The VP fit parameters for the three components identified in \mgii~which traces the low ionization phase, the three component fit for the \siv~which traces the intermediate ionization phase, and the single component fit for the \ovi~tracing a collisionally high ionization phase are given in Table~\ref{tab:HE0153modelpar}, along with the model parameters. The marginalized posterior distributions for the VP fit parameters are shown in Figures~\ref{fig:voigtmgiihe0153}, \ref{fig:voigtsiivhe0153}, and \ref{fig:voigtovihe0153}. The marginalized posterior distributions for the physical properties of clouds are shown in Figures~\ref{fig:MgII0he0153}, \ref{fig:MgII1he0153}, \ref{fig:MgII2he0153}, \ref{fig:SiIV0he0153}, \ref{fig:SiIV1he0153}, \ref{fig:SiIV2he0153}, \ref{fig:OVI0he0153}.

\smallskip

Our first attempt to fit this system revealed a discrepancy in fitting the \ciii~profile with the three clouds optimized on \siiv, with the left side of the \ciii~profile underproduced and the right side overproduced. Though a wavelength shift of the \ciii~profile (within errors in calibration) could have explained this discrepancy, that is not likely because the Lyman series lines at similar wavelengths were not similarly shifted.  Instead, we found that adjusting the $b$ parameters of the \siiv~components, within the errors in our VP fit, could match the \ciii~profile, because the \ciii~is highly dependent on $b$ on its flat part of the curve of growth.  Our adopted values of $b$ for our final model are listed as a separate column in Table~\ref{tab:HE0153modelpar}.

\smallskip

The three low ionization clouds have metallicities ranging over an order of magnitude. The blueward cloud, which has $\log U = -3.5$ and does not \textbf{give} rise to \feii~absorption has $\log Z=-0.8$ in order to fit the left side of the Lyman series profiles. The thickness of this cloud is $\sim$ 300 pc. The redward \mgii~cloud, with $\log U=-4.9$ and $\log Z=0.30$, is unusually broad, with $b=15.4$ \kms, and we note that if there were two narrower components in its place then the metallicity value could be significantly lower. Thus the metallicity of this cloud is a bit uncertain. The middle {\mgii} cloud also has a low $\log U$ value of $-5.2$ because of detected \feii, and a solar metallicity, with a relatively large uncertainty of 0.2 dex because of being in the center of the Lyman series profiles.  Because of their low ionization parameters and high densities ($\log [n(H)/\cc]=-1$), the two redward \mgii~clouds must be quite thin (only $\sim 0.1$ pc).

\smallskip

Most of the \hi~absorption does arise in the low ionization phase, with a total $\log [N(\hi)/\cmsq]$ of $16.8$ from the three clouds. This agrees with the depth of a partial Lyman limit break, observed in the {\fuse} data~\citep{savage2011multiphase}.

\smallskip

The three intermediate ionization clouds, optimized on \siiv, have similar constraints on $\log U \sim -2.5$, corresponding to a density of $\log [n(H)/\cc] = -3.6$.  For these ionization parameters, the observed weak \nv~absorption is produced in these clouds. They also must all have supersolar metallicities in order that they do not exceed the \lya~absorption.  Since the low ionization clouds that are producing the bulk of the \lya~absorption, there is not much room for a contribution to the \hi~column density from these intermediate ionization clouds. They are larger, with thicknesses ranging from 100 pc to 1.7 kpc. In Figure~\ref{fig:ModelsHE0153sysplot}, we show a model (red dashed curve) with solar metallicity in \siiv\_0 cloud which exceeds the absorption seen in {\lya}, \cii, \ciii, \niii, and \nv. 

\smallskip

The collisionally ionized, high ionization phase is found to fit the broad \ovi~profiles and the wings on the \lya~profile with $\log [T/K]=5.64$ and $\log Z=-1.3$.  For this temperature, very little \nv~absorption is produced, which is necessary since the \nv~arose primarily in the \siiv~clouds.

\begin{table*}
\renewcommand\thetable{7}
\begin{center}
\caption{\bf The gas phases in the $z = 0.22596$ absorber towards HE0153-4520}
	\label{tab:HE0153modelpar}  
\begin{tabular}{c@{\hspace{0.25\tabcolsep}}c@{\hspace{0.55\tabcolsep}}c@{\hspace{0.55\tabcolsep}}c@{\hspace{0.55\tabcolsep}}c@{\hspace{0.55\tabcolsep}}@{\vline}c@{\hspace{0.55\tabcolsep}}c@{\hspace{0.55\tabcolsep}}c@{\hspace{0.55\tabcolsep}}c@{\hspace{0.55\tabcolsep}}c@{\hspace{0.55\tabcolsep}}c@{\hspace{0.55\tabcolsep}}@{\vline}@{\hspace{0.55\tabcolsep}}c@{\hspace{0.55\tabcolsep}}c@{\hspace{0.55\tabcolsep}}c@{\hspace{0.55\tabcolsep}}} \hline \hline

& &  & & & & & Lyman lines & & & & & {\lya}-only  & \\  
\hline
Optimized & $z$ & $b$ & $b_{used}$ & $\log \frac{N}{\cmsq}$ & $\log Z$ & $\log U$ & $\log \frac{n(H)}{\cc}$ & $\log \frac{N(\hi)}{\cmsq}$ & $\log \frac{L}{kpc} $ & $\log \frac{T}{K} $ & $\log Z$ & $\log U$ & $\log \frac{N(\hi)}{\cmsq}$\\  
ion & & (\kms) & (\kms) & & & & & & & & & &\\
(1) & (2) & (3) & (4) & (5) & (6) & (7) & (8) & (9) & (10) & (11) & (12) & (13) & (14)\\\hline
\textcolor{blue}{\mgii_0} & 0.22585 & $7.9^{+2.0}_{-1.6}$ & 7.9 & $12.03^{+0.07}_{-0.07}$ & $-0.79^{+0.10}_{-0.11}$ & $-3.49^{+0.08}_{-0.09}$  & $-2.59^{+0.09}_{-0.08}$ & $16.39^{+0.11}_{-0.09}$ & $-0.54^{+0.11}_{-0.15}$ & $4.23^{+0.02}_{-0.02}$ & $-1.54^{+0.32}_{-0.17}$ & $-3.68^{+0.16}_{-0.23}$ & $17.11^{+0.22}_{-0.31}$\\    

\textcolor{blue}{\mgii_1} & 0.22594 & $7.7^{+1.2}_{-1.1}$ & 7.7 & $12.34^{+0.05}_{-0.06}$ & $-0.06^{+0.26}_{-0.25}$ & $-5.24^{+0.25}_{-0.30}$  & $-0.84^{+0.30}_{-0.25}$ & $16.54^{+0.30}_{-0.30}$ & $-3.81^{+0.16}_{-0.10}$ &  $3.63^{+0.12}_{-0.16}$ &$-0.03^{+0.22}_{-0.31}$ & $-5.22^{+0.21}_{-0.39}$ & $16.50^{+0.37}_{-0.25}$\\

\textcolor{blue}{\mgii_2} & 0.22604 & $15.3^{+3.3}_{-2.9}$ & 15.3 & $12.21^{+0.07}_{-0.07}$ & $0.28^{+0.08}_{-0.07}$ & $-4.89^{+0.25}_{-0.13}$  & $-1.19^{+0.14}_{-0.25}$ & $16.00^{+0.08}_{-0.11}$ & $-3.92^{+0.31}_{-0.12}$ &  $3.41^{+0.13}_{-0.11}$ & $0.13^{+0.13}_{-0.16}$ & $-5.07^{+0.28}_{-0.26}$ & $16.19^{+0.21}_{-0.17}$\\

\textcolor{orange}{\siiv_0} & 0.22585 & $11.4^{+4.4}_{-9.1}$ & 12.0 & $12.99^{+0.22}_{-0.31}$ & $1.26^{+0.13}_{-0.11}$ & $-2.40^{+0.04}_{-0.05}$  & $-3.68^{+0.05}_{-0.04}$ & $13.94^{+0.12}_{-0.13}$ & $-0.97^{+0.13}_{-0.19}$ &  $4.08^{+0.06}_{-0.08}$ &$0.89^{+0.18}_{-0.13}$ & $-2.47^{+0.07}_{-0.08}$ & $14.37^{+0.17}_{-0.23}$ \\

\textcolor{orange}{\siiv_1} & 0.22594 & $9.8^{+5.7}_{-7.2}$ & 9.8 &  $13.27^{+0.63}_{-0.23}$ & $0.35^{+0.06}_{-0.06}$ & $-2.58^{+0.04}_{-0.03}$  & $-3.50^{+0.03}_{-0.04}$ & $15.25^{+0.07}_{-0.07}$ & $0.23^{+0.09}_{-0.11}$ &  $4.39^{+0.02}_{-0.02}$ & $0.41^{+0.07}_{-0.07}$ & $-2.54^{+0.05}_{-0.05}$ & $15.17^{+0.09}_{-0.09}$\\

\textcolor{orange}{\siiv_2} & 0.22606 & $8.8^{+4.3}_{-4.3}$ & 7.0 & $13.22^{+0.16}_{-0.12}$ & $0.88^{+0.09}_{-0.08}$ & $-2.49^{+0.04}_{-0.04}$  & $-3.59^{+0.04}_{-0.04}$ & $14.59^{+0.10}_{-0.12}$ & $-0.42^{+0.16}_{-0.12}$ &  $4.23^{+0.04}_{-0.04}$ & $0.85^{+0.11}_{-0.12}$ & $-2.47^{+0.04}_{-0.04}$ & $14.62^{+0.15}_{-0.14}$ \\

\textcolor{darkgreen}{\ovi_0} & 0.22597 & $34.9^{+3.3}_{-3.2}$ & 34.9 & $14.20^{+0.03}_{-0.03}$ & $-1.45^{+0.38}_{-0.19}$ & -  & $-3.90$ & $13.88^{+0.08}_{-0.09}$ & - & $5.59^{+0.13}_{-0.10}$ & $-1.04^{+0.51}_{-0.29}$ & - & $13.54^{+0.20}_{-0.27}$\\

\hline 

\end{tabular} 
\end{center}

Properties of the different gas phases present in $z = 0.22596$ absorber towards HE0153-4520 traced by their respective optimized ions. Notes: (1) Optimized ion tracing a phase; (2) Redshift of the component; (3) Doppler parameter of optimized ion; (4) Adopted Doppler parameter for the optimized ion (5) log column density; (6) Metallicity;  (7) log ionization parameter; (8) log hydrogen number density; (9) log hydrogen column density (10) log thickness in kpc; (11) log temperature in Kelvin. The marginalized posterior values with the median along with the upper and lower bounds associated with 95\% credible interval are given. For the collisionally ionized {\ovi} phase, the quantities $\log U$ and $\log L$ are not determined as they are dependent on the assumed value of $\log n(H)$. The synthetic profiles based on these models are shown in Figure~\ref{fig:ModelsHE0153sysplot}. The marginalized posterior distributions for the VP fit parameters (columns 2, 3, and 5) of the optimized ions are presented in Figures~\ref{fig:voigtmgiihe0153}, \ref{fig:voigtsiivhe0153}, and \ref{fig:voigtovihe0153}. The marginalized posterior distributions for the cloud properties (columns 6, 7, 8, 9, 10, and 11) of absorbers present in this system are shown in Figures~\ref{fig:MgII0he0153}, \ref{fig:MgII1he0153}, \ref{fig:MgII2he0153}, \ref{fig:SiIV0he0153}, \ref{fig:SiIV1he0153}, \ref{fig:SiIV2he0153}, \ref{fig:OVI0he0153}. Columns (12), (13), and (14) describe the marginalized posterior distributions of log metallicity, log ionization parameter, and log hydrogen column density, for a {\lya}-only model described in \S~\ref{subsec:0.22596caveat}.
  
\end{table*}

\subsubsection{Comparison to Previous Work}
\label{subsec:0.22596comparison}
This system has three distinct absorption phases, two photoionized with high densities of $\sim 0.1${\cc} in the low ionization phase, and values much lower $\log [n(H)/\cc]=-3.5$ in the intermediate ionization phase.  The metallicities of the low ionization clouds range from $\log Z=-0.8$ to around the solar value, while those of the intermediate ionization clouds must be supersolar.  The collisionally ionized, high ionization cloud has a temperature of $\sim 10^{5.6}$~K, and a metallicity of $\log Z=-1.45$.

\smallskip

Previous models \citep[e.g.][]{savage2011multiphase,muzahid2018cos,wotta2019cos} derived values of metallicity for the low ionization gas ranging from $\log Z=-0.7$ to $\log Z=-1.0$, depending on the assumed EBR.  At face value, an average of $\log Z=-0.2$ would result from our three separate clouds, however we have noted that there is uncertainty in the redward clouds, and also the value would tend to be lower if a different EBR besides HM12 was assumed as in the other models.

\smallskip

The temperature of our collisionally ionized phase of $400,000$~K is considerably less than the value of over a million degrees which~\citet{savage2011multiphase} derived from the ratio of the measured $b$ parameters of \ovi~and \hi~from VP fitting.  We find that a lower temperature, and a lower $b$ parameter of $\sim 90$ {\kms} is more consistent with the shape of the \lya~profile when combined with the contributions from the other gas phases.

\begin{figure*}
\begin{center}
\includegraphics[width=\linewidth]{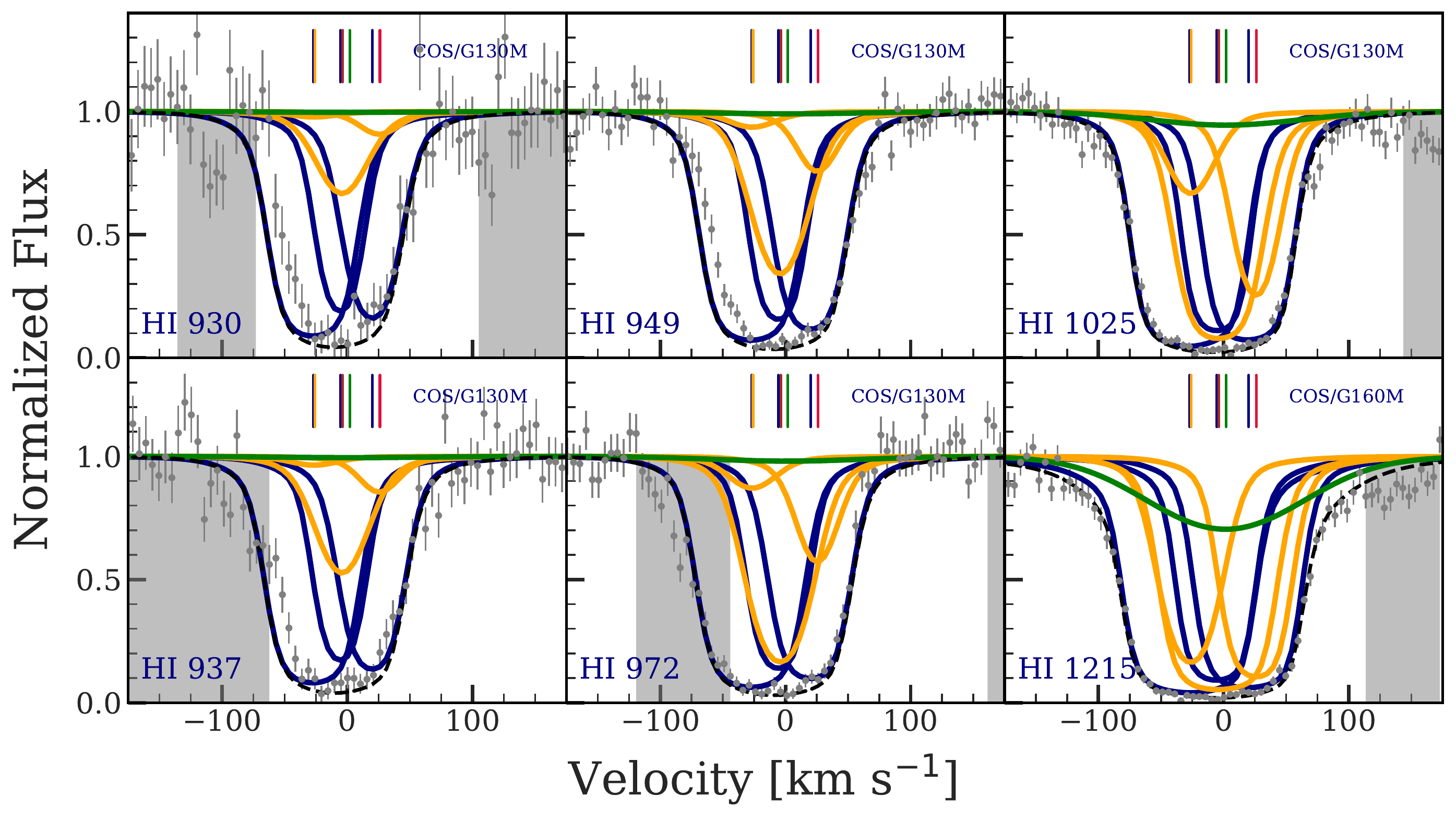}
\caption{{\CLOUDY} models for Lyman lines for the $z=0.22596$ absorber towards HE0153-4520 showing that a ``{\lya} - only'' model does noticeably overproduce the observed Lyman series lines. Symbols, labels, and masked pixels are the same as in Figure~\ref{fig:ModelsHE0153sysplot}.}
\label{fig:ModelsHE0153onlya}
\end{center}
\end{figure*}

\subsubsection{Caveats and Tests}
\label{subsec:0.22596caveat}

We again tried the experiment of considering what constraints would be in the absence of coverage of the Lyman series lines.  Table~\ref{tab:HE0153modelpar} summarizes the results.  The two models closely agree in their ionization parameters and metallicities, with the important exception of the blueward, low ionization cloud. The metallicity of this cloud for the ``\lya - only'' model is $\log Z = -1.54^{+0.32}_{-0.17}$, as compared to $\log Z = -0.8$ for our full model.  This much lower metallicity predicts $\log [N(\hi)/\cmsq] = 17.17$, as compared to a contribution of $\log [N(\hi)/\cmsq] = 16.4$ in our full model.  It is already clear that this is incorrect because we do not observe a full Lyman break from this system.  Thus it is not surprising that the ``{\lya} only'' model over predicts the higher order Lyman series lines.

\smallskip

We now consider the reason for this discrepancy.
Without the constraints from the Lyman series, the wing of the \lya~profile on the blueward side is filled in by the blueward, low ionization cloud, rather than a broad \lya~component related to the broad \ovi~absorption.  The higher order Lyman series lines, starting at \hi~$\lambda$949, make it clear that such a model would overproduce the \hi~absorption on the blueward side as shown in Figure~\ref{fig:ModelsHE0153onlya}.   Despite this change, the inference of the metallicity and temperature of the collisionally ionized phase is not significantly different between the two models.

\smallskip

It is clear that the model that utilizes all the constraints of the Lyman series lines is more accurate, however, we still note that the \lya~line alone does give some very important and correct conclusions.  Even with just the \lya~we can see that two of the low ionization clouds and all three high ionization clouds have solar or higher metallicity.  It is also clear that all six of these clouds, and a broad \ovi~cloud, must exist in order to fit the data.

\smallskip

We show in Figure~\ref{fig:ModelsHE0153sysplot} the expected {\civ} absorption, contributed by the intermediate ionization clouds for this system.  If we had coverage of {\civ} in the observed spectrum then the constraints on these clouds would be quite secure.

\subsubsection{Galaxy Properties and Interpretation}
\label{subsec:0.22596galax}
This absorber is at an impact parameter of 80.2 kpc from a luminous ($L=2.7L^*$) galaxy~\citep{keeney2018galaxy}. However, in this case, a dozen $L>L^*$ galaxies are found within an impact parameter of $4$ Mpc and a velocity of $1000$ {\kms}, suggesting a rich group environment.

\smallskip

This rich galaxy environment would be consistent with the relatively strong {\ovi} absorption associated with collisionally ionized gas, at the boundary between the low ionization layers and the surrounding hot medium \cite{Pointon2017}. The low ionization absorption is produced in higher density gas than in the previous absorbers. These very condensed (sub-parsec) structures may be enriched fragments originating in outflows \cite{mccourt2012thermal}. In this case, the intermediate ionization phase, giving rise to the {\siiv} absorption (and predicted strong {\civ} absorption) in three clouds, seems to be well-aligned in velocity with the low ionization clouds. This would suggest a similar origin for the two phases, perhaps with a higher density collapsed region within a lower density region giving rise to the intermediate ionization absorption. This would also be feasible in winds from dwarf galaxies, which could plausibly produce the high metallicity seen in the intermediate ionization clouds~\citep{Fujita2020}.

\section{DISCUSSION}

\label{sec:discussion}
\begin{figure*}
\begin{center}
\includegraphics[width=\linewidth]{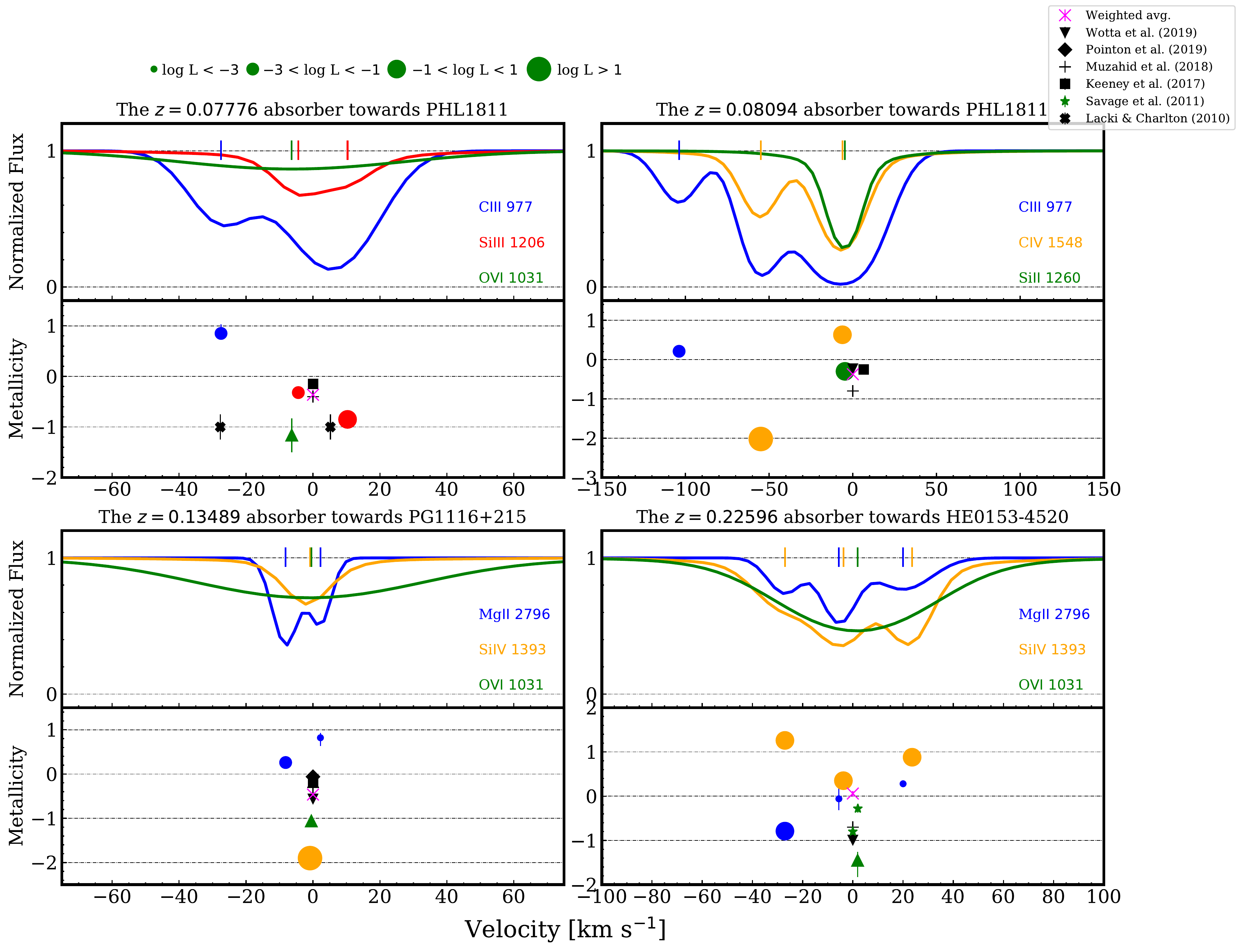}
\caption{Comparison of metallicity values from our CMBM approach with values from the literature. The system is schematically represented by the model curves obtained for the optimized transitions for each phase, indicated by the matching color labels. The $N(\hi)$ weighted average value of metallicity from our work is presented in magenta as a cross. For comparison, we also show the metallicity values obtained from previous studies, represented by various black symbols. The metallicities obtained in our work, corresponding to different optimized ions in an absorption system, are shown as color-coded filled circles and aligned with the respective optimized ion indicated with vertical tick marks, and their sizes are indicative of the line-of-sight thickness (in some cases, data points are larger than the error bars). The thickness ranges are indicated in the key shown above the plot. For the collisionally ionized phase traced by {\ovi} we do not obtain a thickness estimate hence it is presented as a small green triangle. For the $z = 0.07776$ absorber towards PHL1811, \citet{lacki2010z} have also carried out multiphase modeling, but with two components. For the $z = 0.22596$ absorber towards HE0153, \citet{savage2014properties} determine [X/H] = $-0.8 \pm 0.2$ for N, Si, and C. For the {\ovi} and associated BLA, they determine [O/H] = $-0.28^{+0.09}_{-0.08}$. These two data points are shown as green stars.}
\label{fig:comparison}
\end{center}
\end{figure*}

As a test of our methodology, we have applied our CMBM method to four, well-studied, weak, low ionization absorbers. These systems are the $z=0.07776$ absorber towards the quasar PHL1811, the $z=0.08094$ absorber towards the quasar PHL1811, the $z=0.13849$ absorber towards the quasar PG1116+215, and the $z=0.22596$ absorber towards the quasar HE0153-4520.  

Constraints from our modeling of the four system are given in Tables~\ref{tab:phl18110.07776model}, \ref{tab:phl18110.08modelpar}, \ref{tab:pg1116modelpar}, and \ref{tab:HE0153modelpar}. Our best fit models for the four systems are superimposed on the observational data in Figures~\ref{fig:sysplotphl18110.07}, \ref{fig:Modelsphl18110.08}, \ref{fig:sysplotPG1116}, and \ref{fig:ModelsHE0153sysplot}.

\subsection{Metallicity constraints from cloud-by-cloud, multiphase, Bayesian modeling}
\label{sec:metallicityconstraints}
In all four cases we were able to find an adequate fit, with one to four different phases, to all covered radiative transitions. Since metallicity is crucial to interpretations, we summarize, in Figure~\ref{fig:comparison}, the constraints we have placed on the metallicities of each component for the four systems, with each component labeled with the optimized transition for our model, and plotted in velocity space to show the kinematic structure of the system. The results for the individual components are compared to a single value that we derive from averaging our individual components, weighted by $N(\hi)$ and plotted at $0$ \kms.  These values are also compared to constraints on metallicity placed by previous investigators~\citep{lacki2010z,savage2011multiphase,keeney2017characterizing,muzahid2018cos,pointon2019relationship,wotta2019cos}. 
Note that there will be differences of $\sim$ 0.4 dex (for e.g., between HM05 and HM12) in the metallicity derived from models based on their adoption of different EBR amplitudes and shapes~\citep{wotta2019cos}.  We are, however, concerned only with the general, qualitative results that affect our gross interpretations. The sizes of the points in Figure~\ref{fig:comparison} are in proportion to the thickness of the clouds from our models.
We notice the following trends:

1. In three of the four cases (all but the $z=0.22596$ absorber towards HE0153-4520) our mean, $N(\hi)$-weighted, metallicity agrees with those derived in previous studies, taking into account differences in the EBR.

2. In the absorbers at $z=0.07776$ and $z=0.08094$ towards PHL1811 our inferred, mean value is dominated by the strongest, low ionization cloud. However, in the other two cases, several clouds, spanning more than two orders of magnitude in metallicity, strongly influence the mean value.

3. Generally, low ionization phase clouds have high inferred metallicities. In two cases, the $z=0.08094$ absorber towards PHL1811 and the $z=0.22596$ absorber towards HE0153-4520, intermediate ionization ({\siiv} and {\civ}) clouds have even higher metallicity values.

4. For the $z=0.22596$ absorber towards HE0153-4520, the values from previous investigations are an order of magnitude smaller than the $N(\hi)$-weighted mean from our study.  This is due to the two supersolar metallicity {\mgii} clouds.

5. In the three absorbers with detected {\ovi}, the broad {\ovi} profiles can be explained by collisionally ionized gas, which also in all cases makes an important contribution to {\lya} absorption in the wings of the profile (as so-called CBLAs), possibly tracing hot coronal gas and which would otherwise be difficult to explain. In all three cases the metallicity of the collisionally ionized gas is consistent with $\log Z = -1$.

6. In all four cases, the averaging of the metallicities of components and phases, hides important variations along the profile.  For example, supersolar metallicities in some of the components could be a signature of outflowing gas or patchy distribution of metals.  Similarly, low metallicities seen in other components could indicate an origin in pristine, inflowing gas.  It should not be surprising that multiple processes are contributing along the same sightline, which is likely to have contributions to absorption from regions spanning a couple hundred kiloparsecs. 

\smallskip

The CMBM method allows us to dissect the system into these different components.  Because of this refinement, it should be much more likely that trends with galaxy properties will be revealed if modeling is performed on a larger sample of galaxies/absorbers.  For example, in \citet{pointon2019relationship}, no correlation was seen between system metallicity and the inclination and azimuthal angle of the nearest galaxy.  This was surprising because there is an expectation that there will be a higher metallicity along the minor axis due to outflows, and a lower metallicity along the major axis where gas inflowing from the surrounding environment is seen.  This expectation is confirmed by an analysis of the IllustrisTNG and Eagle cosmological hydrodynamic simulations by \citet{Peroux2020}.
Examining possible correlations between the highest and lowest metallicity components of our observed systems could reveal trends that have previously been disguised, and resolve this discrepancy.  There is already an indication of a relationship between metallicity and azimuthal angle using dust ([Zn/Fe]) as a proxy, in a sample of 13 galaxy/{\mgii} pairs \citep{Wendt2020}.

\subsection{Strengths and limitations of cloud-by-cloud, multiphase, Bayesian modeling}
\label{sec:strengthandlimitation}

Generally, for these four weak, low ionization absorbers, we demonstrate that our method is successful in extracting the physical conditions (ionization parameters and metallicities) of the various regions of gas along a sightline.

\smallskip

In \S~\ref{sec:Results_Discussion} of this paper, we gave details of how each model parameter was constrained by specific observed transitions (e.g., if the metallicity of this cloud was any lower, the left side of the {\lya} and {\lyb} profiles would be overproduced).  This lends confidence to the accuracy of the results, within the context of our assumptions about the phase structure.  Similarly, we have considered whether our formal error bars seem reasonable.  For example, in Figure~\ref{fig:ModelsHE0153sysplot} we have shown that a solar metallicity for component {\siiv}\_0 is clearly discrepant as it overproduces {\lya}, \cii, \ciii, \niii, and {\nv} to the blue, thus showing that this component must have a highly supersolar metallicity value.

\smallskip

To test the robustness of our method, we designed a simple test to see if we could recover the input parameters of synthetic spectra.  The synthetic spectra were based on the parameters that we derived for the $z=0.07776$ system towards PHL1811, as listed in Table~\ref{tab:simulandrecov}, with an example shown in Figure~\ref{fig:simulreal}.  For all four synthetic spectra, with a range of metallicities and ionization parameters, we were able to separate the {\ciii} cloud and two {\siiii} clouds and infer their different metallicities, within errors. We are also able to correctly infer the ionization parameters, if they are similar as in the observed system, or if they are different as in our simulated spectra Test C (Table~\ref{tab:simulandrecov}). This builds confidence that we are correctly determining that clouds along the same line of sight, even in the same phase, can have very different densities, metallicities, or thickness.

\smallskip

For weak, low ionization absorbers, including the four in our sample, the {\lya} is often on the flat part of its curve of growth.  That makes it impossible to derive an accurate measurement of $N({\hi})$ via a VP fit.  In \citet{muzahid2018cos}, Table 2 shows that many of the $N({\hi})$ values are quite uncertain for a larger sample of similar systems from which these four were drawn.  Despite this limitation, it is possible to obtain meaningful constraints on the metallicity of the absorbing clouds.  Our method relies on the more accurate measurement of the $b$-parameters of the different components for the optimized metal ion.  From these $b$ parameters, and using the temperatures from \CLOUDY, we can calculate the thermal/turbulent fractions, and derive the value of $b(\hi)$ for each component. One can use the $b$ parameter of a metal line to infer the $b$ parameter of an {\hi} component.  By doing so one can break the degeneracy of VP fitting to even just the {\lya} line and derive accurate metallicities, particularly for the redward and blueward, low ionization components. Interplay between the low and intermediate ionization phases, and even contributions to the wings of {\lya} from the high ionization phase, must be incorporated into the Bayesian optimization. However, even with just a low ionization phase, there is a very strict lower limit on the metallicity derived from the requirement that the left and right sides of the {\lya} profile are not overproduced.  

\smallskip

We emphasize that breaking the degeneracy of measuring $N({\hi})$ on the flat part of the curve of growth is essential to our method.  As a result, it is critical that the $b$ parameters of the optimized transitions are accurate, and are not uncertain either due to low resolution observations or to component blending. We must be aware of the possibility that the best VP fit value for a given optimized transition might result in a {\CLOUDY} model temperature that is inconsistent, and in such a case the adopted $b$ parameter for the optimized transition should be increased. Our method also assumes that the equilibrium temperature from {\CLOUDY} is accurate, and that discrete, constant density, cloud structures are an adequate approximation for the absorbing gas. We considered the effect of a possible 10\% error in the equilibrium temperature, and found it to have negligible effect on inferred metallicity values. We can also test these assumptions in the following way.  In these four absorbers we typically have access to four or five clean (not blended) Lyman series lines in addition to the {\lya}, and in two cases there is coverage of the Lyman break, allowing an accurate and direct constraint on the total $N({\hi})$.  For all four systems we blindly ran our entire CMBM method on the data including the metal lines, but only using {\lya} and not the higher order Lyman series lines.  Despite this limitations, in all cases we were able to correctly infer the metallicity of the dominant component, and in all but one of the systems we extracted qualitatively similar values for all the absorbing cases.  Figures~\ref{fig:Modelsphl18110.07nolyaseries}, \ref{fig:Modelsphl18110.08lyaseries}, \ref{fig:ModelsPG1116newONLYAseries}, and \ref{fig:ModelsHE0153onlya} show how closely the remaining Lyman series models were predicted by our ``{\lya}-only" models.

\smallskip

Of course, when many Lyman series lines are covered we have great confidence that we will accurately derive the metallicities of all of the components. A straightforward way to infer that multiple phases are needed for some systems is to note that a single component fit cannot adequately explain both the shape of the {\lya} line and the strengths of the higher order Lyman series lines.  Narrow, low ionization clouds often account for the higher order Lyman series lines, while broader, intermediate ionization clouds are often needed to match the base/edge of the {\lya} and {\lyb} lines. It also appears that broader, high ionization lines are required to explain the wings of the {\lya} profile.  These wings, which seems to be common in these types of systems, possibly indicate the contribution of the hot coronal gas halos of the galaxies to the {\lya} absorption (CBLAs; \citealt{richter2020}). They do provide, in any case, an important way to determine if photoionization or collisional ionization is dominant for the high ionization phase that produces the {\ovi} absorption.  For example, this provided evidence to favor collisional ionization, instead of photoionization, for the {\ovi} in the $z=0.07776$ system towards PHL1811 and in the $z=0.13849$ system towards PG1116+215.

\smallskip

Just as the {\hi} lines often call for significant contributions from different phases, some metal lines, particularly those with intermediate ionization, also require such contributions.  Once the phase structure is determined, and the optimized transitions are set as input, our method can formally adjust the balance of these phases to produce the best fits to the profiles of {\it all} specified transitions.

\smallskip

With the inferred model parameters it is possible to predict the line profiles of all radiative transitions, even those that were not covered in the data available.  This can predict, for example, whether a system is indeed likely to be a weak {\mgii} absorber even in cases like the two absorbers towards PHL1811 for which wavelength coverage of that doublet was not available.  This allows comparisons to other systems in the same class.

\smallskip

The true power of CMBM method cannot be realized unless there is a way of passing beyond spending weeks laboring over the details of each individual system. This automated Bayesian approach makes it possible to cover all of parameter space in a systematic and automated manner, once the cloud and phase structure are specified.  Building on the experience gained by modeling other systems of the same type, it is expected that most systems can be fully explored within roughly a day.  This puts systematic studies of tens or even 100 absorption systems within reach, which is crucial to compare different absorption properties in the different phases to each other, and to the luminosities, morphologies, and orientations of the absorbing galaxies.

\smallskip

Though there are clear advantages to our methodology, which represents clear progress over previous studies, there are still some limitations, and aspects which require careful attention to individual systems.  Determination of the minimum phase structure that adequately describes an absorber is an iterative process and does require some human intervention.  For example, the low ionization phase can contribute to the intermediate phase absorption, but may not fully produce it.  This contribution would have to be taken into account if an intermediate phase transition is used as an optimized transition, and the optimized column density reduced. In such a case the low ionization phase properties would need to be fixed in the modeling process. Also, it is important to recognize blends with our constraining transitions, and to eliminate the proper regions before running the program.  A discrepancy with a spurious region of spectrum in a single transition can dominate and be a determining factor in model parameters.

\smallskip

Furthermore, though we strive to find an adequate model using the minimum number of phases, such a model is not unique.  It is always possible that an additional phase contributes, and that there is an alternative solution that is equally plausible.  In such a case, predicting additional transitions, or predicting structure in existing observations at higher $S/N$ can provide a helpful way to distinguish between such models should new data acquisition prove possible.  

\smallskip

There is also a limit to conclusions that can be drawn if a variety of radiative transitions, with varying ionization states, are not covered.  Since our method relies on using the $b$ parameter of a metal-line transition to determine $b(\hi)$, and since this is essential to determining metallicity constraints, it is important that at least one metal-line transition in each phase is covered at high resolution and $S/N$, and is unsaturated.  For example, for the $z=0.13849$ absorber toward PG1116+215 the coverage of {\mgii} clearly shows two components.  If we only had coverage of the {\siii} with $HST$/COS resolution we would have tried to fit the system with only one component in place of the two.  We faced this challenge in our modeling of the $z=0.07776$ absorber toward PHL1811, where high resolution ground-based data were not available covering {\mgii} or {\feii} because of the low system redshift.  In this case, we were able to infer that there were likely two blended, narrower components fitting the {\siiii}, however, the advantages of this two-component model were subtle, and some of the model parameters depended on the specific VP fit assumed for the {\siiii}.
This warns us to be cautious to consider the implications of splitting an unresolved component into two when we do not have high resolution coverage.  We also noted above, that though ``{\lya} only'' models often provide reliable constraints, it is also possible that some values, especially for weak clouds, will be inaccurate, as we saw in Figure~\ref{fig:Modelsphl18110.08} for the $z=0.08094$ absorber towards PHL1811.

\smallskip

We simply must be aware of such limitations when not as many transitions have been covered, and our Bayesian optimization can take these systematic uncertainties into account when deriving the range of allowed parameters.  Even if transitions are missing coverage, so that an intermediate or high ionization phase is missed, we can still derive lower limits on the low ionization gas phase metallicity.

\smallskip

The systems in this study do not have coverage of different elements contributing in the same phase to allow meaningful constraints on abundance pattern.  Sometimes more transitions are covered, e.g. ions of O and C spanning several ionization states.  When this is the case, it is possible to extract some basic information about abundance pattern, such as the elevated C/O ratio found in Yadav et al. (in preparation).  

\smallskip

For absorbers that lie within 100 kpc of the star-forming disk of the CGM host galaxies, the local galaxy radiation most likely plays a role. For the Milky Way, \citet{richter09,bouma19} have included the local disk radiation field in their {\CLOUDY} modeling and have discussed this issue in detail and give a perspective what should be done when modeling the photoionization in CGM clouds. We don't incorporate the effects of nearby galaxies as it is beyond the scope of the current study.

\smallskip

A practical limitation on the CMBM method is the need for significant computing power. The number of {\CLOUDY} models per component in order to construct the required $\log Z$ $\times$ $\log U$ grid, is 2700. Every such model takes an average of 5 minutes of compute time on Intel (R) Xeon (R) CPU E5-2650 v4 $@$ 2.20GHz. Even these relatively simple absorbers have 3-7 components in the different phases, and a stronger, low ionization absorber might have several times more. A serial job would take between 28-65 days to obtain the models covering the desired parameter space. Additionally, the need for more exact UV radiation field models including the contribution from the local galaxy disks, would further complicate the modeling and substantially increase the computation time. To generate these {\CLOUDY} models in a reasonable amount of time, it is important to employ parallel processing methods. Our parallel implementation of generating {\CLOUDY} models makes use of the in-house CyberLAMP cluster infrastructure.

\subsection{Application to understanding weak, low ionization absorbers}
\label{sec:applnweakabs}

The focus of this paper has been a demonstration of the CMBM methodology.  We chose to apply it to four, previously well-studied, weak, low-ionization absorbers. As we described in \S~\ref{sec:Intro}, this population of absorber is of particular interest because of their high metallicities despite locations far from known galaxies. In the future, we intend to apply these methods to a systematic determination of the gas phase properties of the larger sample of 34 $z<0.3$ absorbers from \citet{muzahid2018cos}. However, it is still of interest to consider any general conclusions that we can draw about the population based on the properties of these four absorbers in our present study.

We have concluded that:

1) All of the four low ionization weak absorbers have low ionization clouds that must have solar or supersolar metallicities.

2) Of the clouds that produce low ionization ({\mgii} or {\siii}) absorption, six of them have densities, $\log [n(H)/\cc]$, in the range $-2.9$ to $-2.2$, two of them have $\log [n(H)/\cc] \sim -3.2$, and two have much higher densities of $\sim -1$ as constrained by detected {\feii} absorption.

3) Depending on their densities and metallicities, the line-of-sight thickness of the low ionization clouds range from a fraction of a parsec up to a few hundred parsecs, with typical values of a few to tens of parsecs.  The thickness constraints for the different clouds are shown as symbol sizes in Figure~\ref{fig:comparison}. 

4) Only one of the absorbers ($z=0.22596$ toward HE0153-4520) has a distinct and separate intermediate ionization phase aligned with the low ionization clouds in velocity space.  These intermediate ionization clouds have densities of $\sim -3.6$ and thickness of 100 pc to 1 kpc.  Though it is not covered by spectra, we would expect strong {\civ} absorption from this system based upon its model properties. Such phase structure is typical of $z \sim 1$, weak {\mgii} absorbers, and is envisioned as sheetlike or filamentary higher density structures embedded in or adjacent to lower density structures.

5) Three of the absorbers have evidence of multiple galaxies in the fields, indicating a group environment~\citep{jenkins05,keeney2017characterizing,muzahid2018cos}.  The fourth absorber, $z=0.08094$ toward PHL 1811, is found at 35 kpc from a 0.56$L^*$ galaxy that is within tens of kpc of another galaxy, and could plausibly be interacting.

6) Three of the four absorbers have a collisionally ionized phase giving rise to broad, {\ovi} absorption.  It is interesting to note that the case without {\ovi} is the case of the lower impact parameter, interacting galaxy candidate, which has no evidence of a larger, group environment.  In our larger study, we will investigate the hypothesis that warm/hot {\ovi} could be a characteristic of the group environment.

7) In three of the four absorbers we find compelling evidence for broad {\hi} components in the {\lya} profile that point to collisionally ionized gas that either is related to the broad {\ovi} or traces an even hotter gas component, the million-degree coronal gas halo \citep{richter2020}. Our multi-phase Bayesian modelling provides a powerful method to identify such Coronal Broad {\lya} Absorbers (CBLAs) even in complicated, multi-phase CGM absorbers and thus could be of great help for future, systematic studies of CBLAs in low-redshift galaxies.
   
\smallskip
   
Considering the population of weak, low ionization absorbers at low redshift, we note that many of these absorbers do appear to arise in gas in a narrow density range around $\log [n(H)/\cc] \sim -2.5$, and from gas with a solar or supersolar metallicity.  However, we also note that for gas structures with this density, the column density of low ionization transitions scales directly with the metallicity.  Thus if the gas in this density range was in a small structure or if it had a significantly subsolar metallicity it would be unlikely to be detected in low ionization absorption.  If this population of weak, low ionization absorbers are dominated by gas of this density, we would expect a bias towards high metallicities in our sample.  Since these absorbers are commonly observed they must have a large covering factor, and it is still necessary to explain the mystery of the pervasive nature of such high metallicity gas far from galaxies.  However, we also should consider that lower metallicity gas of this same density will likely exist as well, and we should learn to recognize its absorption signatures.  Finally, we note that some of the gas that might exist in that same density range at $z=1$ would tend to have stronger {\civ} absorption and weaker (or undetected) {\mgii} absorption. Our larger, systematic study of weak, low ionization absorbers will consider the redshift evolution of different populations of gas clouds.  


\smallskip

Surely, weak, low ionization absorbers (as with any class of absorbers) have multiple physical origins, and can arise in gas with various combinations of densities and metallicities.  Furthermore, it is always important to remember that the same population of gas clouds is traced by a different population of absorbers at different redshifts, because of the evolution of the EBR, as well as the evolving processes contributing to the gas conditions.  Finally, many different populations of absorption systems are observed at all redshifts, and these are all connected to one another.  The complexity of this situation, and the variety in absorption systems, make it urgent that we explore the physical conditions of the absorbers in a detailed, yet automated way.  We have demonstrated in this paper, with our application to four absorbers and associated tests, that such a study is now possible.

\section{Conclusion and Future Work}
\label{sec:Conclusion}

Our CMBM method provides great detail about the physical conditions, metallicities and densities, of absorption systems.  We have applied the method to four well-studied weak low-ionization absorbers, and demonstrated its power in deriving accurate results, and estimating realistic uncertainties for the physical properties of the gas. This tool has similar potential for studies of absorbers of all types, and for larger samples of absorbers given its semi-automated nature, and its computational efficiency.  By separating the multiple gas phases, in a larger sample, we can explore possible relationships between metallicity, density, phase structure, and temperature of the circumgalactic medium, with galaxy properties such as impact parameter, luminosity, orientation, and galaxy environment.

\smallskip

This methodology hinges on using the Doppler parameters of metal lines to constrain the Doppler parameters of {\hi} lines, thus breaking degeneracies in the curve of growth, and separating out the contributions to the {\lya} and Lyman series lines from multiple phases of gas.  It is particularly effective when there are at least some radiative transitions observed with high spectral resolution, and when there is high quality data covering many different metal-line transitions as well as many Lyman series lines.  Because of this, we expect the method will transfer well to samples of Lyman break and partial Lyman break systems at various redshifts. In future work, we plan to apply this CMBM method to several large samples of systems in order to obtain a much more rigorous understanding of the complex and multiple processes at work in the circumgalactic medium.

\section*{ACKNOWLEDGEMENTS}
We thank the anonymous referee for their insightful review which improved the quality of the paper. Computations for this research were performed on the Pennsylvania State University's Institute for Computational and Data Sciences' Roar supercomputer. J.C.C., S., J.M.N, and C.W.C. acknowledge support by the National Science Foundation under grant No. AST-1517816. G.G.K. and N.M.N.~acknowledge the support of the Australian Research Council through {\it Discovery Project} grant DP170103470. Parts of this research were supported by the Australian Research Council Centre of Excellence for All Sky Astrophysics in 3 Dimensions (ASTRO 3D), through project number CE170100013. S.M. is supported by the Alexander von Humboldt Foundation, Germany, via the Experienced Researchers Fellowship. We acknowledge the work of people involved in the design, construction and deployment of the COS on-board the Hubble Space Telescope, and thank all those who obtained data for the sight-lines studied in this paper.

\section*{Data availability}
The data underlying this article will be shared on reasonable request to the corresponding author.
\clearpage

\bibliographystyle{mnras}
\bibliography{references}

\begin{thebibliography}{}
\makeatletter
\relax
\def\mn@urlcharsother{\let\do\@makeother \do\$\do\&\do\#\do\^\do\_\do\%\do\~}
\def\mn@doi{\begingroup\mn@urlcharsother \@ifnextchar [ {\mn@doi@}
  {\mn@doi@[]}}
\def\mn@doi@[#1]#2{\def\@tempa{#1}\ifx\@tempa\@empty \href
  {http://dx.doi.org/#2} {doi:#2}\else \href {http://dx.doi.org/#2} {#1}\fi
  \endgroup}
\def\mn@eprint#1#2{\mn@eprint@#1:#2::\@nil}
\def\mn@eprint@arXiv#1{\href {http://arxiv.org/abs/#1} {{\tt arXiv:#1}}}
\def\mn@eprint@dblp#1{\href {http://dblp.uni-trier.de/rec/bibtex/#1.xml}
  {dblp:#1}}
\def\mn@eprint@#1:#2:#3:#4\@nil{\def\@tempa {#1}\def\@tempb {#2}\def\@tempc
  {#3}\ifx \@tempc \@empty \let \@tempc \@tempb \let \@tempb \@tempa \fi \ifx
  \@tempb \@empty \def\@tempb {arXiv}\fi \@ifundefined
  {mn@eprint@\@tempb}{\@tempb:\@tempc}{\expandafter \expandafter \csname
  mn@eprint@\@tempb\endcsname \expandafter{\@tempc}}}

\bibitem[\protect\citeauthoryear{Adelberger, Shapley, Steidel, Pettini, Erb  \&
  Reddy}{Adelberger et~al.}{2005}]{adelberger2005connection}
Adelberger K.~L.,  Shapley A.~E.,  Steidel C.~C.,  Pettini M.,  Erb D.~K.,
  Reddy N.~A.,  2005, The Astrophysical Journal, 629, 636

\bibitem[\protect\citeauthoryear{{Angl{\'e}s-Alc{\'a}zar},
  {Faucher-Gigu{\`e}re}, {Kere{\v{s}}}, {Hopkins}, {Quataert}  \&
  {Murray}}{{Angl{\'e}s-Alc{\'a}zar} et~al.}{2017}]{angles2017}
{Angl{\'e}s-Alc{\'a}zar} D.,  {Faucher-Gigu{\`e}re} C.-A.,  {Kere{\v{s}}} D.,
  {Hopkins} P.~F.,  {Quataert} E.,   {Murray} N.,  2017, \mn@doi [\mnras]
  {10.1093/mnras/stx1517}, \href
  {https://ui.adsabs.harvard.edu/abs/2017MNRAS.470.4698A} {470, 4698}

\bibitem[\protect\citeauthoryear{Barber, Dobkin  \& Huhdanpaa}{Barber
  et~al.}{1996}]{barber1996quickhull}
Barber C.~B.,  Dobkin D.~P.,   Huhdanpaa H.,  1996, ACM Transactions on
  Mathematical Software (TOMS), 22, 469

\bibitem[\protect\citeauthoryear{Bergeron}{Bergeron}{1986}]{bergeron1986mg}
Bergeron J.,  1986, Astronomy and Astrophysics, 155, L8

\bibitem[\protect\citeauthoryear{Bergeron \& Boiss{\'e}}{Bergeron \&
  Boiss{\'e}}{1991}]{bergeron1991sample}
Bergeron J.,  Boiss{\'e} P.,  1991, Astronomy and Astrophysics, 243, 344

\bibitem[\protect\citeauthoryear{{Bordoloi} et~al.,}{{Bordoloi}
  et~al.}{2011}]{Bordoloi2011}
{Bordoloi} R.,  et~al., 2011, \mn@doi [\apj] {10.1088/0004-637X/743/1/10},
  \href {https://ui.adsabs.harvard.edu/abs/2011ApJ...743...10B} {743, 10}

\bibitem[\protect\citeauthoryear{Bordoloi et~al.,}{Bordoloi
  et~al.}{2014}]{bordoloi2014cos}
Bordoloi R.,  et~al., 2014, The Astrophysical Journal, 796, 136

\bibitem[\protect\citeauthoryear{{Borthakur} et~al.,}{{Borthakur}
  et~al.}{2015}]{Borthakur2015}
{Borthakur} S.,  et~al., 2015, \mn@doi [\apj] {10.1088/0004-637X/813/1/46},
  \href {https://ui.adsabs.harvard.edu/abs/2015ApJ...813...46B} {813, 46}

\bibitem[\protect\citeauthoryear{{Bouch{\'e}}, {Hohensee}, {Vargas},
  {Kacprzak}, {Martin}, {Cooke}  \& {Churchill}}{{Bouch{\'e}}
  et~al.}{2012}]{bouche2012}
{Bouch{\'e}} N.,  {Hohensee} W.,  {Vargas} R.,  {Kacprzak} G.~G.,  {Martin}
  C.~L.,  {Cooke} J.,   {Churchill} C.~W.,  2012, \mn@doi [\mnras]
  {10.1111/j.1365-2966.2012.21114.x}, \href
  {https://ui.adsabs.harvard.edu/abs/2012MNRAS.426..801B} {426, 801}

\bibitem[\protect\citeauthoryear{{Bouma}, {Richter}  \& {Fechner}}{{Bouma}
  et~al.}{2019}]{bouma19}
{Bouma} S.~J.~D.,  {Richter} P.,   {Fechner} C.,  2019, \mn@doi [\aap]
  {10.1051/0004-6361/201935078}, \href
  {https://ui.adsabs.harvard.edu/abs/2019A&A...627A..20B} {627, A20}

\bibitem[\protect\citeauthoryear{Buchner et~al.,}{Buchner
  et~al.}{2014}]{buchner2014x}
Buchner J.,  et~al., 2014, Astronomy \& Astrophysics, 564, A125

\bibitem[\protect\citeauthoryear{Charlton, Mellon, Rigby  \&
  Churchill}{Charlton et~al.}{2000}]{charlton2000anticipating}
Charlton J.~C.,  Mellon R.~R.,  Rigby J.~R.,   Churchill C.~W.,  2000, The
  Astrophysical Journal, 545, 635

\bibitem[\protect\citeauthoryear{Charlton, Ding, Zonak, Churchill, Bond  \&
  Rigby}{Charlton et~al.}{2003}]{charlton2003high}
Charlton J.~C.,  Ding J.,  Zonak S.~G.,  Churchill C.~W.,  Bond N.~A.,   Rigby
  J.~R.,  2003, The Astrophysical Journal, 589, 111

\bibitem[\protect\citeauthoryear{Chatzikos et~al.,}{Chatzikos
  et~al.}{2015}]{chatzikos2015implications}
Chatzikos M.,  et~al., 2015, Monthly Notices of the Royal Astronomical Society,
  446, 1234

\bibitem[\protect\citeauthoryear{Chen \& Mulchaey}{Chen \&
  Mulchaey}{2009}]{chen2009probing}
Chen H.-W.,  Mulchaey J.~S.,  2009, The Astrophysical Journal, 701, 1219

\bibitem[\protect\citeauthoryear{Chen, Lanzetta  \& Webb}{Chen
  et~al.}{2001}]{chen2001origin}
Chen H.-W.,  Lanzetta K.~M.,   Webb J.~K.,  2001, The Astrophysical Journal,
  556, 158

\bibitem[\protect\citeauthoryear{Churchill \& Charlton}{Churchill \&
  Charlton}{1999}]{churchill1999multiple}
Churchill C.~W.,  Charlton J.~C.,  1999, The Astronomical Journal, 118, 59

\bibitem[\protect\citeauthoryear{{Churchill}, {Steidel}  \& {Vogt}}{{Churchill}
  et~al.}{1996}]{Churchill1996}
{Churchill} C.~W.,  {Steidel} C.~C.,   {Vogt} S.~S.,  1996, \mn@doi [\apj]
  {10.1086/177960}, \href
  {https://ui.adsabs.harvard.edu/abs/1996ApJ...471..164C} {471, 164}

\bibitem[\protect\citeauthoryear{Churchill, Rigby, Charlton  \& Vogt}{Churchill
  et~al.}{1999}]{churchill1999population}
Churchill C.~W.,  Rigby J.~R.,  Charlton J.~C.,   Vogt S.~S.,  1999, The
  Astrophysical Journal Supplement Series, 120, 51

\bibitem[\protect\citeauthoryear{{Churchill}, {Mellon}, {Charlton}, {Jannuzi},
  {Kirhakos}, {Steidel}  \& {Schneider}}{{Churchill}
  et~al.}{2000}]{Churchill2000}
{Churchill} C.~W.,  {Mellon} R.~R.,  {Charlton} J.~C.,  {Jannuzi} B.~T.,
  {Kirhakos} S.,  {Steidel} C.~C.,   {Schneider} D.~P.,  2000, \mn@doi [\apj]
  {10.1086/317120}, \href
  {https://ui.adsabs.harvard.edu/abs/2000ApJ...543..577C} {543, 577}

\bibitem[\protect\citeauthoryear{{Churchill}, {Vander Vliet}, {Trujillo-Gomez},
  {Kacprzak}  \& {Klypin}}{{Churchill} et~al.}{2015}]{Churchill2015}
{Churchill} C.~W.,  {Vander Vliet} J.~R.,  {Trujillo-Gomez} S.,  {Kacprzak}
  G.~G.,   {Klypin} A.,  2015, \mn@doi [\apj] {10.1088/0004-637X/802/1/10},
  \href {https://ui.adsabs.harvard.edu/abs/2015ApJ...802...10C} {802, 10}

\bibitem[\protect\citeauthoryear{{Churchill}, {Evans}, {Stemock}, {Nielsen},
  {Kacprzak}  \& {Murphy}}{{Churchill} et~al.}{2020}]{Churchill2020}
{Churchill} C.~W.,  {Evans} J.~L.,  {Stemock} B.,  {Nielsen} N.~M.,  {Kacprzak}
  G.~G.,   {Murphy} M.~T.,  2020, arXiv e-prints, \href
  {https://ui.adsabs.harvard.edu/abs/2020arXiv200808487C} {p. arXiv:2008.08487}

\bibitem[\protect\citeauthoryear{Danforth et~al.,}{Danforth
  et~al.}{2016}]{danforth2016HST}
Danforth C.~W.,  et~al., 2016, The Astrophysical Journal, 817, 111

\bibitem[\protect\citeauthoryear{{Danovich}, {Dekel}, {Hahn}, {Ceverino}  \&
  {Primack}}{{Danovich} et~al.}{2015}]{Danovich2015}
{Danovich} M.,  {Dekel} A.,  {Hahn} O.,  {Ceverino} D.,   {Primack} J.,  2015,
  \mn@doi [\mnras] {10.1093/mnras/stv270}, \href
  {https://ui.adsabs.harvard.edu/abs/2015MNRAS.449.2087D} {449, 2087}

\bibitem[\protect\citeauthoryear{Dekker, D’Odorico, Kaufer, Delabre,
  Kotzlowski, Iye  \& Moorwood}{Dekker et~al.}{2000}]{dekker2000optical}
Dekker H.,  D’Odorico S.,  Kaufer A.,  Delabre B.,  Kotzlowski H.,  Iye M.,
  Moorwood A.,  2000, in Proc. SPIE. p.~534

\bibitem[\protect\citeauthoryear{Ding, Charlton, Bond, Zonak  \&
  Churchill}{Ding et~al.}{2003a}]{ding2003quadruple}
Ding J.,  Charlton J.~C.,  Bond N.~A.,  Zonak S.~G.,   Churchill C.~W.,  2003a,
  The Astrophysical Journal, 587, 551

\bibitem[\protect\citeauthoryear{Ding, Charlton, Churchill  \& Palma}{Ding
  et~al.}{2003b}]{ding2003multiphase}
Ding J.,  Charlton J.~C.,  Churchill C.~W.,   Palma C.,  2003b, The
  Astrophysical Journal, 590, 746

\bibitem[\protect\citeauthoryear{Ding, Charlton  \& Churchill}{Ding
  et~al.}{2005}]{ding2005absorption}
Ding J.,  Charlton J.~C.,   Churchill C.~W.,  2005, The Astrophysical Journal,
  621, 615

\bibitem[\protect\citeauthoryear{Ferland et~al.,}{Ferland
  et~al.}{2017}]{cloudy17}
Ferland G.,  et~al., 2017, Revista mexicana de astronom{\'\i}a y
  astrof{\'\i}sica, 53

\bibitem[\protect\citeauthoryear{Feroz, Hobson  \& Bridges}{Feroz
  et~al.}{2009}]{feroz2009multinest}
Feroz F.,  Hobson M.,   Bridges M.,  2009, Monthly Notices of the Royal
  Astronomical Society, 398, 1601

\bibitem[\protect\citeauthoryear{Ford, Oppenheimer, Dav{\'e}, Katz, Kollmeier
  \& Weinberg}{Ford et~al.}{2013}]{ford2013hydrogen}
Ford A.~B.,  Oppenheimer B.~D.,  Dav{\'e} R.,  Katz N.,  Kollmeier J.~A.,
  Weinberg D.~H.,  2013, Monthly Notices of the Royal Astronomical Society,
  432, 89

\bibitem[\protect\citeauthoryear{{Ford}, {Dav{\'e}}, {Oppenheimer}, {Katz},
  {Kollmeier}, {Thompson}  \& {Weinberg}}{{Ford} et~al.}{2014}]{Ford2014}
{Ford} A.~B.,  {Dav{\'e}} R.,  {Oppenheimer} B.~D.,  {Katz} N.,  {Kollmeier}
  J.~A.,  {Thompson} R.,   {Weinberg} D.~H.,  2014, \mn@doi [\mnras]
  {10.1093/mnras/stu1418}, \href
  {https://ui.adsabs.harvard.edu/abs/2014MNRAS.444.1260F} {444, 1260}

\bibitem[\protect\citeauthoryear{{Fujita}, {Misawa}, {Charlton}, {Meiksin}  \&
  {Mac Low}}{{Fujita} et~al.}{2020}]{Fujita2020}
{Fujita} A.,  {Misawa} T.,  {Charlton} J.~C.,  {Meiksin} A.,   {Mac Low} M.-M.,
   2020, arXiv e-prints, \href
  {https://ui.adsabs.harvard.edu/abs/2020arXiv200909954F} {p. arXiv:2009.09954}

\bibitem[\protect\citeauthoryear{Gnat \& Sternberg}{Gnat \&
  Sternberg}{2007}]{gnat2007time}
Gnat O.,  Sternberg A.,  2007, The Astrophysical Journal Supplement Series,
  168, 213

\bibitem[\protect\citeauthoryear{{Grevesse}, {Asplund}, {Sauval}  \&
  {Scott}}{{Grevesse} et~al.}{2010}]{grevesse2011chemical}
{Grevesse} N.,  {Asplund} M.,  {Sauval} A.~J.,   {Scott} P.,  2010, \mn@doi
  [\apss] {10.1007/s10509-010-0288-z}, \href
  {https://ui.adsabs.harvard.edu/abs/2010Ap&SS.328..179G} {328, 179}

\bibitem[\protect\citeauthoryear{{Haardt} \& {Madau}}{{Haardt} \&
  {Madau}}{2001}]{Haardt2001}
{Haardt} F.,  {Madau} P.,  2001, in {Neumann} D.~M.,  {Tran} J.~T.~V.,  eds,
  Clusters of Galaxies and the High Redshift Universe Observed in X-rays. p.~64
  (\mn@eprint {arXiv} {astro-ph/0106018})

\bibitem[\protect\citeauthoryear{Haardt \& Madau}{Haardt \&
  Madau}{2012}]{haardt2012radiative}
Haardt F.,  Madau P.,  2012, The Astrophysical Journal, 746, 125

\bibitem[\protect\citeauthoryear{{Hafen} et~al.,}{{Hafen}
  et~al.}{2019}]{Hafen2019}
{Hafen} Z.,  et~al., 2019, \mn@doi [\mnras] {10.1093/mnras/stz1773}, \href
  {https://ui.adsabs.harvard.edu/abs/2019MNRAS.488.1248H} {488, 1248}

\bibitem[\protect\citeauthoryear{{Jenkins}, {Bowen}, {Tripp}, {Sembach},
  {Leighly}, {Halpern}  \& {Lauroesch}}{{Jenkins}
  et~al.}{2003}]{jenkins2003absorption}
{Jenkins} E.~B.,  {Bowen} D.~V.,  {Tripp} T.~M.,  {Sembach} K.~R.,  {Leighly}
  K.~M.,  {Halpern} J.~P.,   {Lauroesch} J.~T.,  2003, \mn@doi [\aj]
  {10.1086/375321}, \href
  {https://ui.adsabs.harvard.edu/abs/2003AJ....125.2824J} {125, 2824}

\bibitem[\protect\citeauthoryear{{Jenkins}, {Bowen}, {Tripp}  \&
  {Sembach}}{{Jenkins} et~al.}{2005}]{jenkins05}
{Jenkins} E.~B.,  {Bowen} D.~V.,  {Tripp} T.~M.,   {Sembach} K.~R.,  2005,
  \mn@doi [\apj] {10.1086/428878}, \href
  {https://ui.adsabs.harvard.edu/abs/2005ApJ...623..767J} {623, 767}

\bibitem[\protect\citeauthoryear{Johnson, Chen  \& Mulchaey}{Johnson
  et~al.}{2015}]{johnson2015possible}
Johnson S.~D.,  Chen H.-W.,   Mulchaey J.~S.,  2015, Monthly Notices of the
  Royal Astronomical Society, 449, 3263

\bibitem[\protect\citeauthoryear{Jones, Misawa, Charlton, Mshar  \&
  Ferland}{Jones et~al.}{2010}]{jones2010bare}
Jones T.~M.,  Misawa T.,  Charlton J.~C.,  Mshar A.~C.,   Ferland G.~J.,  2010,
  The Astrophysical Journal, 715, 1497

\bibitem[\protect\citeauthoryear{{Kaaret} et~al.,}{{Kaaret}
  et~al.}{2020}]{Kaaret2020}
{Kaaret} P.,  et~al., 2020, \mn@doi [Nature Astronomy]
  {10.1038/s41550-020-01215-w}, \href
  {https://ui.adsabs.harvard.edu/abs/2020NatAs...4.1072K} {4, 1072}

\bibitem[\protect\citeauthoryear{{Kacprzak}, {Churchill}, {Steidel}  \&
  {Murphy}}{{Kacprzak} et~al.}{2008}]{kacprzak2008halo}
{Kacprzak} G.~G.,  {Churchill} C.~W.,  {Steidel} C.~C.,   {Murphy} M.~T.,
  2008, \mn@doi [\aj] {10.1088/0004-6256/135/3/922}, \href
  {https://ui.adsabs.harvard.edu/abs/2008AJ....135..922K} {135, 922}

\bibitem[\protect\citeauthoryear{Kacprzak, Churchill  \& Nielsen}{Kacprzak
  et~al.}{2012}]{kacprzak2012tracing}
Kacprzak G.~G.,  Churchill C.~W.,   Nielsen N.~M.,  2012, The Astrophysical
  Journal Letters, 760, L7

\bibitem[\protect\citeauthoryear{{Kacprzak}, {Muzahid}, {Churchill}, {Nielsen}
  \& {Charlton}}{{Kacprzak} et~al.}{2015}]{kacprzak2015azimuthal}
{Kacprzak} G.~G.,  {Muzahid} S.,  {Churchill} C.~W.,  {Nielsen} N.~M.,
  {Charlton} J.~C.,  2015, \mn@doi [\apj] {10.1088/0004-637X/815/1/22}, \href
  {https://ui.adsabs.harvard.edu/abs/2015ApJ...815...22K} {815, 22}

\bibitem[\protect\citeauthoryear{{Kacprzak}, {Pointon}, {Nielsen}, {Churchill},
  {Muzahid}  \& {Charlton}}{{Kacprzak} et~al.}{2019}]{Kacprzak2019ApJ}
{Kacprzak} G.~G.,  {Pointon} S.~K.,  {Nielsen} N.~M.,  {Churchill} C.~W.,
  {Muzahid} S.,   {Charlton} J.~C.,  2019, \mn@doi [\apj]
  {10.3847/1538-4357/ab4c3c}, \href
  {https://ui.adsabs.harvard.edu/abs/2019ApJ...886...91K} {886, 91}

\bibitem[\protect\citeauthoryear{Keeney et~al.,}{Keeney
  et~al.}{2017}]{keeney2017characterizing}
Keeney B.~A.,  et~al., 2017, The Astrophysical Journal Supplement Series, 230,
  6

\bibitem[\protect\citeauthoryear{Keeney et~al.,}{Keeney
  et~al.}{2018}]{keeney2018galaxy}
Keeney B.~A.,  et~al., 2018, The Astrophysical Journal Supplement Series, 237,
  11

\bibitem[\protect\citeauthoryear{{Khaire} \& {Srianand}}{{Khaire} \&
  {Srianand}}{2015}]{KS15}
{Khaire} V.,  {Srianand} R.,  2015, \mn@doi [\apj]
  {10.1088/0004-637X/805/1/33}, \href
  {https://ui.adsabs.harvard.edu/abs/2015ApJ...805...33K} {805, 33}

\bibitem[\protect\citeauthoryear{Krogager}{Krogager}{2018}]{krogager2018voigtfit}
Krogager J.-K.,  2018, arXiv preprint arXiv:1803.01187

\bibitem[\protect\citeauthoryear{Lacki \& Charlton}{Lacki \&
  Charlton}{2010}]{lacki2010z}
Lacki B.~C.,  Charlton J.~C.,  2010, Monthly Notices of the Royal Astronomical
  Society, 403, 1556

\bibitem[\protect\citeauthoryear{{Lanzetta} \& {Bowen}}{{Lanzetta} \&
  {Bowen}}{1992}]{Lanzetta1992}
{Lanzetta} K.~M.,  {Bowen} D.~V.,  1992, \mn@doi [\apj] {10.1086/171325}, \href
  {https://ui.adsabs.harvard.edu/abs/1992ApJ...391...48L} {391, 48}

\bibitem[\protect\citeauthoryear{Lehner et~al.,}{Lehner
  et~al.}{2013}]{lehner2013bimodal}
Lehner N.,  et~al., 2013, The Astrophysical Journal, 770, 138

\bibitem[\protect\citeauthoryear{{Liang} \& {Chen}}{{Liang} \&
  {Chen}}{2014}]{liangchen2014}
{Liang} C.~J.,  {Chen} H.-W.,  2014, \mn@doi [\mnras] {10.1093/mnras/stu1901},
  \href {https://ui.adsabs.harvard.edu/abs/2014MNRAS.445.2061L} {445, 2061}

\bibitem[\protect\citeauthoryear{Liang \& Remming}{Liang \&
  Remming}{2020}]{liang2020model}
Liang C.~J.,  Remming I.,  2020, Monthly Notices of the Royal Astronomical
  Society, 491, 5056

\bibitem[\protect\citeauthoryear{Lynch \& Charlton}{Lynch \&
  Charlton}{2007}]{lynch2007physical}
Lynch R.~S.,  Charlton J.~C.,  2007, The Astrophysical Journal, 666, 64

\bibitem[\protect\citeauthoryear{{Martin}, {Shapley}, {Coil}, {Kornei},
  {Bundy}, {Weiner}, {Noeske}  \& {Schiminovich}}{{Martin}
  et~al.}{2012}]{martin2012}
{Martin} C.~L.,  {Shapley} A.~E.,  {Coil} A.~L.,  {Kornei} K.~A.,  {Bundy} K.,
  {Weiner} B.~J.,  {Noeske} K.~G.,   {Schiminovich} D.,  2012, \mn@doi [\apj]
  {10.1088/0004-637X/760/2/127}, \href
  {https://ui.adsabs.harvard.edu/abs/2012ApJ...760..127M} {760, 127}

\bibitem[\protect\citeauthoryear{Masiero, Charlton, Ding, Churchill  \&
  Kacprzak}{Masiero et~al.}{2005}]{masiero2005models}
Masiero J.~R.,  Charlton J.~C.,  Ding J.,  Churchill C.~W.,   Kacprzak G.,
  2005, The Astrophysical Journal, 623, 57

\bibitem[\protect\citeauthoryear{Massa et~al.}{Massa
  et~al.}{2013}]{massa2013cos}
Massa D.,  et~al., 2013, COS Data Handbook, HST Data Handbooks

\bibitem[\protect\citeauthoryear{McCourt, Sharma, Quataert  \& Parrish}{McCourt
  et~al.}{2012}]{mccourt2012thermal}
McCourt M.,  Sharma P.,  Quataert E.,   Parrish I.~J.,  2012, Monthly Notices
  of the Royal Astronomical Society, 419, 3319

\bibitem[\protect\citeauthoryear{Milutinovi{\'c}, Rigby, Masiero, Lynch, Palma
  \& Charlton}{Milutinovi{\'c} et~al.}{2006}]{milutinovic2006nature}
Milutinovi{\'c} N.,  Rigby J.~R.,  Masiero J.~R.,  Lynch R.~S.,  Palma C.,
  Charlton J.~C.,  2006, The Astrophysical Journal, 641, 190

\bibitem[\protect\citeauthoryear{Misawa, Charlton  \& Narayanan}{Misawa
  et~al.}{2008}]{misawa2008supersolar}
Misawa T.,  Charlton J.~C.,   Narayanan A.,  2008, The Astrophysical Journal,
  679, 220

\bibitem[\protect\citeauthoryear{{Murphy}}{{Murphy}}{2016}]{murphy2016}
{Murphy} M.,  2016, {Uves\_Popler: Uves\_Popler: Post-Pipeline Echelle
  Reduction Software}, \mn@doi{10.5281/zenodo.56158}

\bibitem[\protect\citeauthoryear{{Murphy}, {Kacprzak}, {Savorgnan}  \&
  {Carswell}}{{Murphy} et~al.}{2019}]{murphy2019}
{Murphy} M.~T.,  {Kacprzak} G.~G.,  {Savorgnan} G. A.~D.,   {Carswell} R.~F.,
  2019, \mn@doi [\mnras] {10.1093/mnras/sty2834}, \href
  {https://ui.adsabs.harvard.edu/abs/2019MNRAS.482.3458M} {482, 3458}

\bibitem[\protect\citeauthoryear{Muzahid, Kacprzak, Churchill, Charlton,
  Nielsen, Mathes  \& Trujillo-Gomez}{Muzahid
  et~al.}{2015}]{muzahid2015extreme}
Muzahid S.,  Kacprzak G.~G.,  Churchill C.~W.,  Charlton J.~C.,  Nielsen N.~M.,
   Mathes N.~L.,   Trujillo-Gomez S.,  2015, The Astrophysical Journal, 811,
  132

\bibitem[\protect\citeauthoryear{Muzahid, Fonseca, Roberts, Rosenwasser,
  Richter, Narayanan, Churchill  \& Charlton}{Muzahid
  et~al.}{2018}]{muzahid2018cos}
Muzahid S.,  Fonseca G.,  Roberts A.,  Rosenwasser B.,  Richter P.,  Narayanan
  A.,  Churchill C.,   Charlton J.,  2018, Monthly Notices of the Royal
  Astronomical Society, 476, 4965

\bibitem[\protect\citeauthoryear{Narayanan, Charlton, Masiero  \&
  Lynch}{Narayanan et~al.}{2005}]{narayanan2005survey}
Narayanan A.,  Charlton J.~C.,  Masiero J.~R.,   Lynch R.,  2005, The
  Astrophysical Journal, 632, 92

\bibitem[\protect\citeauthoryear{Narayanan, Misawa, Charlton  \& Kim}{Narayanan
  et~al.}{2007}]{narayanan2007survey}
Narayanan A.,  Misawa T.,  Charlton J.~C.,   Kim T.-S.,  2007, The
  Astrophysical Journal, 660, 1093

\bibitem[\protect\citeauthoryear{Narayanan, Charlton, Misawa, Green  \&
  Kim}{Narayanan et~al.}{2008}]{narayanan2008chemical}
Narayanan A.,  Charlton J.~C.,  Misawa T.,  Green R.~E.,   Kim T.-S.,  2008,
  The Astrophysical Journal, 689, 782

\bibitem[\protect\citeauthoryear{{Nielsen}, {Churchill}, {Kacprzak}  \&
  {Murphy}}{{Nielsen} et~al.}{2013a}]{Nielsen20131}
{Nielsen} N.~M.,  {Churchill} C.~W.,  {Kacprzak} G.~G.,   {Murphy} M.~T.,
  2013a, \mn@doi [\apj] {10.1088/0004-637X/776/2/114}, \href
  {https://ui.adsabs.harvard.edu/abs/2013ApJ...776..114N} {776, 114}

\bibitem[\protect\citeauthoryear{{Nielsen}, {Churchill}  \&
  {Kacprzak}}{{Nielsen} et~al.}{2013b}]{Nielsen20132}
{Nielsen} N.~M.,  {Churchill} C.~W.,   {Kacprzak} G.~G.,  2013b, \mn@doi [\apj]
  {10.1088/0004-637X/776/2/115}, \href
  {https://ui.adsabs.harvard.edu/abs/2013ApJ...776..115N} {776, 115}

\bibitem[\protect\citeauthoryear{{Nielsen}, {Churchill}, {Kacprzak}, {Murphy}
  \& {Evans}}{{Nielsen} et~al.}{2015}]{Nielsen2015}
{Nielsen} N.~M.,  {Churchill} C.~W.,  {Kacprzak} G.~G.,  {Murphy} M.~T.,
  {Evans} J.~L.,  2015, \mn@doi [\apj] {10.1088/0004-637X/812/1/83}, \href
  {https://ui.adsabs.harvard.edu/abs/2015ApJ...812...83N} {812, 83}

\bibitem[\protect\citeauthoryear{{Nielsen}, {Kacprzak}, {Pointon}, {Churchill}
  \& {Murphy}}{{Nielsen} et~al.}{2018}]{nielsen2018magiicat}
{Nielsen} N.~M.,  {Kacprzak} G.~G.,  {Pointon} S.~K.,  {Churchill} C.~W.,
  {Murphy} M.~T.,  2018, \mn@doi [\apj] {10.3847/1538-4357/aaedbd}, \href
  {https://ui.adsabs.harvard.edu/abs/2018ApJ...869..153N} {869, 153}

\bibitem[\protect\citeauthoryear{{Oppenheimer} et~al.,}{{Oppenheimer}
  et~al.}{2016}]{Oppenheimer2016}
{Oppenheimer} B.~D.,  et~al., 2016, \mn@doi [\mnras] {10.1093/mnras/stw1066},
  \href {https://ui.adsabs.harvard.edu/abs/2016MNRAS.460.2157O} {460, 2157}

\bibitem[\protect\citeauthoryear{Peeples et~al.,}{Peeples
  et~al.}{2019}]{peeples2019figuring}
Peeples M.~S.,  et~al., 2019, The Astrophysical Journal, 873, 129

\bibitem[\protect\citeauthoryear{{Peroux}, {Nelson}, {van de Voort},
  {Pillepich}, {Marinacci}, {Vogelsberger}  \& {Hernquist}}{{Peroux}
  et~al.}{2020}]{Peroux2020}
{Peroux} C.,  {Nelson} D.,  {van de Voort} F.,  {Pillepich} A.,  {Marinacci}
  F.,  {Vogelsberger} M.,   {Hernquist} L.,  2020, arXiv e-prints, \href
  {https://ui.adsabs.harvard.edu/abs/2020arXiv200907809P} {p. arXiv:2009.07809}

\bibitem[\protect\citeauthoryear{Peterson, Wanders, Bertram, Hunley, Pogge  \&
  Wagner}{Peterson et~al.}{1998}]{peterson1998optical}
Peterson B.~M.,  Wanders I.,  Bertram R.,  Hunley J.~F.,  Pogge R.~W.,   Wagner
  R.~M.,  1998, ApJ, 501, 82

\bibitem[\protect\citeauthoryear{{Pointon}, {Nielsen}, {Kacprzak}, {Muzahid},
  {Churchill}  \& {Charlton}}{{Pointon} et~al.}{2017}]{Pointon2017}
{Pointon} S.~K.,  {Nielsen} N.~M.,  {Kacprzak} G.~G.,  {Muzahid} S.,
  {Churchill} C.~W.,   {Charlton} J.~C.,  2017, \mn@doi [\apj]
  {10.3847/1538-4357/aa7743}, \href
  {https://ui.adsabs.harvard.edu/abs/2017ApJ...844...23P} {844, 23}

\bibitem[\protect\citeauthoryear{Pointon, Kacprzak, Nielsen, Muzahid, Murphy,
  Churchill  \& Charlton}{Pointon et~al.}{2019}]{pointon2019relationship}
Pointon S.~K.,  Kacprzak G.~G.,  Nielsen N.~M.,  Muzahid S.,  Murphy M.~T.,
  Churchill C.~W.,   Charlton J.~C.,  2019, The Astrophysical Journal, 883, 78

\bibitem[\protect\citeauthoryear{Pradeep, Sankar, Umasree, Narayanan, Khaire,
  Gebhardt, Sameer  \& Charlton}{Pradeep et~al.}{2020}]{jayadev20}
Pradeep J.,  Sankar S.,  Umasree T.~M.,  Narayanan A.,  Khaire V.,  Gebhardt
  M.,  Sameer  Charlton J.~C.,  2020, \mn@doi [Monthly Notices of the Royal
  Astronomical Society] {10.1093/mnras/staa184}, 493, 250

\bibitem[\protect\citeauthoryear{{Prochaska}, {Weiner}, {Chen}, {Mulchaey}  \&
  {Cooksey}}{{Prochaska} et~al.}{2011}]{Prochaska2011}
{Prochaska} J.~X.,  {Weiner} B.,  {Chen} H.~W.,  {Mulchaey} J.,   {Cooksey} K.,
   2011, \mn@doi [\apj] {10.1088/0004-637X/740/2/91}, \href
  {https://ui.adsabs.harvard.edu/abs/2011ApJ...740...91P} {740, 91}

\bibitem[\protect\citeauthoryear{{Richter}}{{Richter}}{2020}]{richter2020}
{Richter} P.,  2020, \mn@doi [\apj] {10.3847/1538-4357/ab7937}, \href
  {https://ui.adsabs.harvard.edu/abs/2020ApJ...892...33R} {892, 33}

\bibitem[\protect\citeauthoryear{{Richter}, {Charlton}, {Fangano}, {Bekhti}  \&
  {Masiero}}{{Richter} et~al.}{2009}]{richter09}
{Richter} P.,  {Charlton} J.~C.,  {Fangano} A. P.~M.,  {Bekhti} N.~B.,
  {Masiero} J.~R.,  2009, \mn@doi [\apj] {10.1088/0004-637X/695/2/1631}, \href
  {https://ui.adsabs.harvard.edu/abs/2009ApJ...695.1631R} {695, 1631}

\bibitem[\protect\citeauthoryear{{Richter}, {Krause}, {Fechner}, {Charlton}  \&
  {Murphy}}{{Richter} et~al.}{2011}]{Richter2011}
{Richter} P.,  {Krause} F.,  {Fechner} C.,  {Charlton} J.~C.,   {Murphy} M.~T.,
   2011, \mn@doi [\aap] {10.1051/0004-6361/201015566}, \href
  {https://ui.adsabs.harvard.edu/abs/2011A&A...528A..12R} {528, A12}

\bibitem[\protect\citeauthoryear{{Richter}, {Wakker}, {Fechner}, {Herenz},
  {Tepper-Garc{\'\i}a}  \& {Fox}}{{Richter} et~al.}{2016}]{richter2016}
{Richter} P.,  {Wakker} B.~P.,  {Fechner} C.,  {Herenz} P.,
  {Tepper-Garc{\'\i}a} T.,   {Fox} A.~J.,  2016, \mn@doi [\aap]
  {10.1051/0004-6361/201527038}, \href
  {https://ui.adsabs.harvard.edu/abs/2016A&A...590A..68R} {590, A68}

\bibitem[\protect\citeauthoryear{{Richter} et~al.,}{{Richter}
  et~al.}{2018}]{richter2018}
{Richter} P.,  et~al., 2018, \mn@doi [\apj] {10.3847/1538-4357/aae838}, \href
  {https://ui.adsabs.harvard.edu/abs/2018ApJ...868..112R} {868, 112}

\bibitem[\protect\citeauthoryear{Rigby, Charlton  \& Churchill}{Rigby
  et~al.}{2002}]{rigby2002population}
Rigby J.~R.,  Charlton J.~C.,   Churchill C.~W.,  2002, The Astrophysical
  Journal, 565, 743

\bibitem[\protect\citeauthoryear{Rosenwasser, Muzahid, Charlton, Kacprzak,
  Wakker  \& Churchill}{Rosenwasser
  et~al.}{2018}]{rosenwasser2018understanding}
Rosenwasser B.,  Muzahid S.,  Charlton J.~C.,  Kacprzak G.~G.,  Wakker B.~P.,
  Churchill C.~W.,  2018, Monthly Notices of the Royal Astronomical Society,
  476, 2258

\bibitem[\protect\citeauthoryear{{Rubin}, {Prochaska}, {Koo}  \&
  {Phillips}}{{Rubin} et~al.}{2012}]{Rubin2012}
{Rubin} K. H.~R.,  {Prochaska} J.~X.,  {Koo} D.~C.,   {Phillips} A.~C.,  2012,
  \mn@doi [\apjl] {10.1088/2041-8205/747/2/L26}, \href
  {https://ui.adsabs.harvard.edu/abs/2012ApJ...747L..26R} {747, L26}

\bibitem[\protect\citeauthoryear{{Rubin}, {Prochaska}, {Koo}, {Phillips},
  {Martin}  \& {Winstrom}}{{Rubin} et~al.}{2014}]{Rubin2014}
{Rubin} K. H.~R.,  {Prochaska} J.~X.,  {Koo} D.~C.,  {Phillips} A.~C.,
  {Martin} C.~L.,   {Winstrom} L.~O.,  2014, \mn@doi [\apj]
  {10.1088/0004-637X/794/2/156}, \href
  {https://ui.adsabs.harvard.edu/abs/2014ApJ...794..156R} {794, 156}

\bibitem[\protect\citeauthoryear{Savage, Narayanan, Lehner  \& Wakker}{Savage
  et~al.}{2011}]{savage2011multiphase}
Savage B.,  Narayanan A.,  Lehner N.,   Wakker B.,  2011, The Astrophysical
  Journal, 731, 14

\bibitem[\protect\citeauthoryear{Savage, Kim, Wakker, Keeney, Shull, Stocke  \&
  Green}{Savage et~al.}{2014}]{savage2014properties}
Savage B.,  Kim T.-S.,  Wakker B.,  Keeney B.,  Shull J.,  Stocke J.,   Green
  J.,  2014, The Astrophysical Journal Supplement Series, 212, 8

\bibitem[\protect\citeauthoryear{{Schroetter} et~al.,}{{Schroetter}
  et~al.}{2016}]{Schroetter2016}
{Schroetter} I.,  et~al., 2016, \mn@doi [\apj] {10.3847/1538-4357/833/1/39},
  \href {https://ui.adsabs.harvard.edu/abs/2016ApJ...833...39S} {833, 39}

\bibitem[\protect\citeauthoryear{{Schroetter} et~al.,}{{Schroetter}
  et~al.}{2019}]{Schroetter2019}
{Schroetter} I.,  et~al., 2019, \mn@doi [\mnras] {10.1093/mnras/stz2822}, \href
  {https://ui.adsabs.harvard.edu/abs/2019MNRAS.490.4368S} {490, 4368}

\bibitem[\protect\citeauthoryear{Shaw, Bridges  \& Hobson}{Shaw
  et~al.}{2007}]{shaw2007efficient}
Shaw J.,  Bridges M.,   Hobson M.,  2007, Monthly Notices of the Royal
  Astronomical Society, 378, 1365

\bibitem[\protect\citeauthoryear{Shen, Madau, Guedes, Mayer, Prochaska  \&
  Wadsley}{Shen et~al.}{2013}]{shen2013circumgalactic}
Shen S.,  Madau P.,  Guedes J.,  Mayer L.,  Prochaska J.~X.,   Wadsley J.,
  2013, The Astrophysical Journal, 765, 89

\bibitem[\protect\citeauthoryear{Shull}{Shull}{2014}]{shull2014galaxies}
Shull J.~M.,  2014, The Astrophysical Journal, 784, 142

\bibitem[\protect\citeauthoryear{Steidel \& Sargent}{Steidel \&
  Sargent}{1992}]{steidel1992mg}
Steidel C.~C.,  Sargent W.~L.,  1992, The Astrophysical Journal Supplement
  Series, 80, 1

\bibitem[\protect\citeauthoryear{Steidel, Erb, Shapley, Pettini, Reddy,
  Bogosavljevi{\'c}, Rudie  \& Rakic}{Steidel
  et~al.}{2010}]{steidel2010structure}
Steidel C.~C.,  Erb D.~K.,  Shapley A.~E.,  Pettini M.,  Reddy N.,
  Bogosavljevi{\'c} M.,  Rudie G.~C.,   Rakic O.,  2010, The Astrophysical
  Journal, 717, 289

\bibitem[\protect\citeauthoryear{Stinson et~al.,}{Stinson
  et~al.}{2012}]{stinson2012magicc}
Stinson G.,  et~al., 2012, Monthly Notices of the Royal Astronomical Society,
  425, 1270

\bibitem[\protect\citeauthoryear{Stocke, Keeney, McLin, Rosenberg, Weymann  \&
  Giroux}{Stocke et~al.}{2004}]{stocke2004discovery}
Stocke J.~T.,  Keeney B.~A.,  McLin K.~M.,  Rosenberg J.~L.,  Weymann R.,
  Giroux M.~L.,  2004, The Astrophysical Journal, 609, 94

\bibitem[\protect\citeauthoryear{Stocke, Keeney, Danforth, Shull, Froning,
  Green, Penton  \& Savage}{Stocke et~al.}{2013}]{stocke2013characterizing}
Stocke J.~T.,  Keeney B.~A.,  Danforth C.~W.,  Shull J.~M.,  Froning C.~S.,
  Green J.~C.,  Penton S.~V.,   Savage B.~D.,  2013, The Astrophysical Journal,
  763, 148

\bibitem[\protect\citeauthoryear{{Stocke} et~al.,}{{Stocke}
  et~al.}{2014}]{Stocke2014}
{Stocke} J.~T.,  et~al., 2014, \mn@doi [\apj] {10.1088/0004-637X/791/2/128},
  \href {https://ui.adsabs.harvard.edu/abs/2014ApJ...791..128S} {791, 128}

\bibitem[\protect\citeauthoryear{Suresh, Rubin, Kannan, Werk, Hernquist  \&
  Vogelsberger}{Suresh et~al.}{2016}]{suresh2016ovi}
Suresh J.,  Rubin K.~H.,  Kannan R.,  Werk J.~K.,  Hernquist L.,   Vogelsberger
  M.,  2016, Monthly Notices of the Royal Astronomical Society, p. stw2499

\bibitem[\protect\citeauthoryear{Tepper-García}{Tepper-García}{2006}]{garcia2006voigt}
Tepper-García T.,  2006, \mn@doi [Monthly Notices of the Royal Astronomical
  Society] {10.1111/j.1365-2966.2006.10450.x}, 369, 2025

\bibitem[\protect\citeauthoryear{Tumlinson et~al.,}{Tumlinson
  et~al.}{2011}]{tumlinson2011large}
Tumlinson J.,  et~al., 2011, Science, 334, 948

\bibitem[\protect\citeauthoryear{Tumlinson, Peeples  \& Werk}{Tumlinson
  et~al.}{2017}]{tumlinson2017circumgalactic}
Tumlinson J.,  Peeples M.~S.,   Werk J.~K.,  2017, Annual Review of Astronomy
  and Astrophysics, 55, 389

\bibitem[\protect\citeauthoryear{Turner, Schaye, Steidel, Rudie  \&
  Strom}{Turner et~al.}{2014}]{turner2014metal}
Turner M.~L.,  Schaye J.,  Steidel C.~C.,  Rudie G.~C.,   Strom A.~L.,  2014,
  Monthly Notices of the Royal Astronomical Society, 445, 794

\bibitem[\protect\citeauthoryear{Wakker}{Wakker}{2006}]{wakker2006fuse}
Wakker B.,  2006, The Astrophysical Journal Supplement Series, 163, 282

\bibitem[\protect\citeauthoryear{{Wendt}, {Bouch{\'e}}, {Zabl}, {Schroetter}
  \& {Muzahid}}{{Wendt} et~al.}{2020}]{Wendt2020}
{Wendt} M.,  {Bouch{\'e}} N.~F.,  {Zabl} J.,  {Schroetter} I.,   {Muzahid} S.,
  2020, arXiv e-prints, \href
  {https://ui.adsabs.harvard.edu/abs/2020arXiv200908464W} {p. arXiv:2009.08464}

\bibitem[\protect\citeauthoryear{Werk, Prochaska, Thom, Tumlinson, Tripp,
  O'Meara  \& Peeples}{Werk et~al.}{2013}]{werk2013cos}
Werk J.~K.,  Prochaska J.~X.,  Thom C.,  Tumlinson J.,  Tripp T.~M.,  O'Meara
  J.~M.,   Peeples M.~S.,  2013, The Astrophysical Journal Supplement Series,
  204, 17

\bibitem[\protect\citeauthoryear{Wotta, Lehner, Howk, O’Meara, Oppenheimer
  \& Cooksey}{Wotta et~al.}{2019}]{wotta2019cos}
Wotta C.~B.,  Lehner N.,  Howk J.~C.,  O’Meara J.~M.,  Oppenheimer B.~D.,
  Cooksey K.~L.,  2019, The Astrophysical Journal, 872, 81

\bibitem[\protect\citeauthoryear{Zahedy, Chen, Johnson, Pierce, Rauch, Huang,
  Weiner  \& Gauthier}{Zahedy et~al.}{2019}]{zahedy2019characterizing}
Zahedy F.~S.,  Chen H.-W.,  Johnson S.~D.,  Pierce R.~M.,  Rauch M.,  Huang
  Y.-H.,  Weiner B.~J.,   Gauthier J.-R.,  2019, Monthly Notices of the Royal
  Astronomical Society, 484, 2257

\bibitem[\protect\citeauthoryear{Zonak, Charlton, Ding  \& Churchill}{Zonak
  et~al.}{2004}]{zonak2004absorption}
Zonak S.~G.,  Charlton J.~C.,  Ding J.,   Churchill C.~W.,  2004, The
  Astrophysical Journal, 606, 196

\makeatother
\end{thebibliography}
\bsp
\newpage
\appendix 
\section{Plots for PHL1811 0.07776 absorber}
\label{appendix:phl18110.07}
\subsection{Posterior distributions for the Voigt Profile fit parameters}
\label{appendix:paramsvoigtphl18110.07}
\begin{figure*}
\begin{center}
\includegraphics[width=\linewidth]{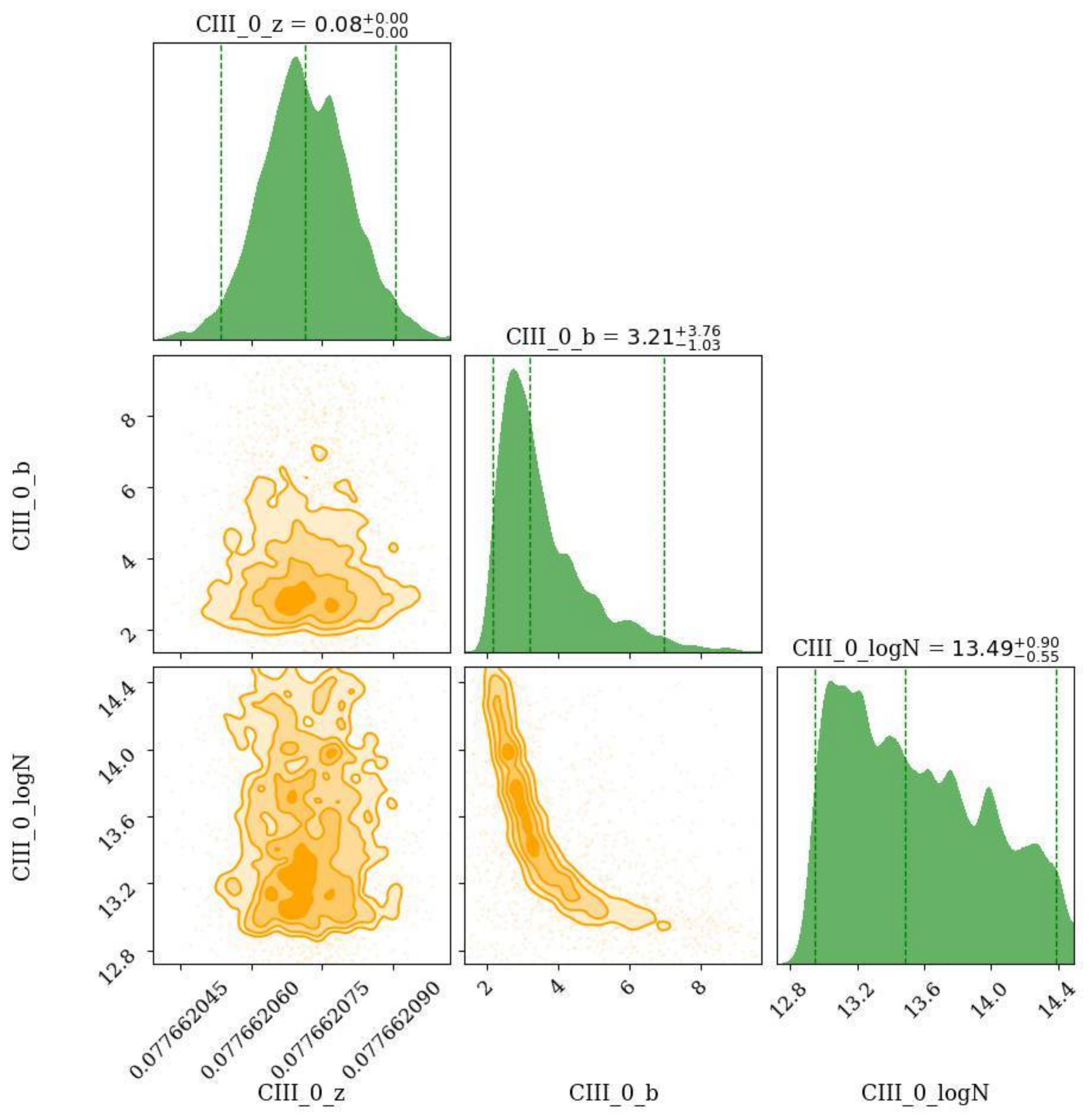}
\caption{The corner plot showing the marginalized posterior distributions of the redshift ($z$), Doppler parameter ($b$), and column density ($\log N$) for the blueward {\ciii} cloud of the $z=0.07776$ absorber towards PHL1811. The over-plotted vertical lines in the posterior distribution span the 95\% credible interval. The contours indicate 0.5$\sigma$, 1$\sigma$, 1.5$\sigma$, and 2$\sigma$ levels. The model results are summarised in Table~\ref{tab:phl18110.07776model}.}
\label{fig:voigtCIII0phl18110.07}
\end{center}
\end{figure*}

\begin{figure*}
\begin{center}
\includegraphics[width=\linewidth]{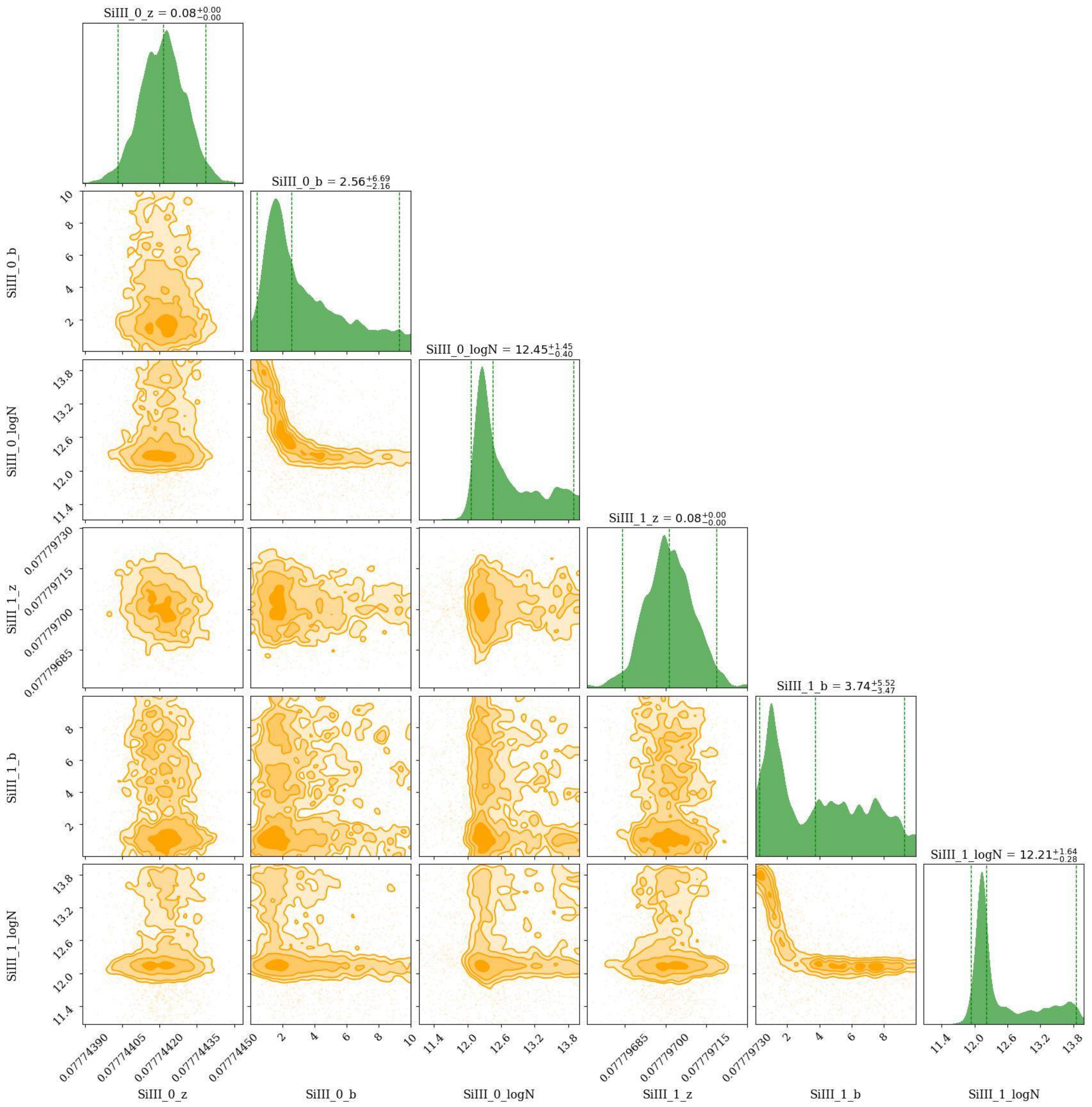}
\caption{The corner plot showing the marginalized posterior distributions of the redshift ($z$), Doppler parameter ($b$), and column density ($\log N$) for the blended redwards {\siiii} clouds of the $z=0.07776$ absorber towards PHL1811. The over-plotted vertical lines in the posterior distribution span the 95\% credible interval. The contours indicate 0.5$\sigma$, 1$\sigma$, 1.5$\sigma$, and 2$\sigma$ levels. The model results are summarised in Table~\ref{tab:phl18110.07776model}.}
\label{fig:voigtSiIIIphl18110.07}
\end{center}
\end{figure*}

\begin{figure*}
\begin{center}
\includegraphics[width=\linewidth]{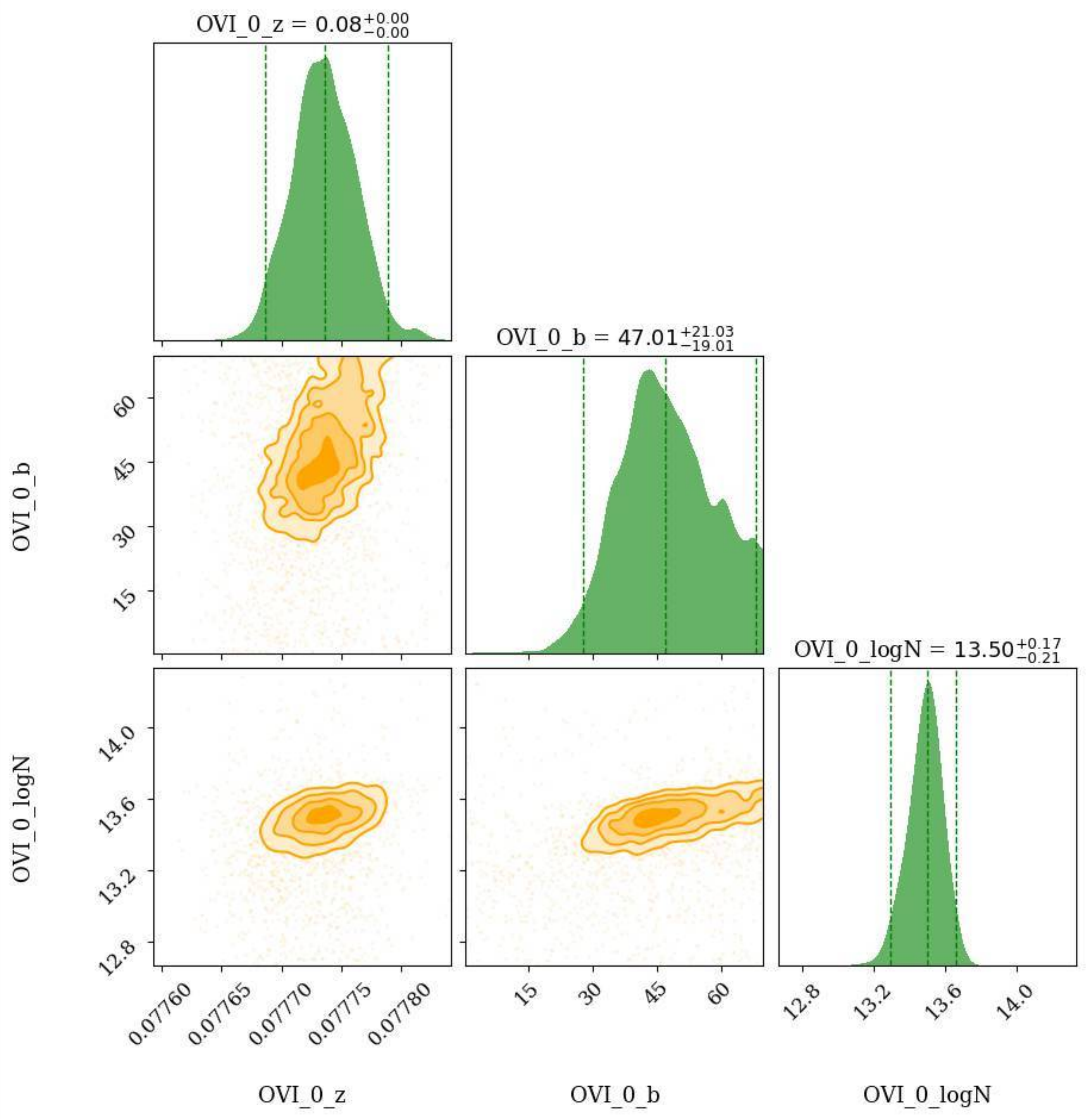}
\caption{The corner plot showing the marginalized posterior distributions of the redshift ($z$), Doppler parameter ($b$), and column density ($\log N$) for the {\ovi} cloud of the $z=0.07776$ absorber towards PHL1811. The over-plotted vertical lines in the posterior distribution span the 95\% credible interval. The contours indicate 0.5$\sigma$, 1$\sigma$, 1.5$\sigma$, and 2$\sigma$ levels. The model results are summarised in Table~\ref{tab:phl18110.07776model}.}
\label{fig:voigtOVIphl18110.07}
\end{center}
\end{figure*}

\subsection{Posterior distributions for the absorber properties}
\label{appendix:paramsphl18110.07}
\begin{figure*}

\begin{center}
\includegraphics[width=\linewidth]{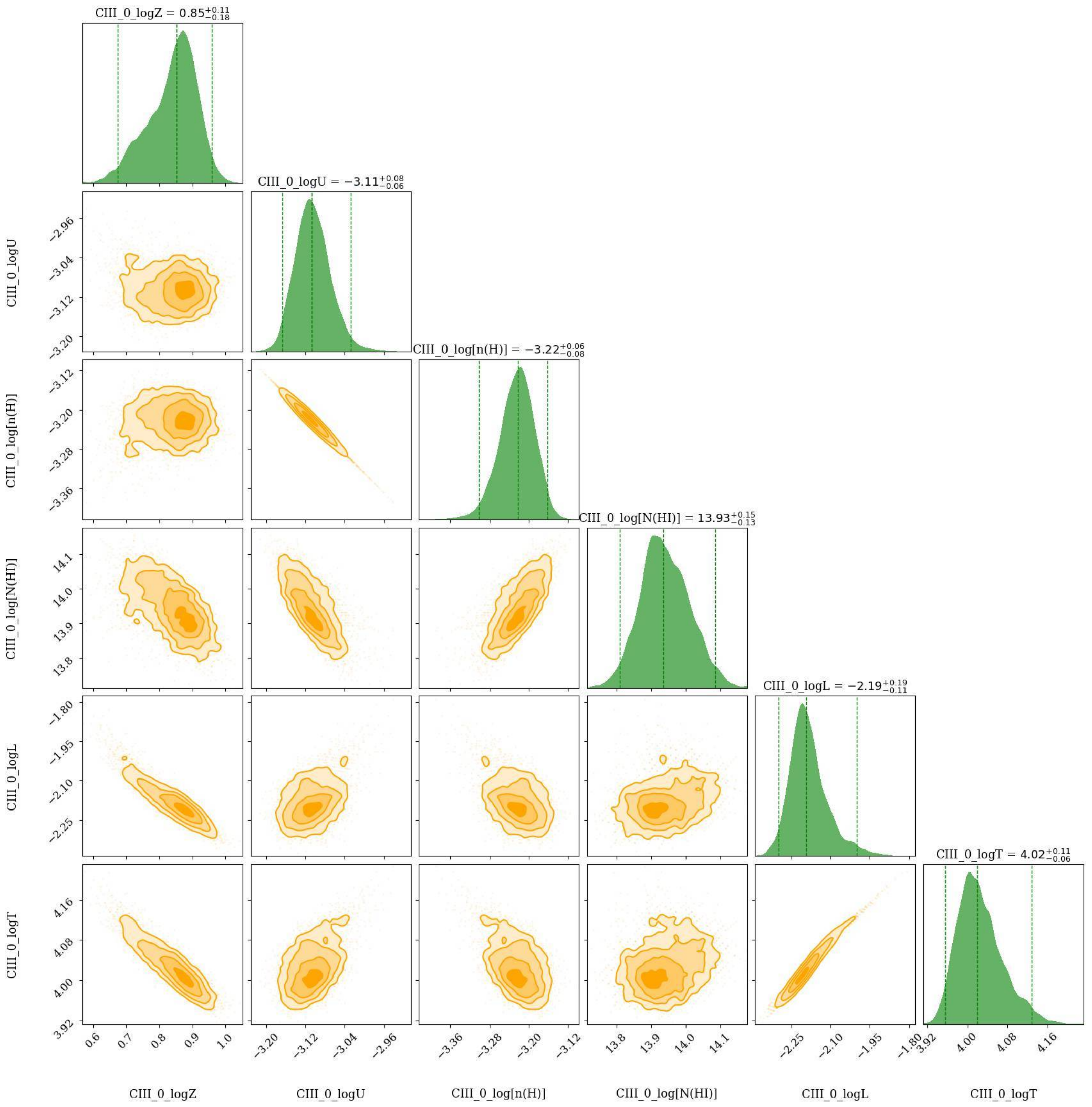}
\caption{The corner plot showing the marginalized posterior distributions for the metallicity ($\log Z$), ionization parameter ($\log U$), and other physical properties of the low ionization phase traced by the blueward {\ciii} cloud of the $z=0.07776$ absorber towards PHL1811. The over-plotted vertical lines in the posterior distribution span the 95\% credible interval. The contours indicate 0.5$\sigma$, 1$\sigma$, 1.5$\sigma$, and 2$\sigma$ levels. The model results are summarised in Table~\ref{tab:phl18110.07776model}, and the synthetic profiles based on these models are shown in Figure~\ref{fig:sysplotphl18110.07}.}
\label{fig:CIII0PHL18110.07}
\end{center}
\end{figure*}

\begin{figure*}
\begin{center}
\includegraphics[width=\linewidth]{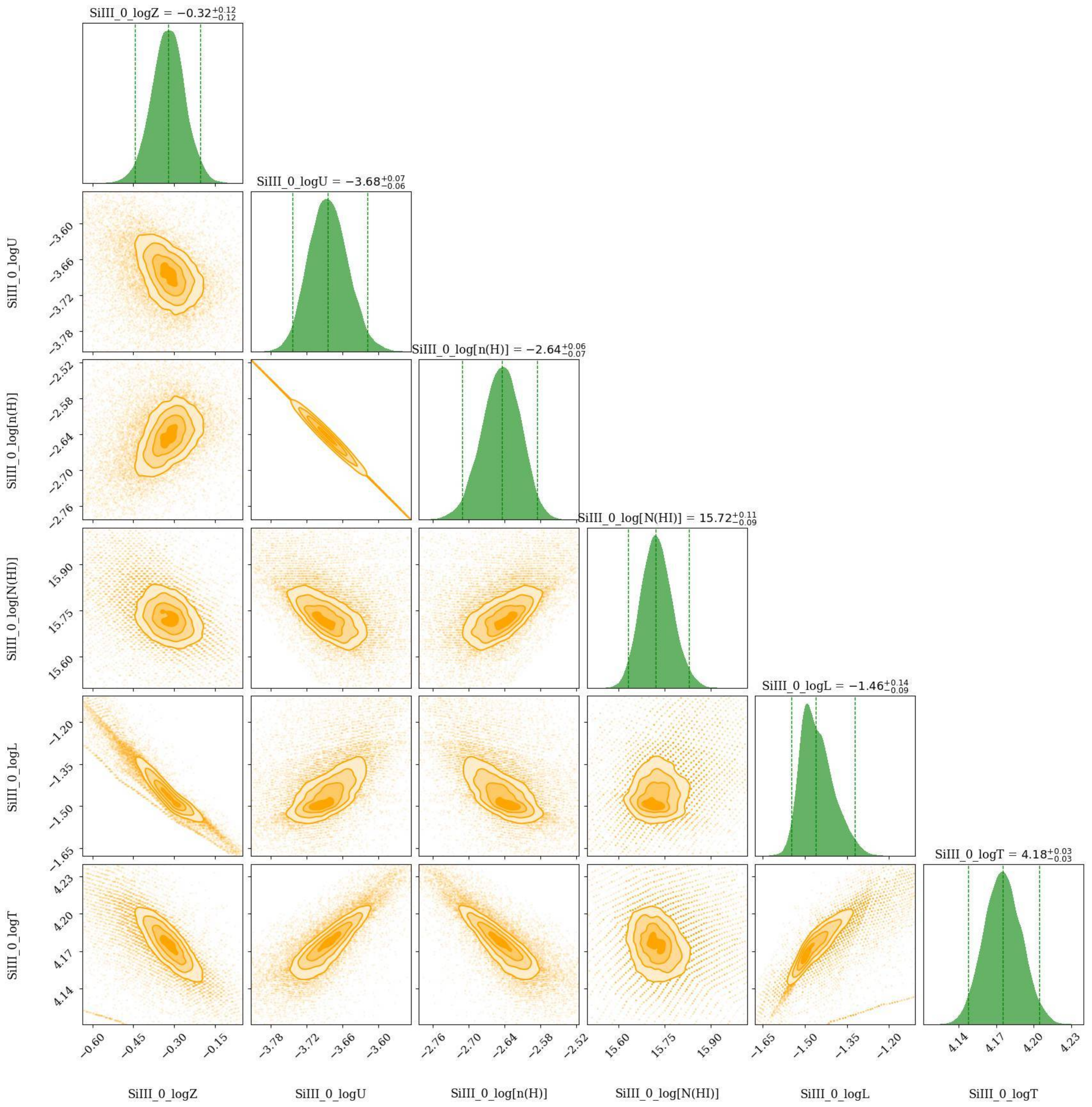}
\caption{The corner plot showing the marginalized posterior distributions for the metallicity ($\log Z$), ionization parameter ($\log U$), and other physical properties of the low ionization phase traced by the blueward blended {\siiii} cloud of the $z=0.07776$ absorber towards PHL1811. The over-plotted vertical lines in the posterior distribution span the 95\% credible interval. The contours indicate 0.5$\sigma$, 1$\sigma$, 1.5$\sigma$, and 2$\sigma$ levels. The model results are summarised in Table~\ref{tab:phl18110.07776model}, and the synthetic profiles based on these models are shown in Figure~\ref{fig:sysplotphl18110.07}.}
\label{fig:SiIII0PHL18110.07}
\end{center}
\end{figure*}

\begin{figure*}
\begin{center}
\includegraphics[width=\linewidth]{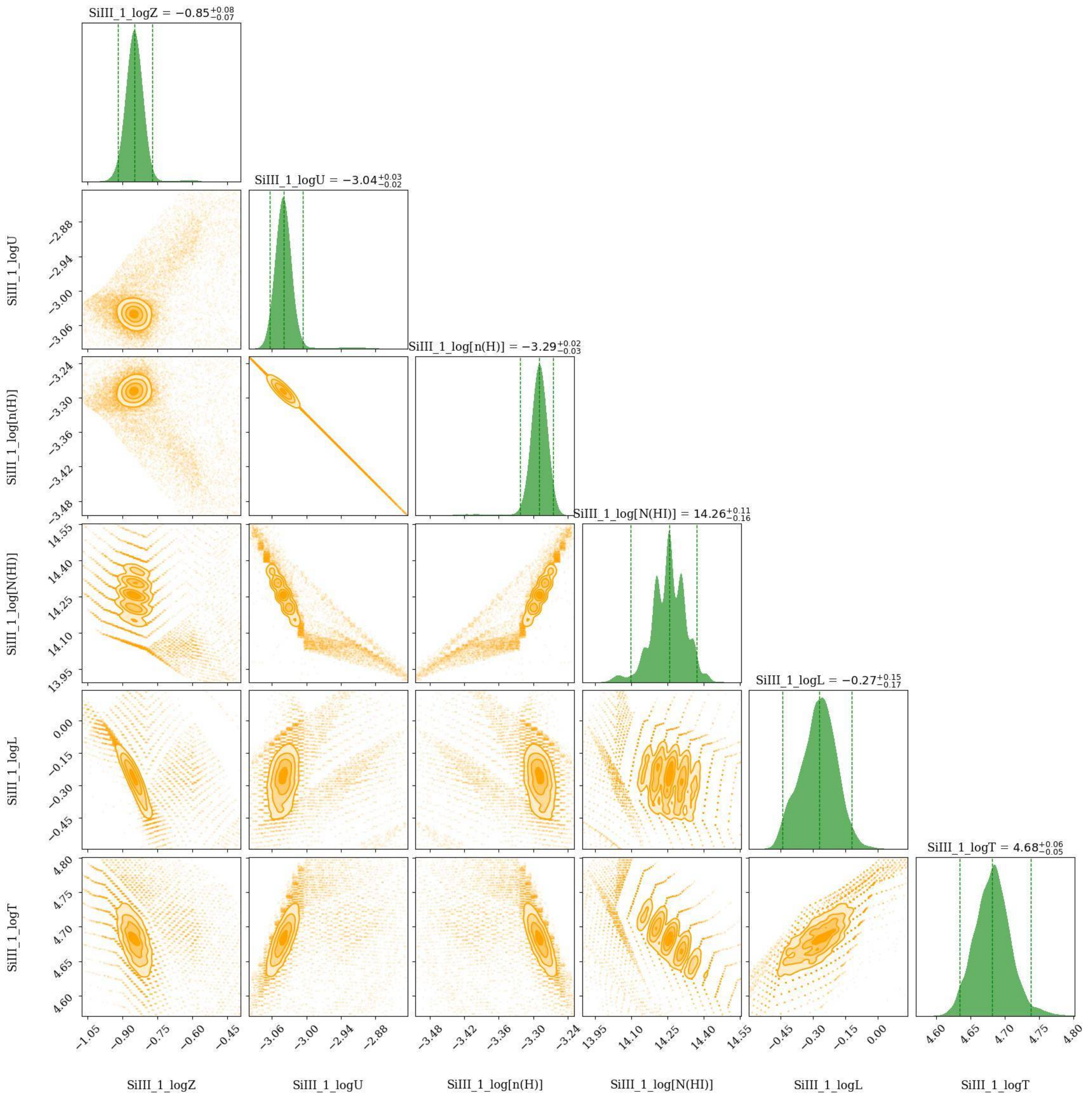}
\caption{The corner plot showing the marginalized posterior distributions for the metallicity ($\log Z$), ionization parameter ($\log U$), and other physical properties of the low ionization phase traced by the redward blended {\siiii} cloud of the $z=0.07776$ absorber towards PHL1811. The over-plotted vertical lines in the posterior distribution span the 95\% credible interval. The contours indicate 0.5$\sigma$, 1$\sigma$, 1.5$\sigma$, and 2$\sigma$ levels. The model results are summarised in Table~\ref{tab:phl18110.07776model}, and the synthetic profiles based on these models are shown in Figure~\ref{fig:sysplotphl18110.07}.}
\label{fig:SiIII1PHL18110.07}
\end{center}
\end{figure*}

\begin{figure*}
\begin{center}
\includegraphics[width=\linewidth]{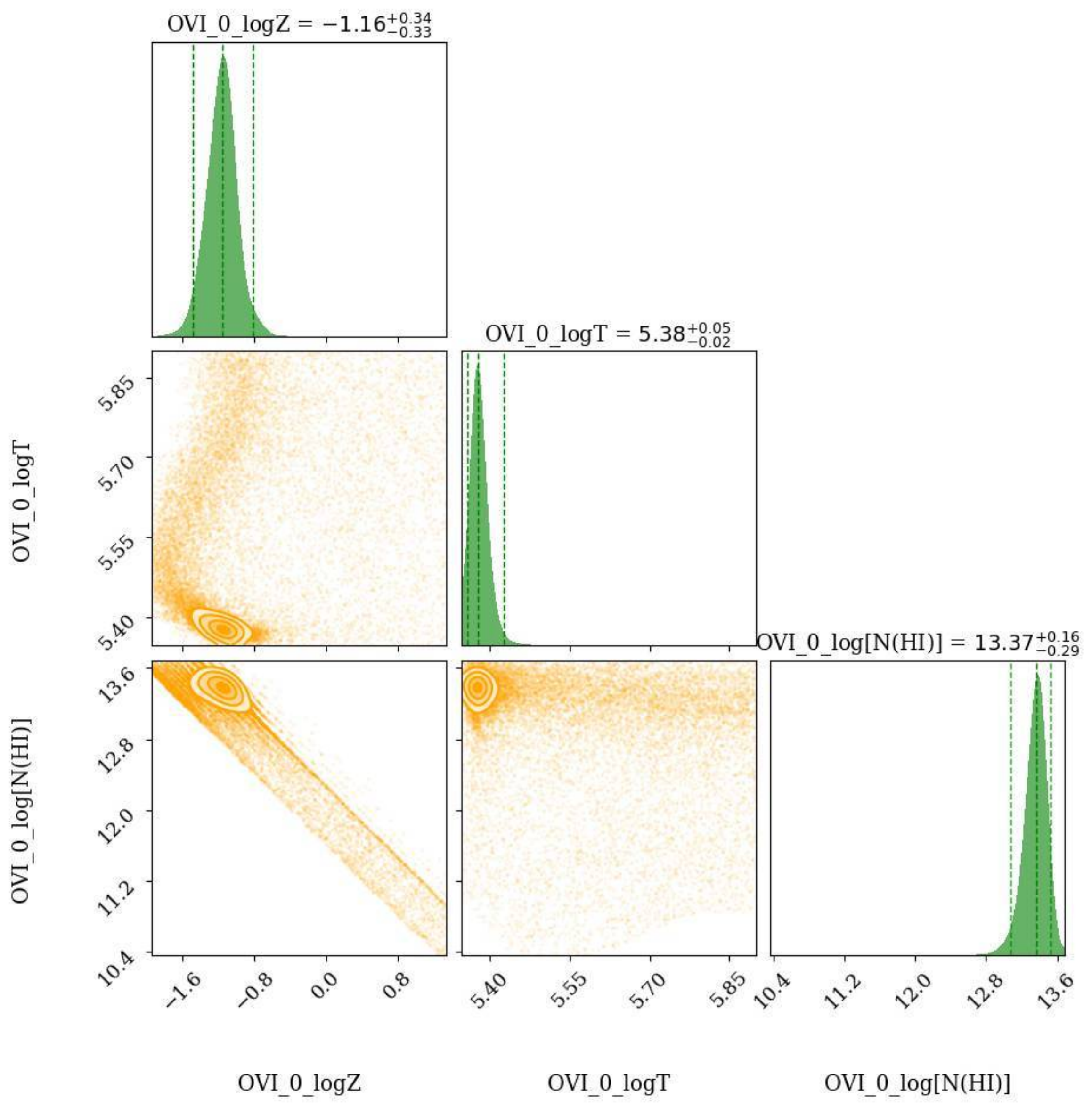}
\caption{The corner plot showing the marginalized posterior distributions for the metallicity ($\log Z$), ionization parameter ($\log U$), and other physical properties of the collisionally ionized phase traced by {\ovi} of the $z=0.07776$ absorber towards PHL1811. The over-plotted vertical lines in the posterior distribution span the 95\% credible interval. The contours indicate 0.5$\sigma$, 1$\sigma$, 1.5$\sigma$, and 2$\sigma$ levels. The model results are summarised in Table~\ref{tab:phl18110.07776model}, and the synthetic profiles based on these models are shown in Figure~\ref{fig:sysplotphl18110.07}.}
\label{fig:OVI0PHL18110.07}
\end{center}
\end{figure*}

\section{Plots for PHL1811 0.08094 absorber}
\label{appendix:phl18110.08}

\subsection{Posterior distributions for the Voigt Profile fit parameters}
\label{appendix:paramsvoigtphl18110.08}
\begin{figure*}
\begin{center}
\includegraphics[width=\linewidth]{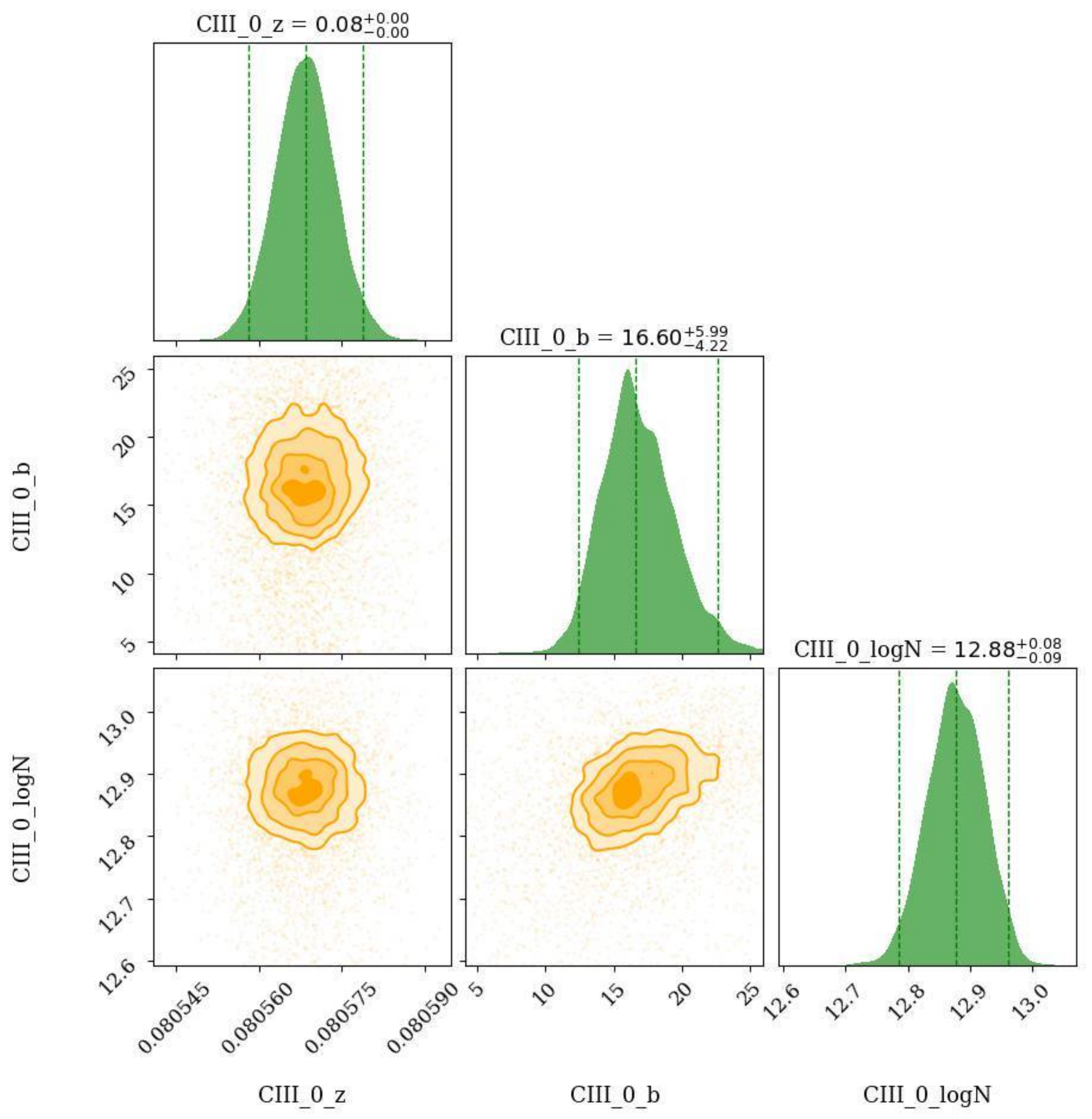}
\caption{The corner plot showing the marginalized posterior distributions of the redshift ($z$), Doppler parameter ($b$), and column density ($\log N$) for the blueward {\ciii} of the $z=0.08094$ absorber towards PHL1811. The over-plotted vertical lines in the posterior distribution span the 95\% credible interval. The contours indicate 0.5$\sigma$, 1$\sigma$, 1.5$\sigma$, and 2$\sigma$ levels. The model results are summarised in Table~\ref{tab:phl18110.08modelpar}.}
\label{fig:voigtCIII0phl18110.08}
\end{center}
\end{figure*}

\begin{figure*}
\begin{center}
\includegraphics[width=\linewidth]{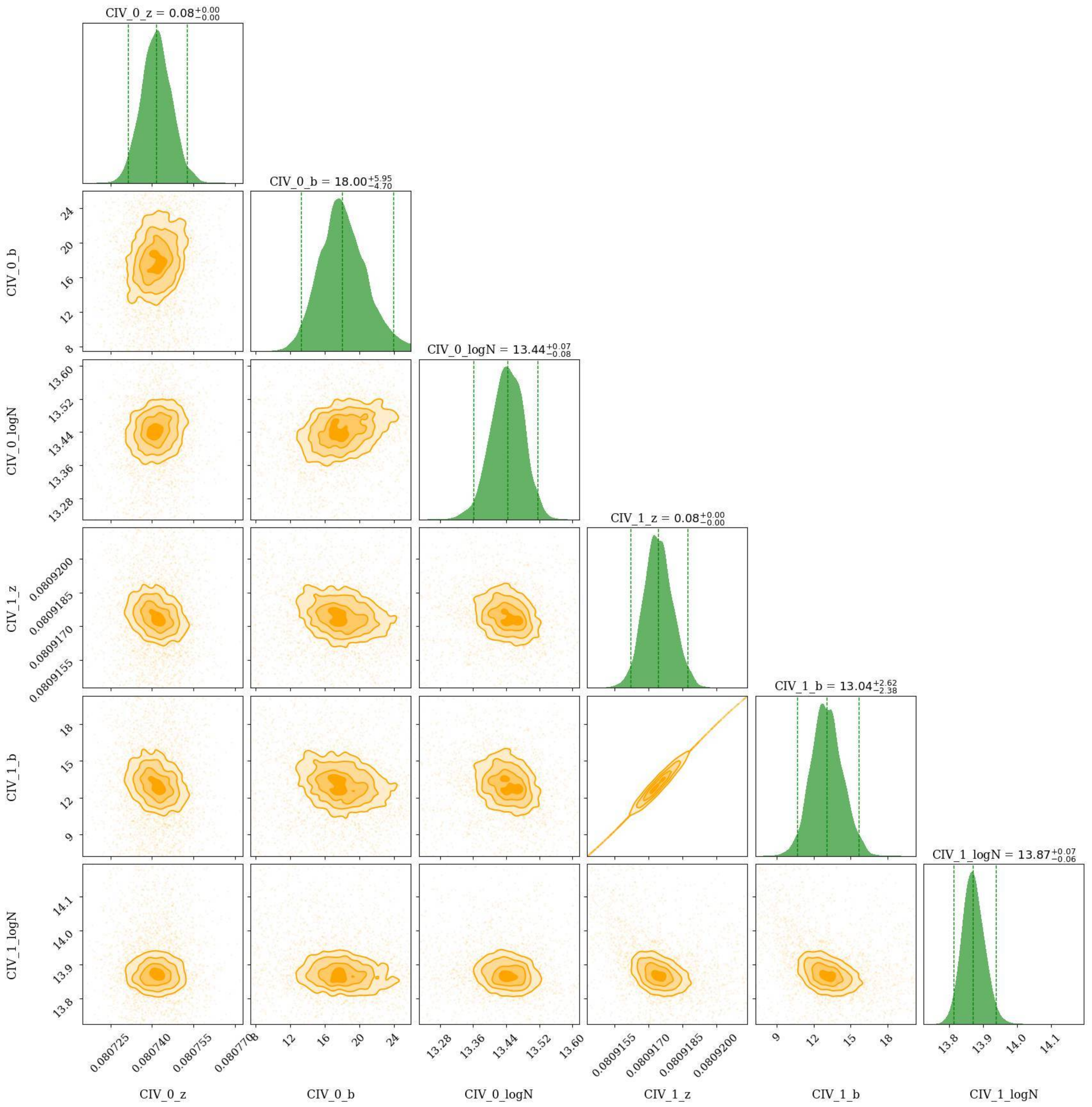}
\caption{The corner plot showing the marginalized posterior distributions of the redshift ($z$), Doppler parameter ($b$), and column density ($\log N$) for the two redward {\civ} clouds of the $z=0.08094$ absorber towards PHL1811. The over-plotted vertical lines in the posterior distribution span the 95\% credible interval. The contours indicate 0.5$\sigma$, 1$\sigma$, 1.5$\sigma$, and 2$\sigma$ levels. The model results are summarised in Table~\ref{tab:phl18110.08modelpar}.}
\label{fig:voigtCIVphl18110.08}
\end{center}
\end{figure*}

\begin{figure*}
\begin{center}
\includegraphics[width=\linewidth]{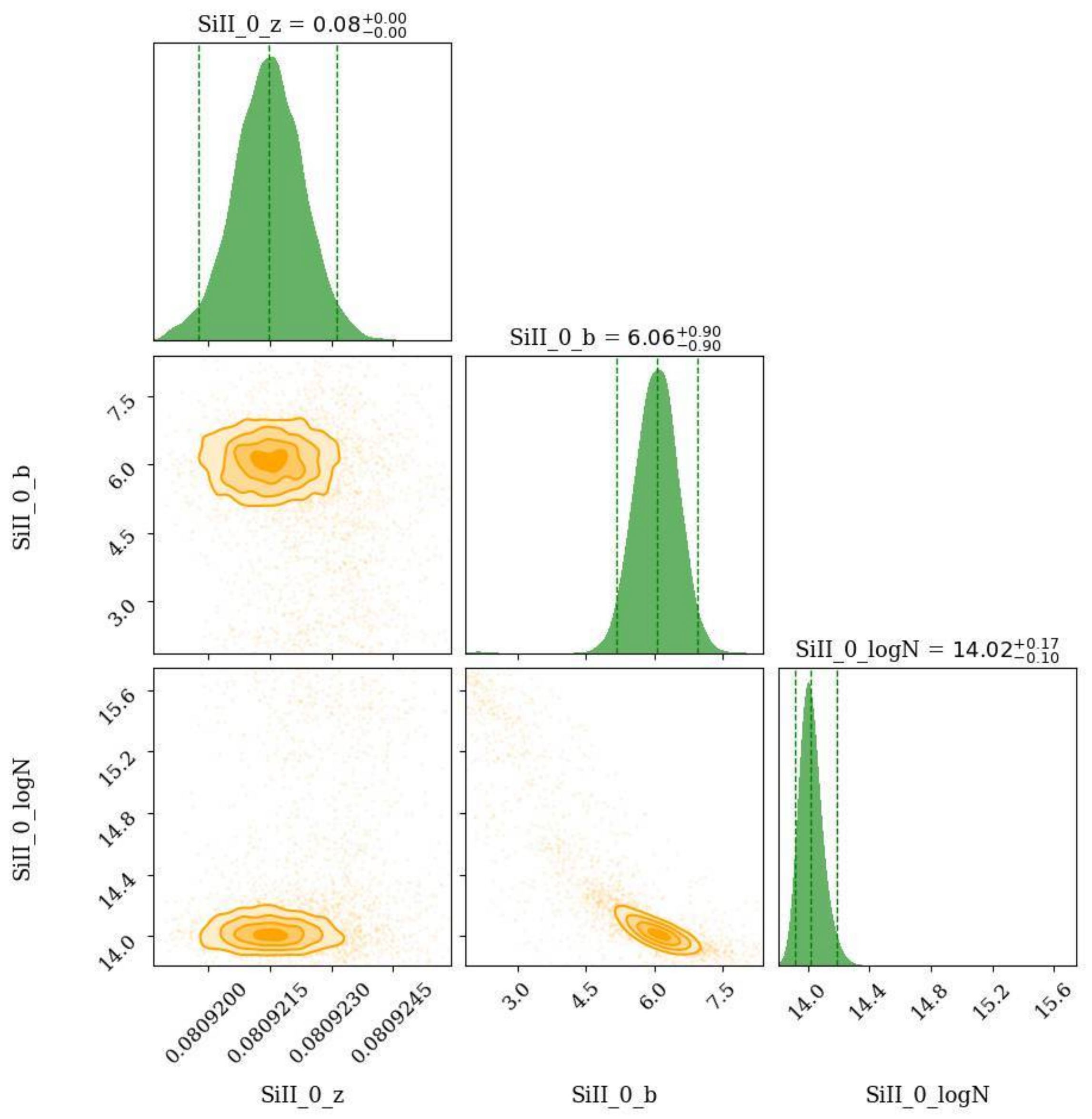}
\caption{The corner plot showing the marginalized posterior distributions of the redshift ($z$), Doppler parameter ($b$), and column density for the redward {\siii} cloud of the $z=0.08094$ absorber towards PHL1811. The over-plotted vertical lines in the posterior distribution span the 95\% credible interval. The contours indicate 0.5$\sigma$, 1$\sigma$, 1.5$\sigma$, and 2$\sigma$ levels. The model results are summarised in Table~\ref{tab:phl18110.08modelpar}.}
\label{fig:voigtSiIIphl18110.08}
\end{center}
\end{figure*}

\subsection{Posterior distributions for the absorber properties}
\label{appendix:paramsphl18110.08}
\begin{figure*}
\begin{center}
\includegraphics[width=\linewidth]{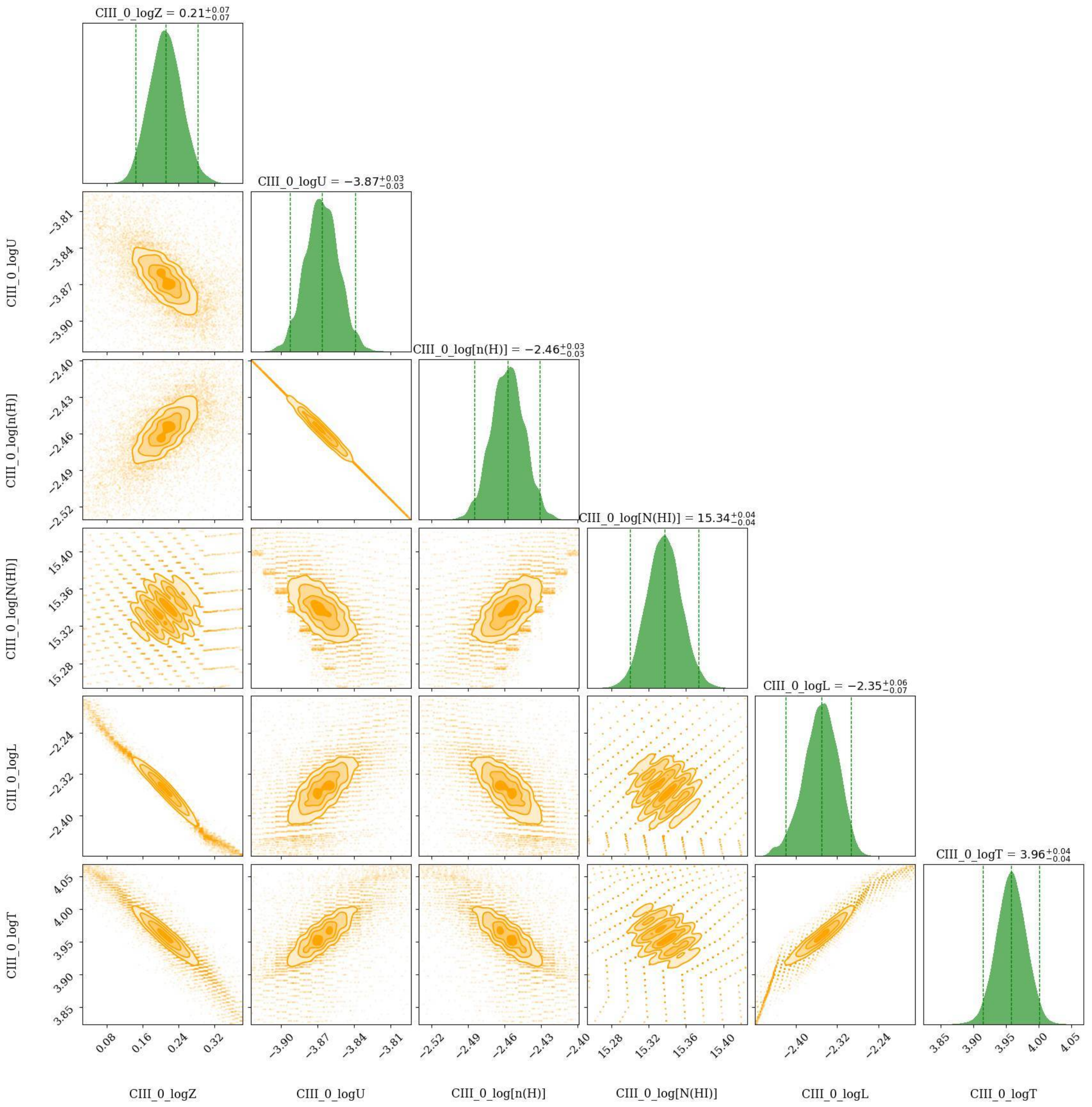}
\caption{The corner plot showing the marginalized posterior distributions for the metallicity ($\log Z$), ionization parameter ($\log U$), and other physical properties of blueward low ionization phase traced by {\ciii} of the $z=0.08094$ absorber towards PHL1811. The over-plotted vertical lines in the posterior distribution span the 95\% credible interval. The contours indicate 0.5$\sigma$, 1$\sigma$, 1.5$\sigma$, and 2$\sigma$ levels. The model results are summarised in Table~\ref{tab:phl18110.08modelpar}, and the synthetic profiles based on these models are shown in Figure~\ref{fig:Modelsphl18110.08}.}
\label{fig:CIII0PHL18110.08}
\end{center}
\end{figure*}

\begin{figure*}
\begin{center}
\includegraphics[width=\linewidth]{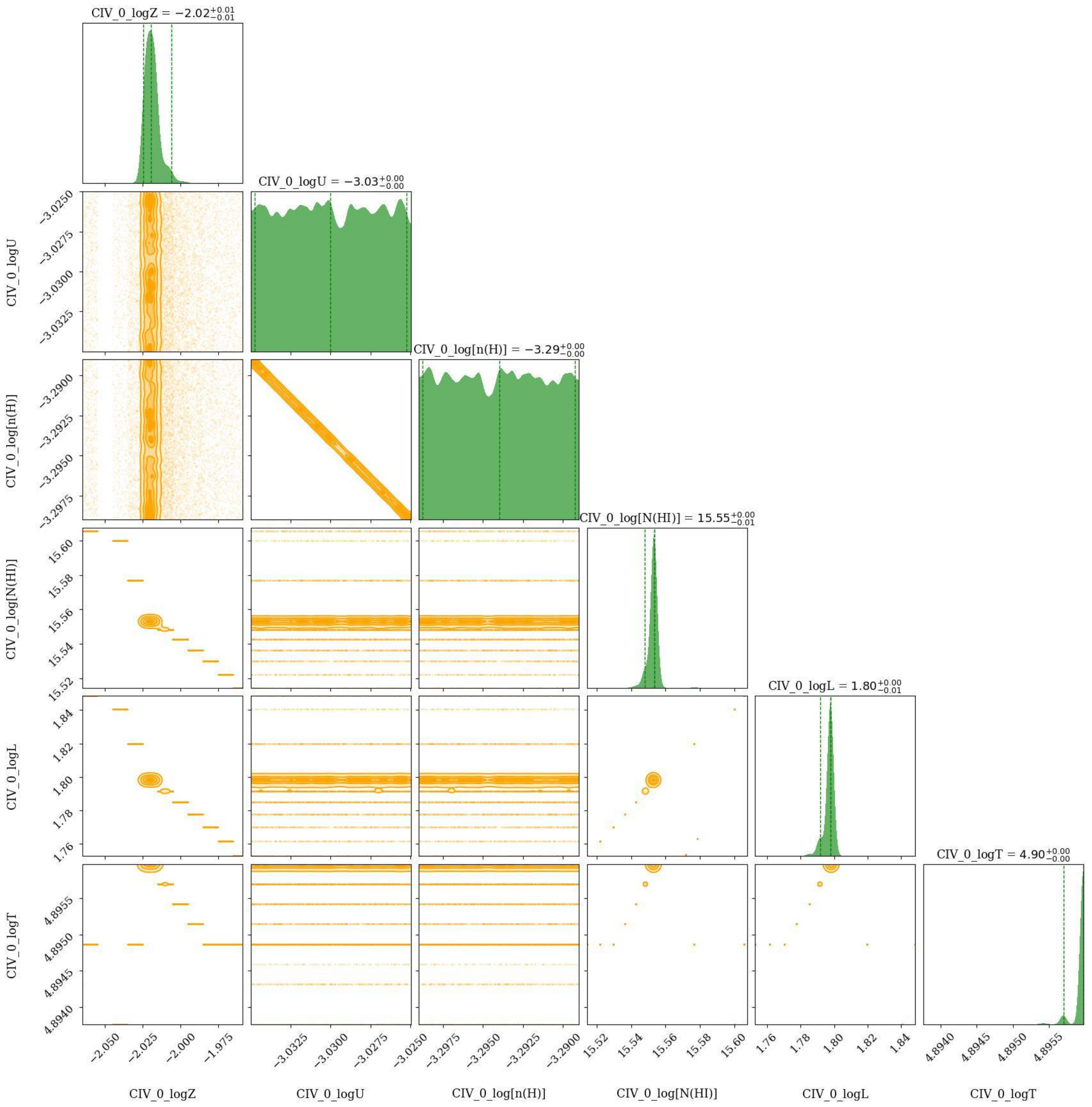}
\caption{The corner plot showing the marginalized posterior distributions for the metallicity ($\log Z$), ionization parameter ($\log U$), and other physical properties of the blueward high ionization phase traced by {\civ} of the $z=0.08094$ absorber towards PHL1811. The over-plotted vertical lines in the posterior distribution span the 95\% credible interval. The contours indicate 0.5$\sigma$, 1$\sigma$, 1.5$\sigma$, and 2$\sigma$ levels. The model results are summarised in Table~\ref{tab:phl18110.08modelpar}, and the synthetic profiles based on these models are shown in Figure~\ref{fig:Modelsphl18110.08}. We found that the solution for this phase exist only for a very narrow range of values, hence the uniform distribution in allowed range of values for some parameters.}
\label{fig:CIV0PHL18110.08}
\end{center}
\end{figure*}

\begin{figure*}
\begin{center}
\includegraphics[width=\linewidth]{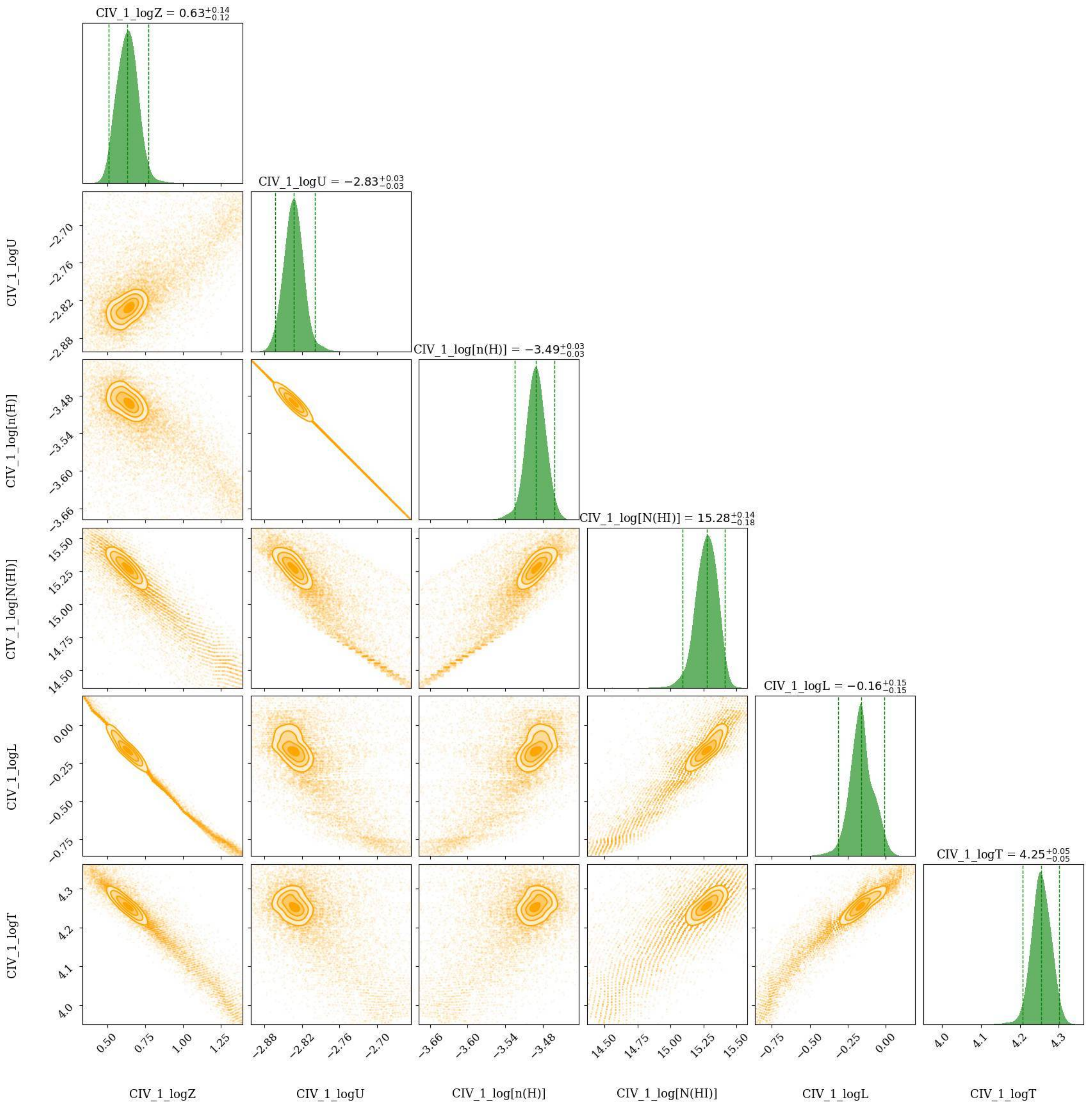}
\caption{The corner plot showing the marginalized posterior distributions for the metallicity ($\log Z$), ionization parameter ($\log U$), and other physical properties of redward high ionization phase traced by {\civ} of the $z=0.08094$ absorber towards PHL1811. The over-plotted vertical lines in the posterior distribution span the 95\% credible interval. The contours indicate 0.5$\sigma$, 1$\sigma$, 1.5$\sigma$, and 2$\sigma$ levels. The model results are summarised in Table~\ref{tab:phl18110.08modelpar}, and the synthetic profiles based on these models are shown in Figure~\ref{fig:Modelsphl18110.08}.}
\label{fig:CIV1PHL18110.08}
\end{center}
\end{figure*}

\begin{figure*}
\begin{center}
\includegraphics[width=\linewidth]{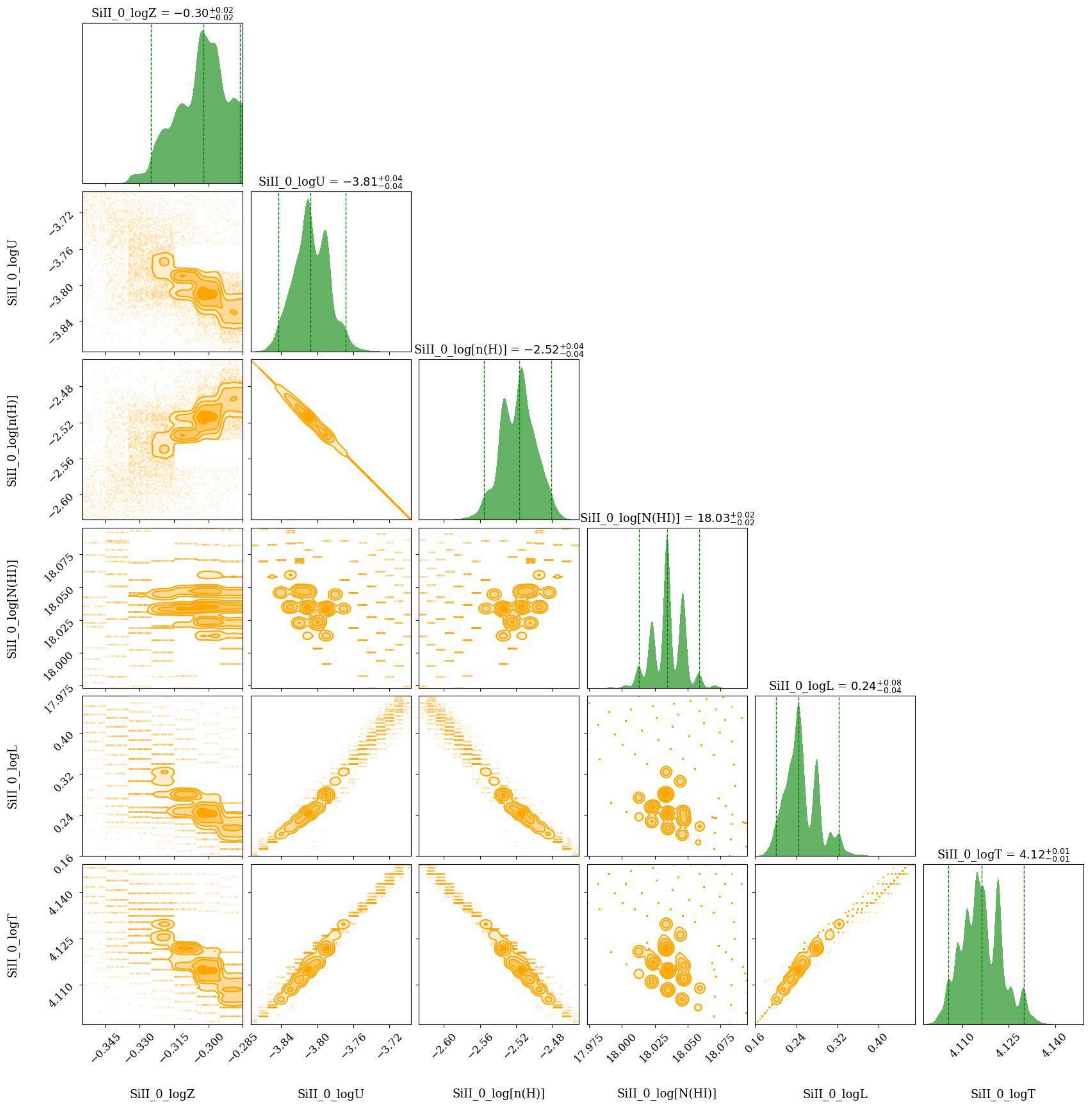}
\caption{The corner plot showing the marginalized posterior distributions for the metallicity ($\log Z$), ionization parameter ($\log U$), and other physical properties of the redward low ionization phase traced by {\siii} of the $z=0.08094$ absorber towards PHL1811. The over-plotted vertical lines in the posterior distribution span the 95\% credible interval. The contours indicate 0.5$\sigma$, 1$\sigma$, 1.5$\sigma$, and 2$\sigma$ levels. The model results are summarised in Table~\ref{tab:phl18110.08modelpar}, and the synthetic profiles based on these models are shown in Figure~\ref{fig:Modelsphl18110.08}.}
\label{fig:SiIIPHL18110.08}
\end{center}
\end{figure*}

\section{Plots for PG1116+215 0.13849 absorber}
\label{appendix:pg1116}

\subsection{Posterior distributions for the Voigt Profile fit parameters}
\label{appendix:paramsvoigtpg1116}
\begin{figure*}
\begin{center}
\includegraphics[width=\linewidth]{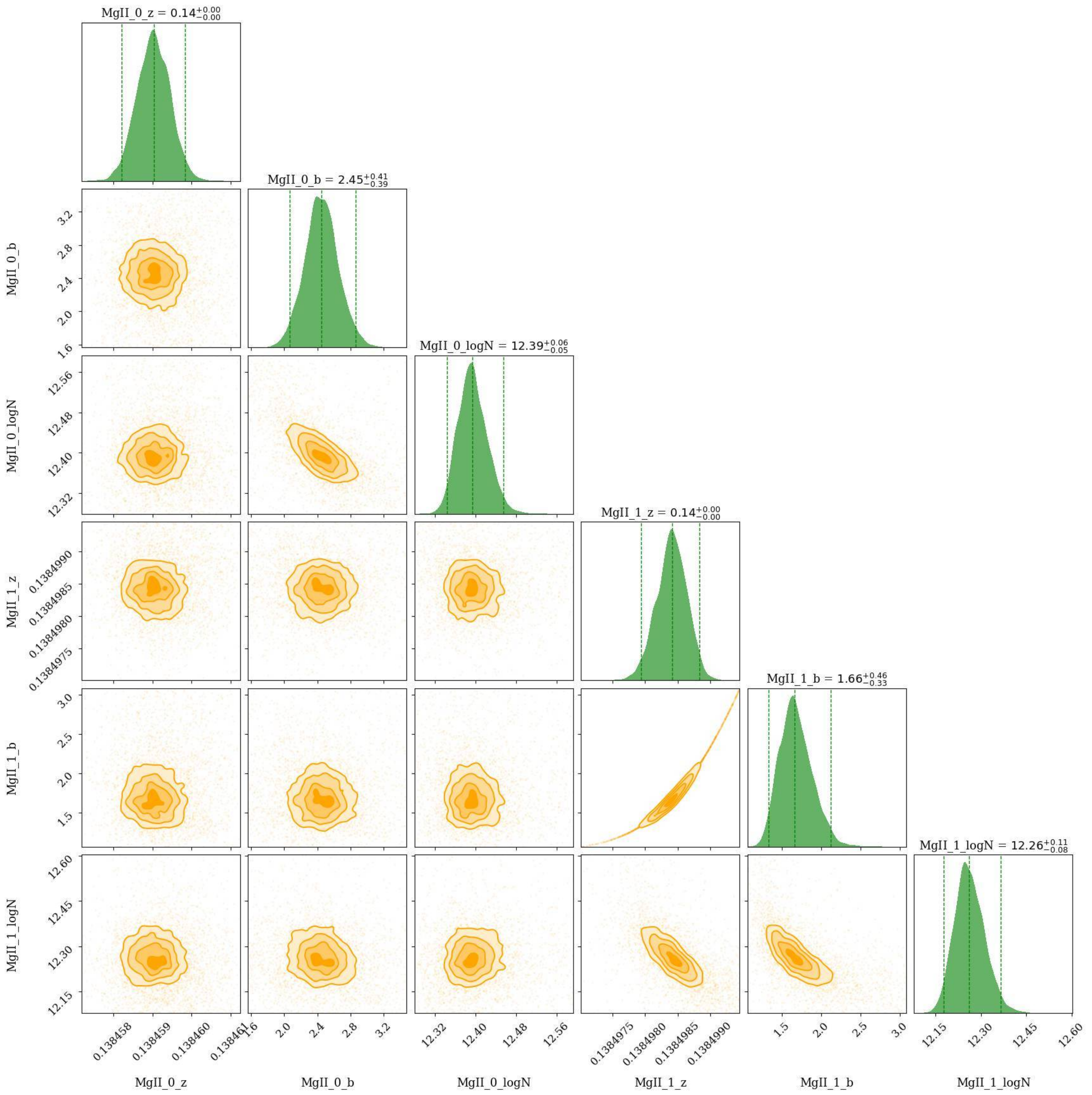}
\caption{The corner plot showing the marginalized posterior distributions of the redshift ($z$), Doppler parameter ($b$), and column density ($\log N$) for the low ionization clouds traced by {\mgii} of the $z=0.13849$ absorber towards PG1116+215. The over-plotted vertical lines in the posterior distribution span the 95\% credible interval. The contours indicate 0.5$\sigma$, 1$\sigma$, 1.5$\sigma$, and 2$\sigma$ levels. The model results are summarised in Table~\ref{tab:pg1116modelpar}.}
\label{fig:voigtMgIIpg1116}
\end{center}
\end{figure*}

\begin{figure*}
\begin{center}
\includegraphics[width=\linewidth]{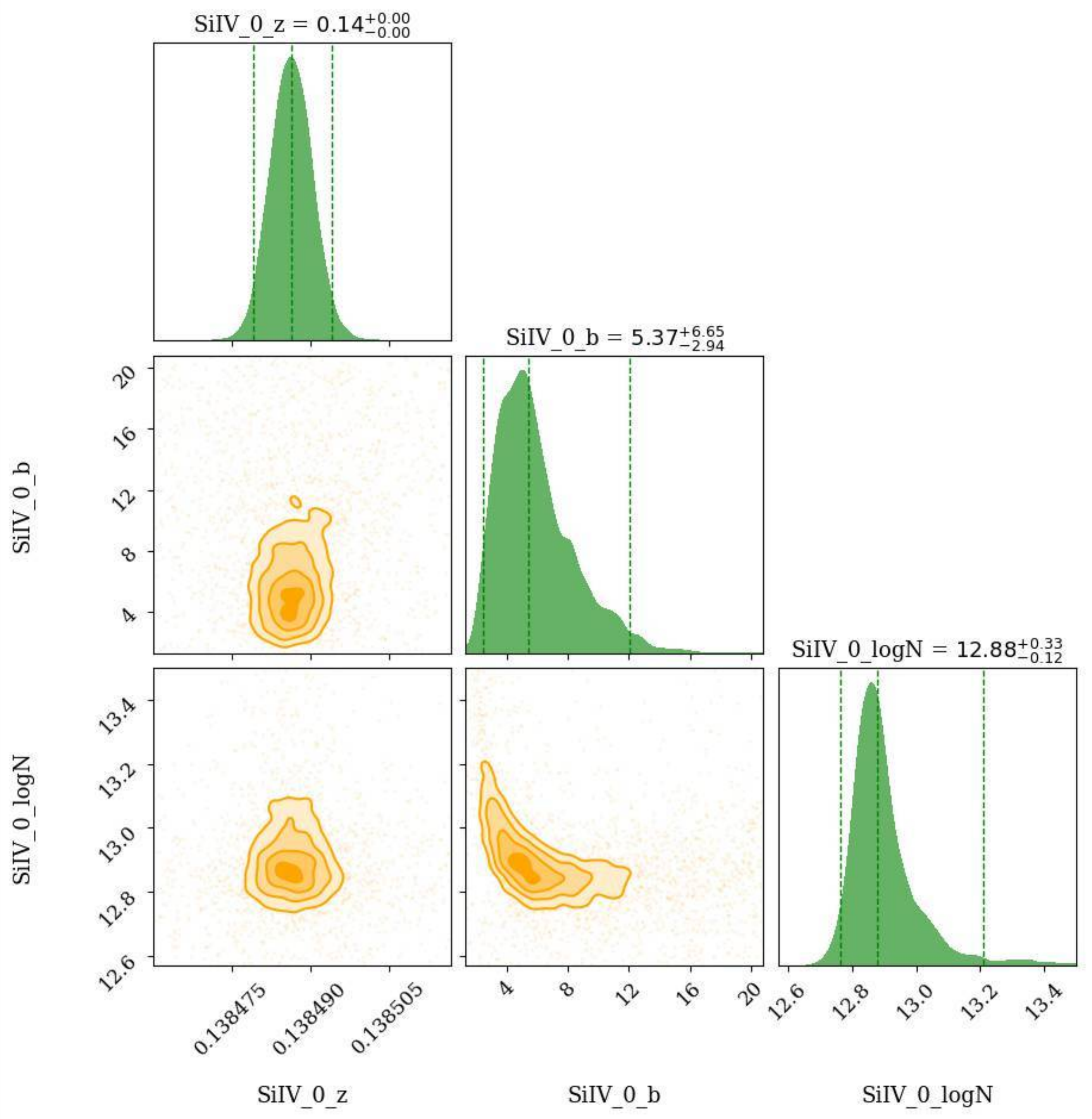}
\caption{The corner plot showing the marginalized posterior distributions of the redshift ($z$), Doppler parameter ($b$), and column density ($\log N$) for the intermediate ionization cloud traced by {\siiv} of the $z=0.13849$ absorber towards PG1116+215. The over-plotted vertical lines in the posterior distribution span the 95\% credible interval. The contours indicate 0.5$\sigma$, 1$\sigma$, 1.5$\sigma$, and 2$\sigma$ levels. The model results are summarised in Table~\ref{tab:pg1116modelpar}.}
\label{fig:voigtSiIVpg1116}
\end{center}
\end{figure*}

\begin{figure*}
\begin{center}
\includegraphics[width=\linewidth]{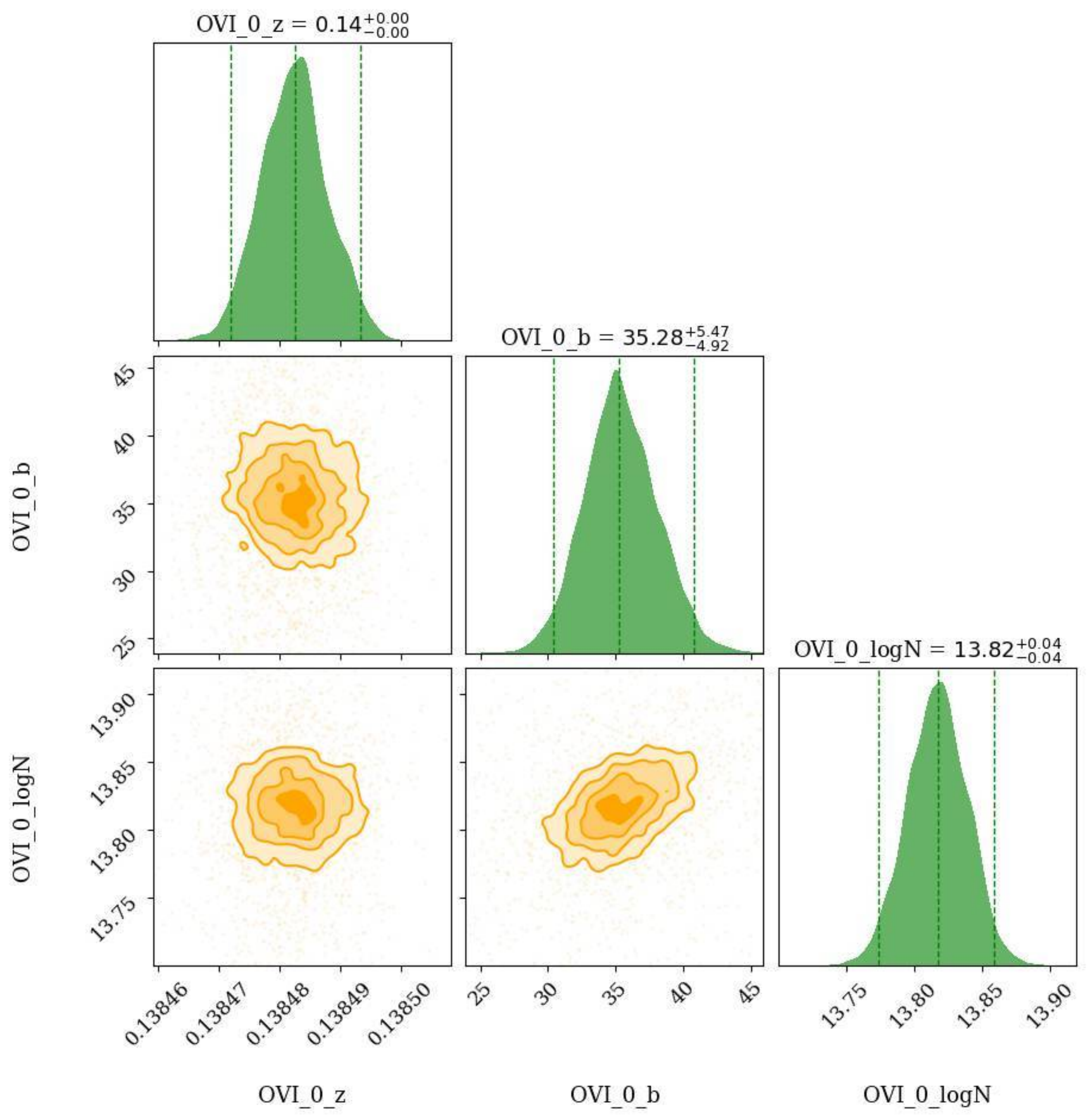}
\caption{The corner plot showing the marginalized posterior distributions of the redshift ($z$), Doppler parameter ($b$), and column density ($\log N$) for the collisionally ionized cloud traced by {\ovi} of the $z=0.13849$ absorber towards PG1116+215. The over-plotted vertical lines in the posterior distribution span the 95\% credible interval. The contours indicate 0.5$\sigma$, 1$\sigma$, 1.5$\sigma$, and 2$\sigma$ levels. The model results are summarised in Table~\ref{tab:pg1116modelpar}.}
\label{fig:voigtOVIpg1116}
\end{center}
\end{figure*}

\subsection{Posterior distributions for the absorber properties}
\label{appendix:paramspg1116}
\begin{figure*}
\begin{center}
\includegraphics[width=\linewidth]{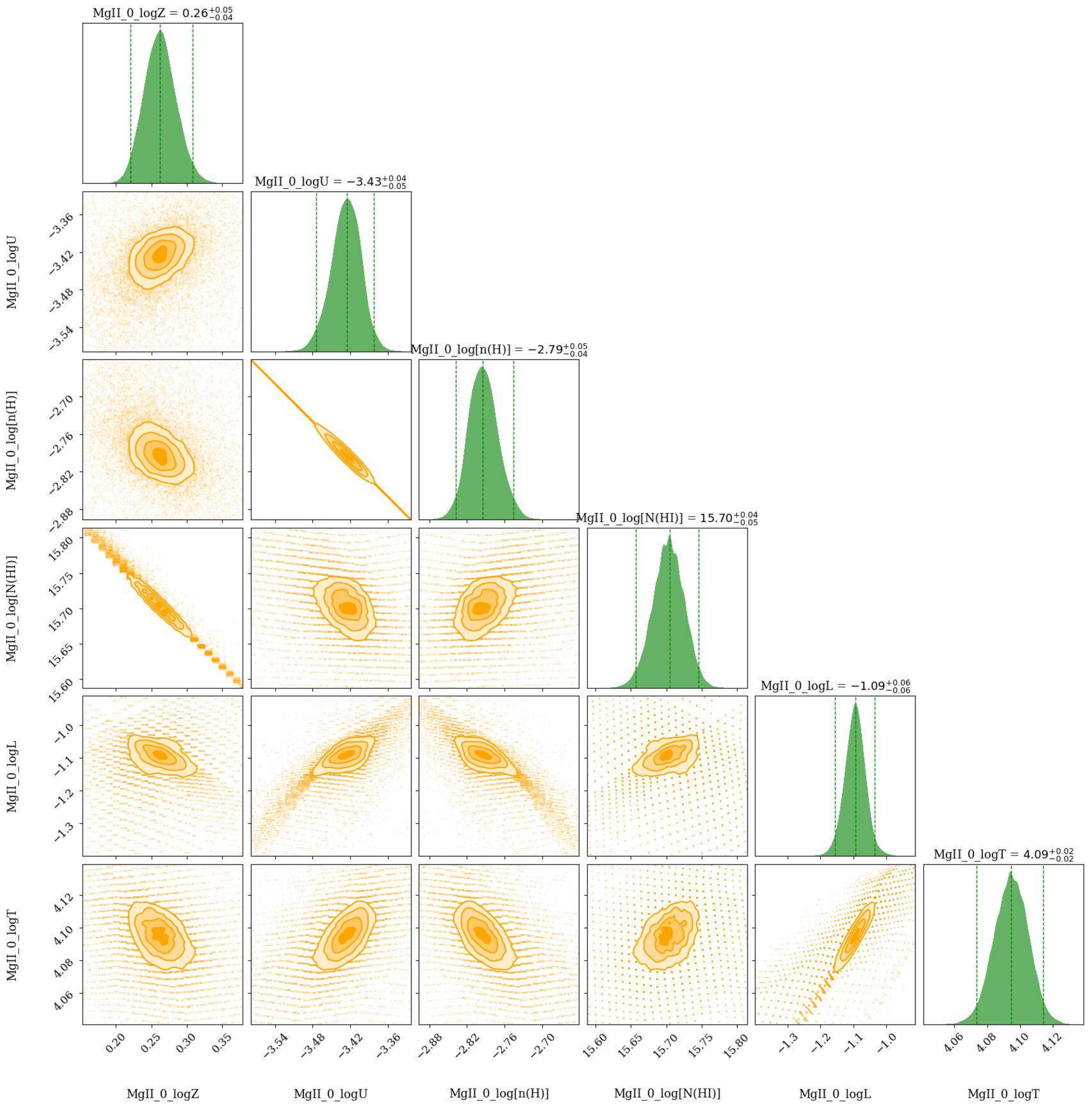}
\caption{The corner plot showing the marginalized posterior distributions for the metallicity ($\log Z$), ionization parameter ($\log U$), and other physical properties of blueward low phase traced by {\mgii} of the $z=0.13849$ absorber towards PG1116+215. The over-plotted vertical lines in the posterior distribution span the 95\% credible interval. The contours indicate 0.5$\sigma$, 1$\sigma$, 1.5$\sigma$, and 2$\sigma$ levels. The model results are summarised in Table~\ref{tab:pg1116modelpar}, and the synthetic profiles based on these models are shown in Figure~\ref{fig:sysplotPG1116}.}
\label{fig:mgii0pg1116}
\end{center}
\end{figure*}

\begin{figure*}
\begin{center}
\includegraphics[width=\linewidth]{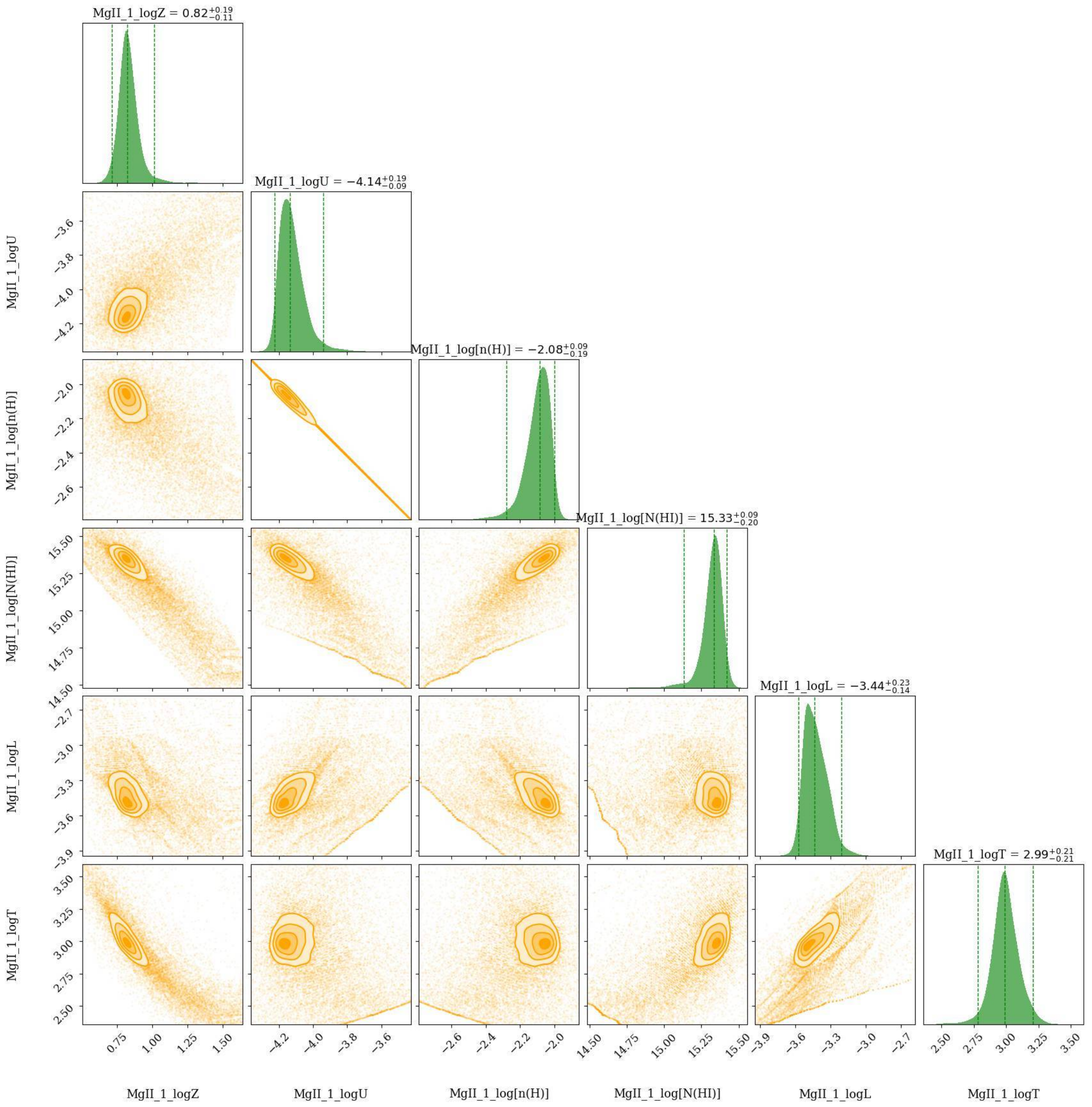}
\caption{The corner plot showing the marginalized posterior distributions for the metallicity ($\log Z$), ionization parameter ($\log U$), and other physical properties of redward low phase traced by {\mgii} of the $z=0.13849$ absorber towards PG1116+215. The over-plotted vertical lines in the posterior distribution span the 95\% credible interval. The contours indicate 0.5$\sigma$, 1$\sigma$, 1.5$\sigma$, and 2$\sigma$ levels. The model results are summarised in Table~\ref{tab:pg1116modelpar}, and the synthetic profiles based on these models are shown in Figure~\ref{fig:sysplotPG1116}.}
\label{fig:mgii1pg1116}
\end{center}
\end{figure*}

\begin{figure*}
\begin{center}
\includegraphics[width=\linewidth]{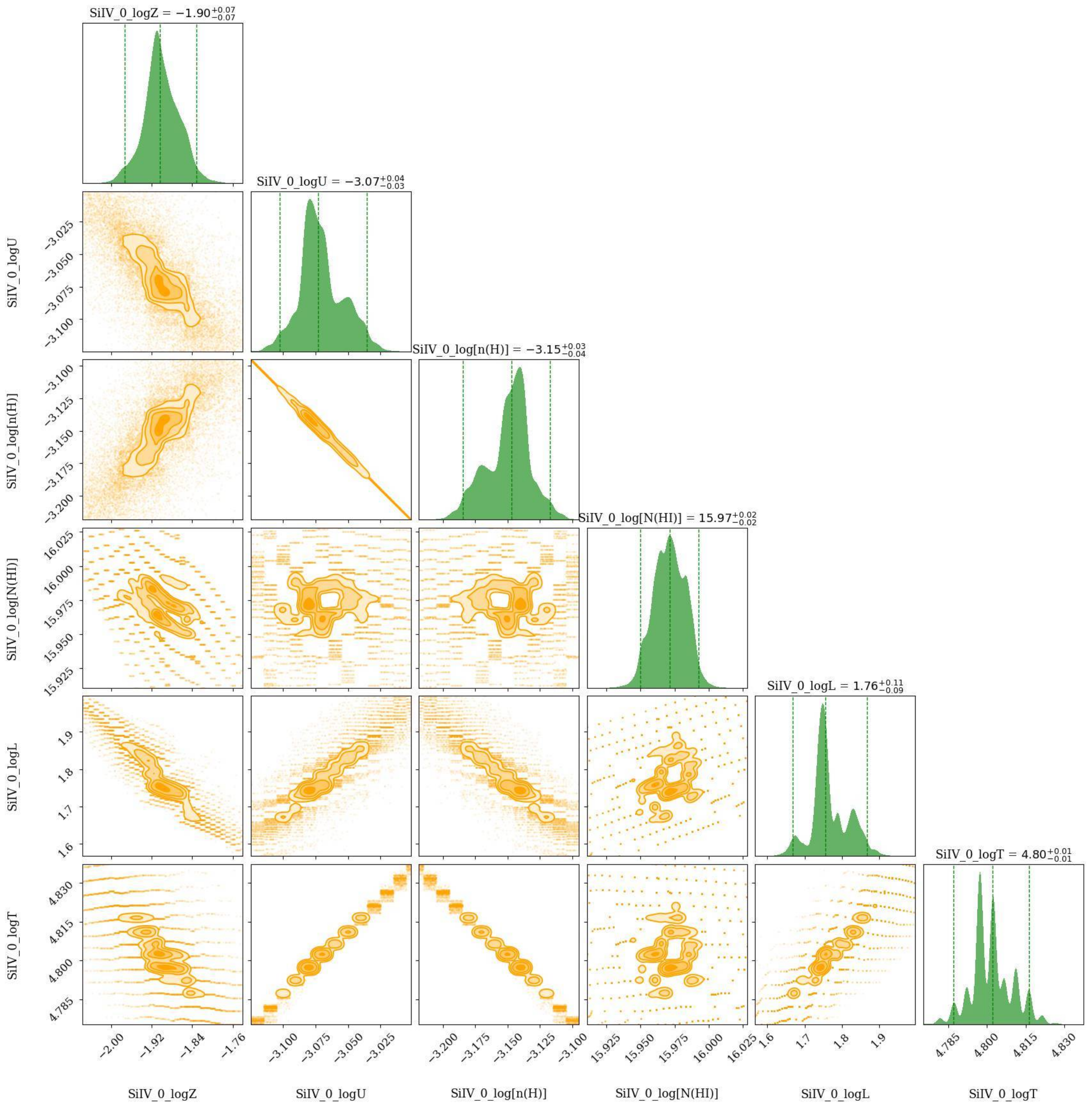}
\caption{The corner plot showing the marginalized posterior distributions for the metallicity ($\log Z$), ionization parameter ($\log U$), and other physical properties of intermediate ionization phase traced by {\siiv} of the $z=0.13849$ absorber towards PG1116+215. The over-plotted vertical lines in the posterior distribution span the 95\% credible interval. The contours indicate 0.5$\sigma$, 1$\sigma$, 1.5$\sigma$, and 2$\sigma$ levels. The model results are summarised in Table~\ref{tab:pg1116modelpar}, and the synthetic profiles based on these models are shown in Figure~\ref{fig:sysplotPG1116}.}
\label{fig:siivpg1116}
\end{center}
\end{figure*}

\begin{figure*}
\begin{center}
\includegraphics[width=\linewidth]{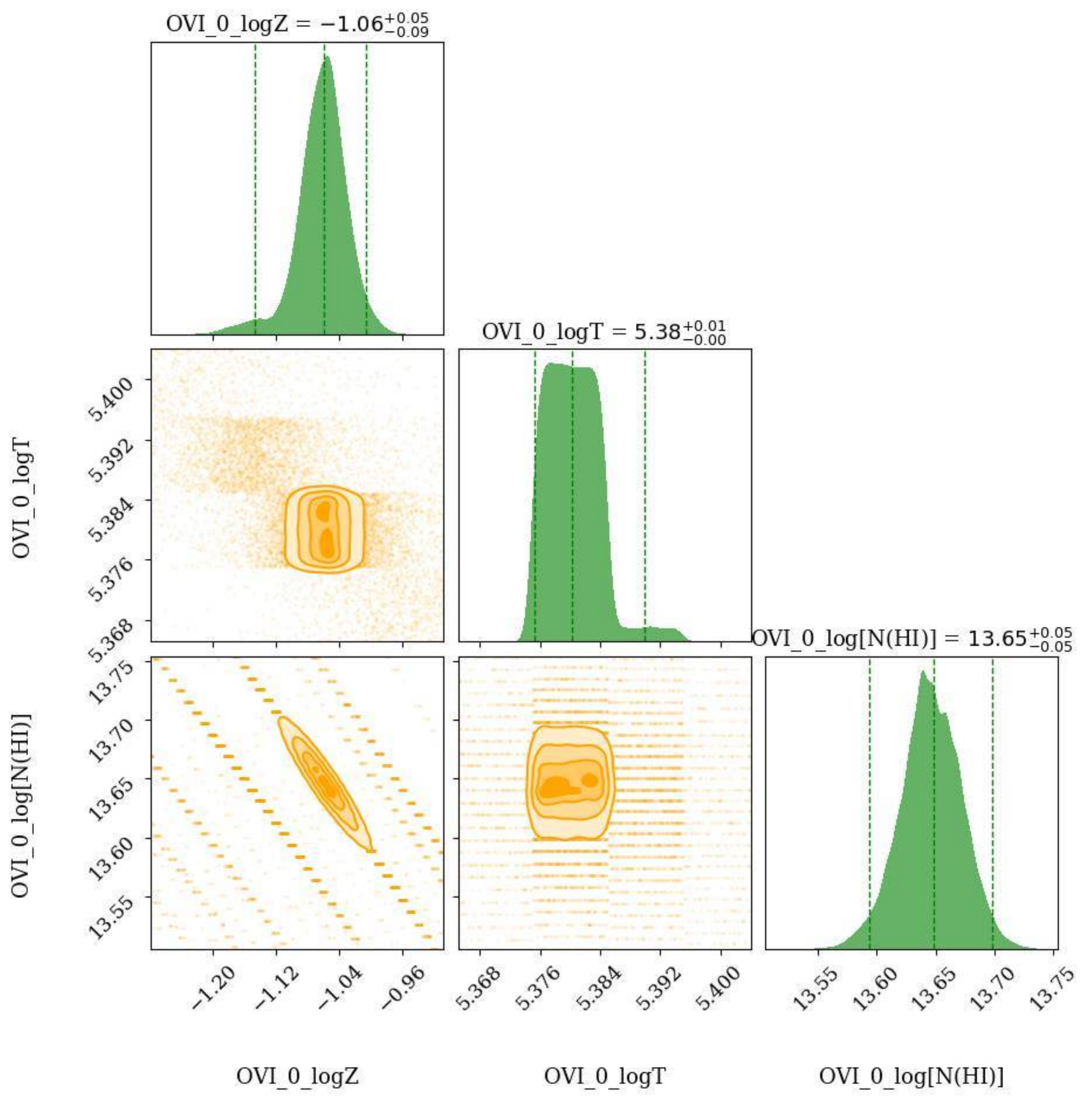}
\caption{The corner plot showing the marginalized posterior distributions for the metallicity ($\log Z$), ionization parameter ($\log U$), and other physical properties of collisionally ionized phase traced by {\ovi} of the $z=0.13849$ absorber towards PG1116+215. The over-plotted vertical lines in the posterior distribution span the 95\% credible interval. The contours indicate 0.5$\sigma$, 1$\sigma$, 1.5$\sigma$, and 2$\sigma$ levels. The model results are summarised in Table~\ref{tab:pg1116modelpar}, and the synthetic profiles based on these models are shown in Figure~\ref{fig:sysplotPG1116}.}
\label{fig:ovipg1116}
\end{center}
\end{figure*}

\section{Plots for HE0153-4520 0.22596 absorber}
\label{appendix:he0153}

\subsection{Posterior distributions for the Voigt Profile fit parameters}
\label{appendix:paramsvoigthe0153}

\begin{figure*}
\begin{center}
\includegraphics[width=\linewidth]{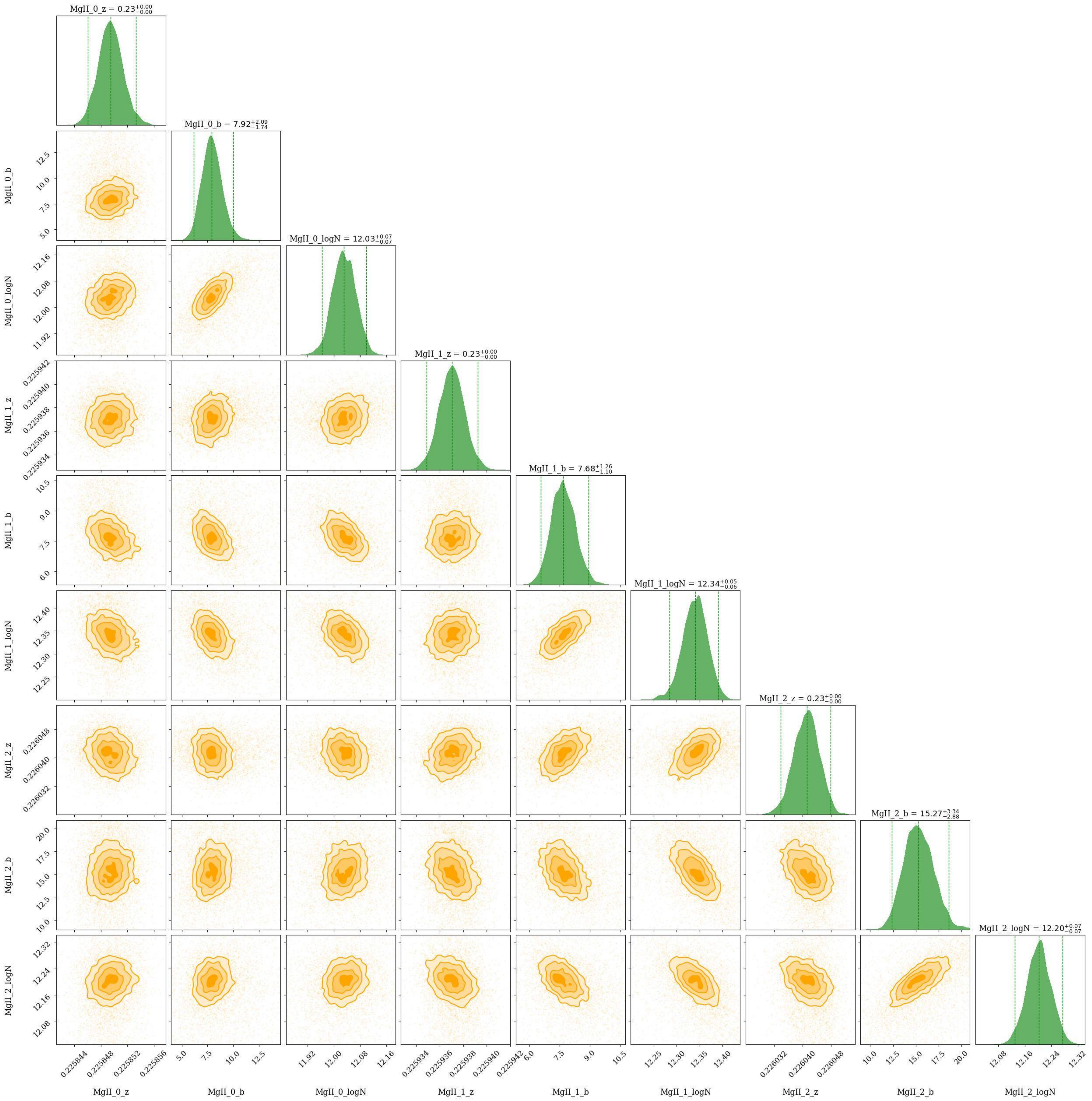}
\caption{The corner plot showing the marginalized posterior distributions of the redshift ($z$), Doppler parameter ($b$), and column density ($\log N$) for the low ionization clouds of {\mgii} of the $z=0.22596$ absorber towards HE0153-4520. The over-plotted vertical lines in the posterior distribution span the 95\% credible interval. The contours indicate 0.5$\sigma$, 1$\sigma$, 1.5$\sigma$, and 2$\sigma$ levels. The model results are summarised in Table~\ref{tab:HE0153modelpar}.}
\label{fig:voigtmgiihe0153}
\end{center}
\end{figure*}

\begin{figure*}
\begin{center}
\includegraphics[width=\linewidth]{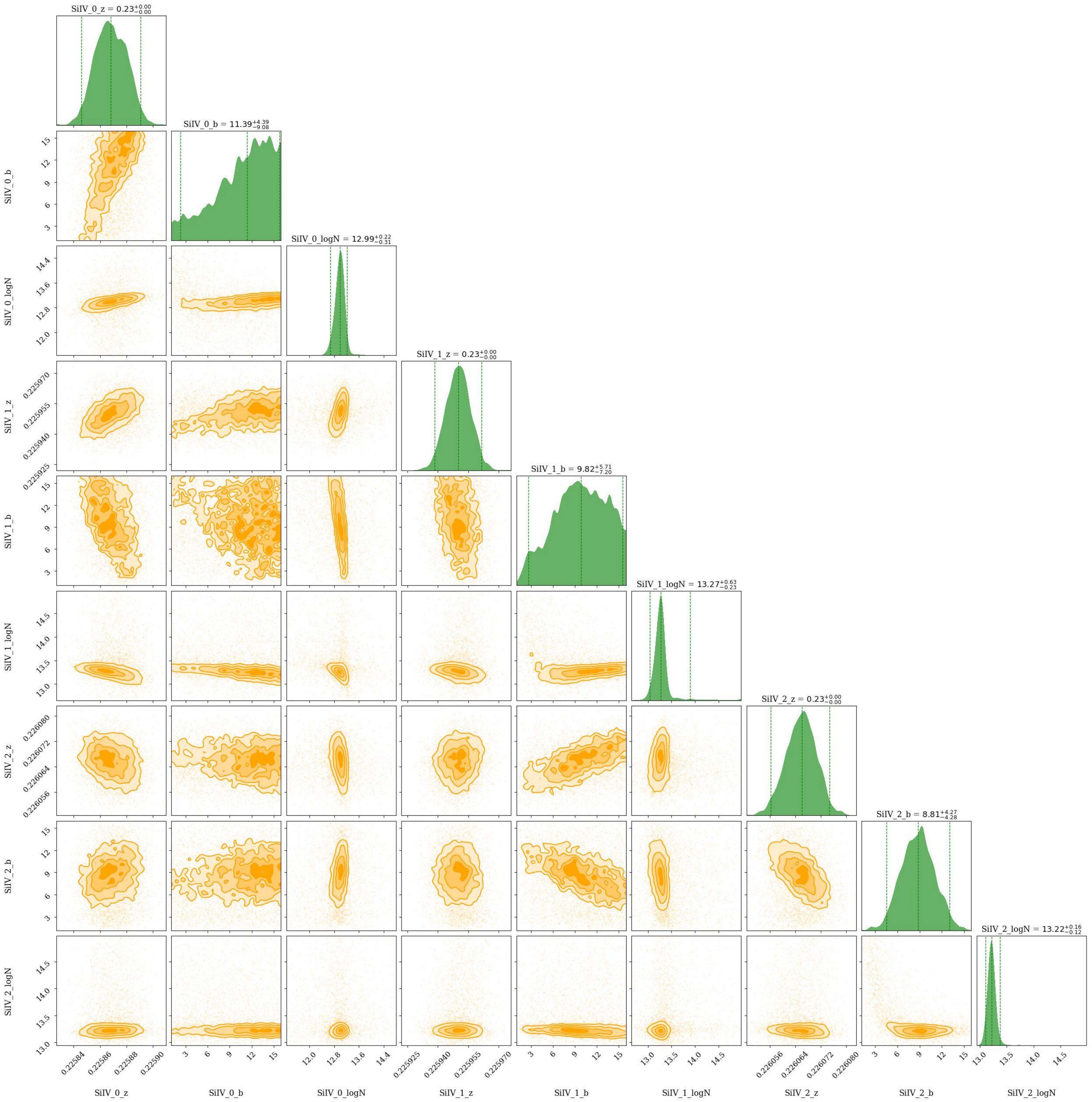}
\caption{The corner plot showing the marginalized posterior distributions of the redshift ($z$), Doppler parameter ($b$), and column density ($\log N$) for the intermediate ionization clouds of {\siiv} of the $z=0.22596$ absorber towards HE0153-4520. The over-plotted vertical lines in the posterior distribution span the 95\% credible interval. The contours indicate 0.5$\sigma$, 1$\sigma$, 1.5$\sigma$, and 2$\sigma$ levels. The model results are summarised in Table~\ref{tab:HE0153modelpar}.}
\label{fig:voigtsiivhe0153}
\end{center}
\end{figure*}

\begin{figure*}
\begin{center}
\includegraphics[width=\linewidth]{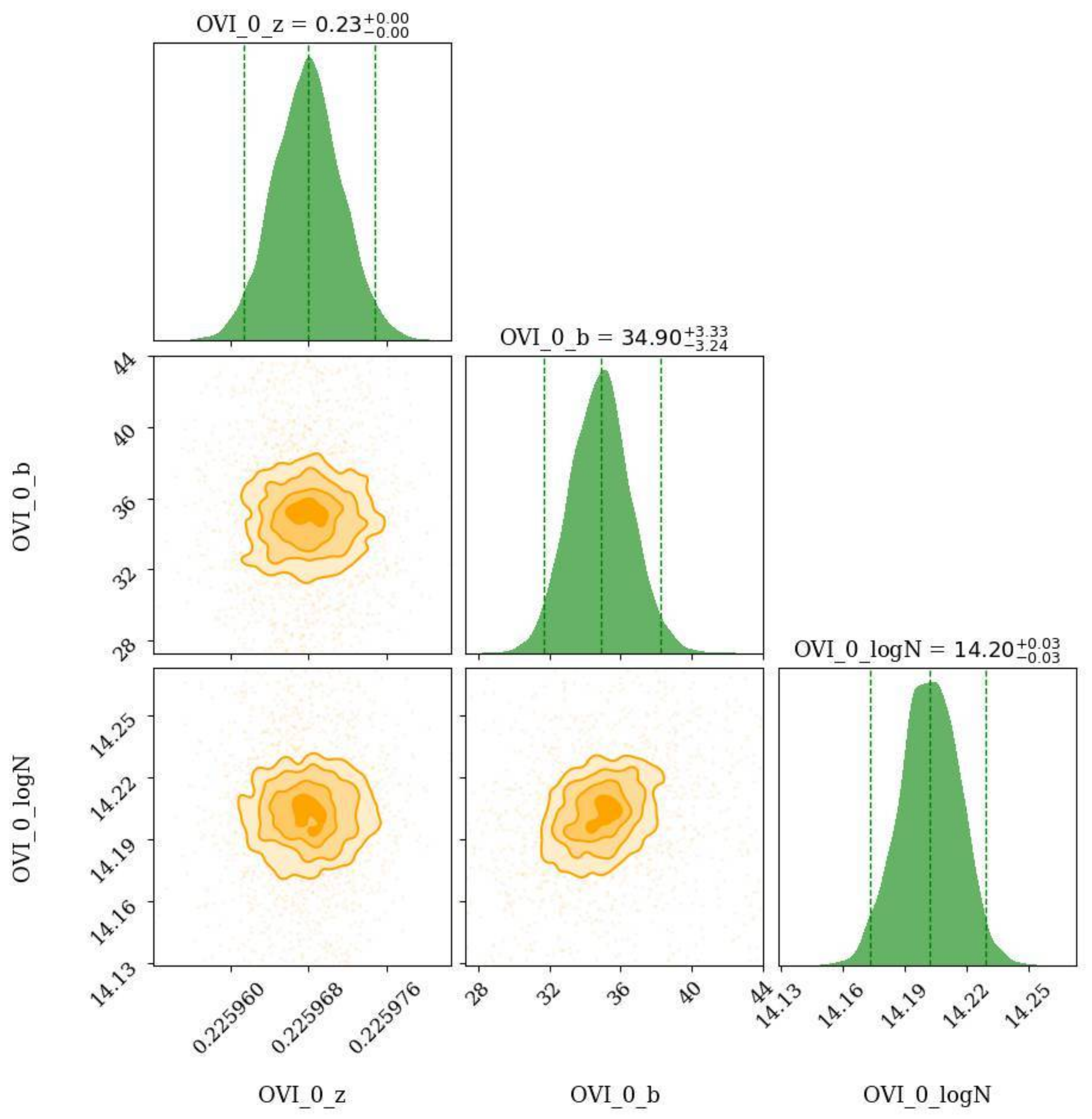}
\caption{The corner plot showing the marginalized posterior distributions of the redshift ($z$), Doppler parameter ($b$), and column density ($\log N$) for the collisionally ionized cloud of {\ovi} of the $z=0.22596$ absorber towards HE0153-4520. The over-plotted vertical lines in the posterior distribution span the 95\% credible interval. The contours indicate 0.5$\sigma$, 1$\sigma$, 1.5$\sigma$, and 2$\sigma$ levels. The model results are summarised in Table~\ref{tab:HE0153modelpar}.}
\label{fig:voigtovihe0153}
\end{center}
\end{figure*}

\subsection{Posterior distributions for the absorber properties}
\label{appendix:paramshe0153}

\begin{figure*}
\begin{center}
\includegraphics[width=\linewidth]{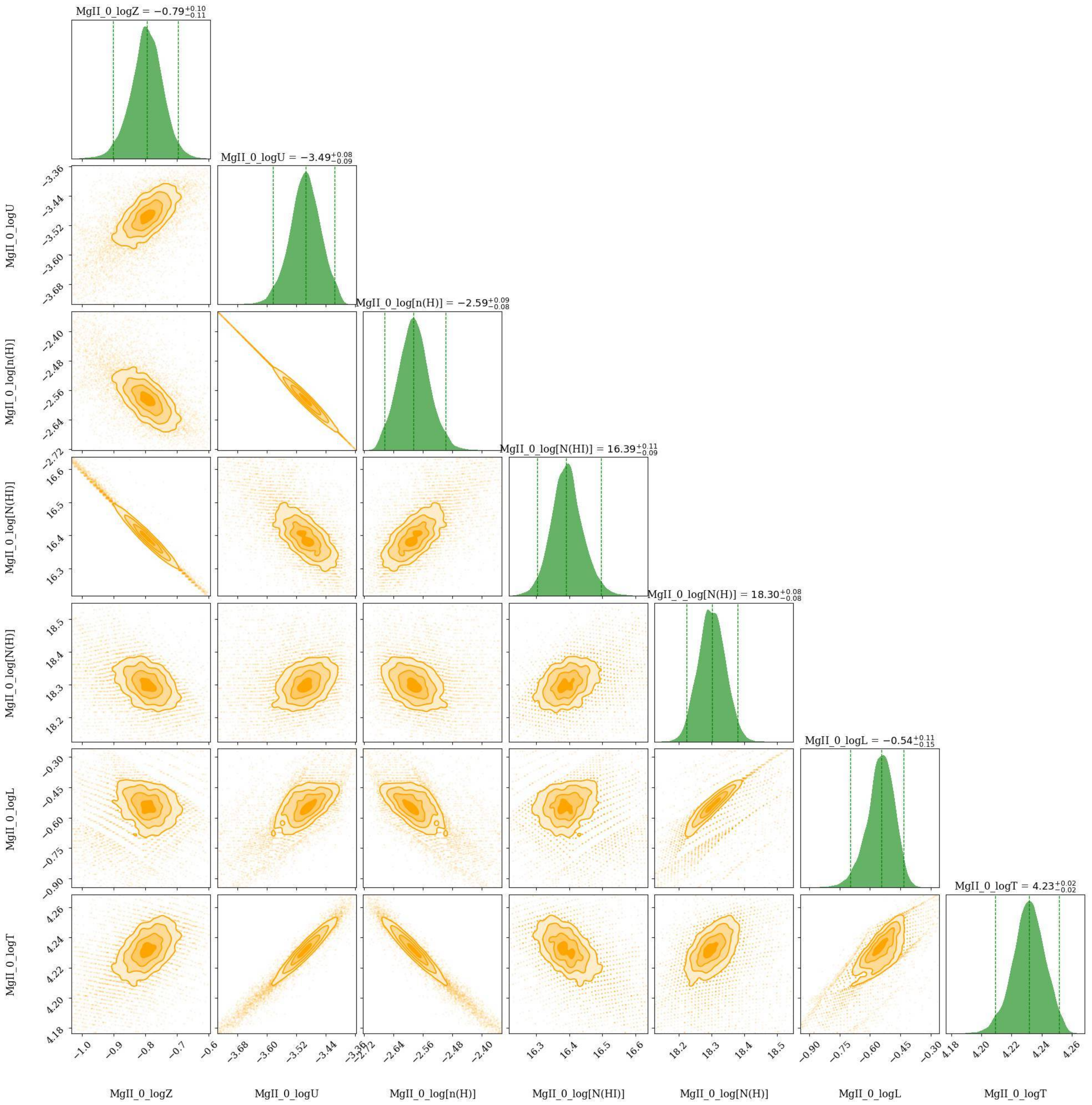}
\caption{The corner plot showing the marginalized posterior distributions for the metallicity ($\log Z$), ionization parameter ($\log U$), and other physical properties of the most blueward low phase traced by {\mgii} of the $z=0.22596$ absorber towards HE0153-4520. The over-plotted vertical lines in the posterior distribution span the 95\% credible interval. The contours indicate 0.5$\sigma$, 1$\sigma$, 1.5$\sigma$, and 2$\sigma$ levels. The model results are summarised in Table~\ref{tab:HE0153modelpar}, and the synthetic profiles based on these models are shown in Figure~\ref{fig:ModelsHE0153sysplot}.}
\label{fig:MgII0he0153}
\end{center}
\end{figure*}

\begin{figure*}
\begin{center}
\includegraphics[width=\linewidth]{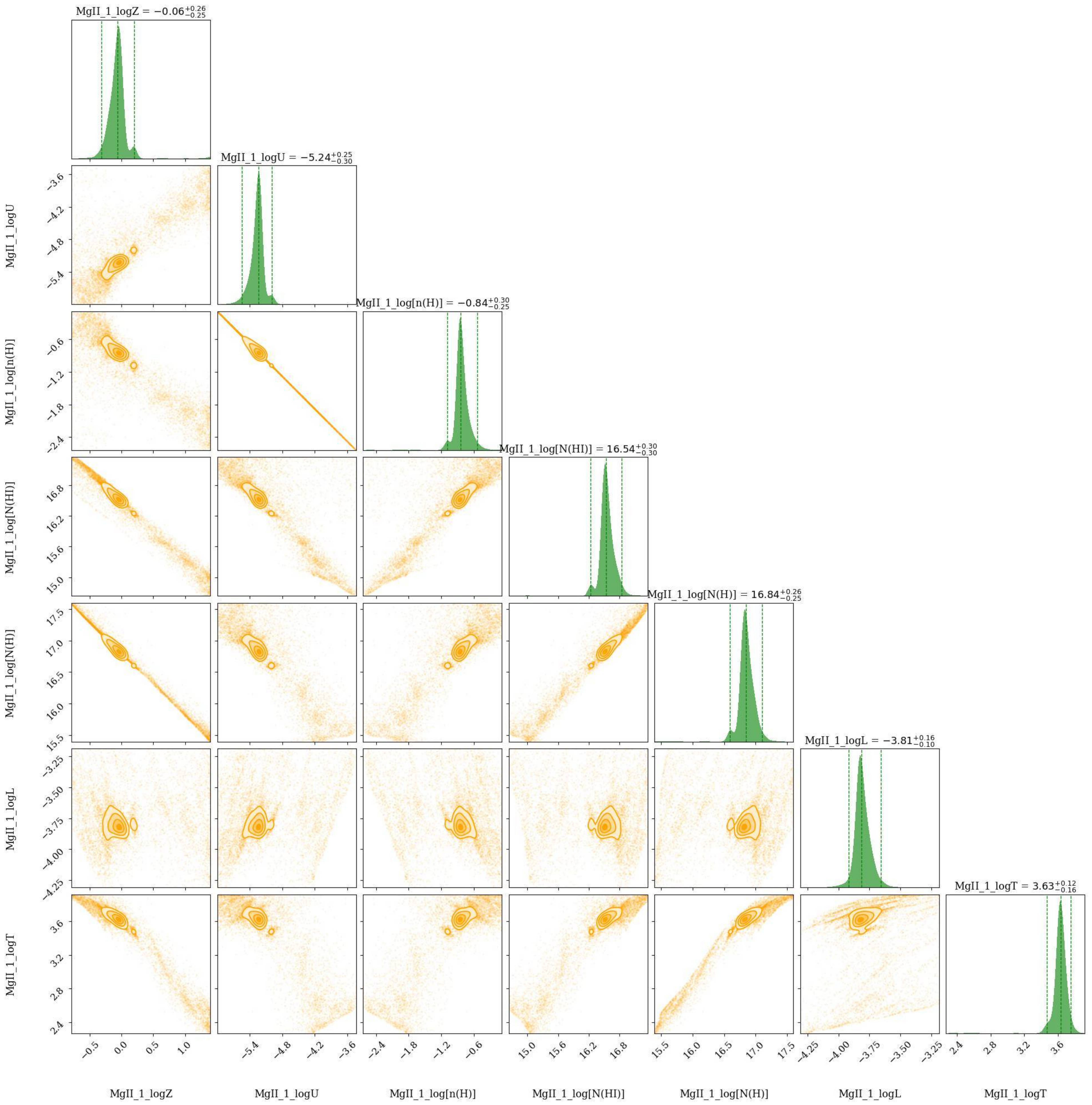}
\caption{The corner plot showing the marginalized posterior distributions for the metallicity ($\log Z$), ionization parameter ($\log U$), and other physical properties of the middle low phase traced by {\mgii} of the $z=0.22596$ absorber towards HE0153-4520. The over-plotted vertical lines in the posterior distribution span the 95\% credible interval. The contours indicate 0.5$\sigma$, 1$\sigma$, 1.5$\sigma$, and 2$\sigma$ levels. The model results are summarised in Table~\ref{tab:HE0153modelpar}, and the synthetic profiles based on these models are shown in Figure~\ref{fig:ModelsHE0153sysplot}.}
\label{fig:MgII1he0153}
\end{center}
\end{figure*}

\begin{figure*}
\begin{center}
\includegraphics[width=\linewidth]{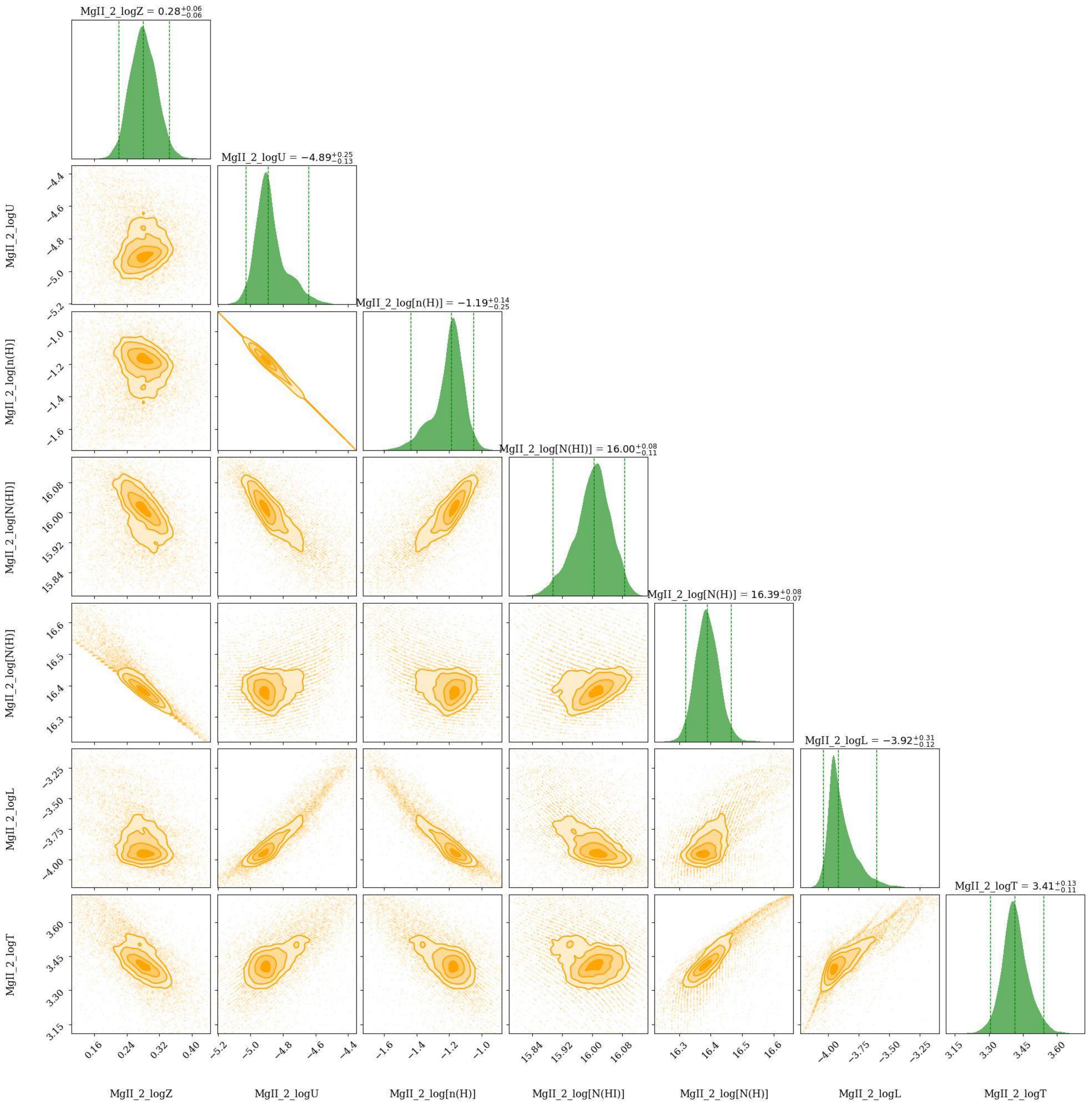}
\caption{The corner plot showing the marginalized posterior distributions for the metallicity ($\log Z$), ionization parameter ($\log U$), and other physical properties of the redward low phase traced by {\mgii} of the $z=0.22596$ absorber towards HE0153-4520. The over-plotted vertical lines in the posterior distribution span the 95\% credible interval. The contours indicate 0.5$\sigma$, 1$\sigma$, 1.5$\sigma$, and 2$\sigma$ levels. The model results are summarised in Table~\ref{tab:HE0153modelpar}, and the synthetic profiles based on these models are shown in Figure~\ref{fig:ModelsHE0153sysplot}.}
\label{fig:MgII2he0153}
\end{center}
\end{figure*}

\begin{figure*}
\begin{center}
\includegraphics[width=\linewidth]{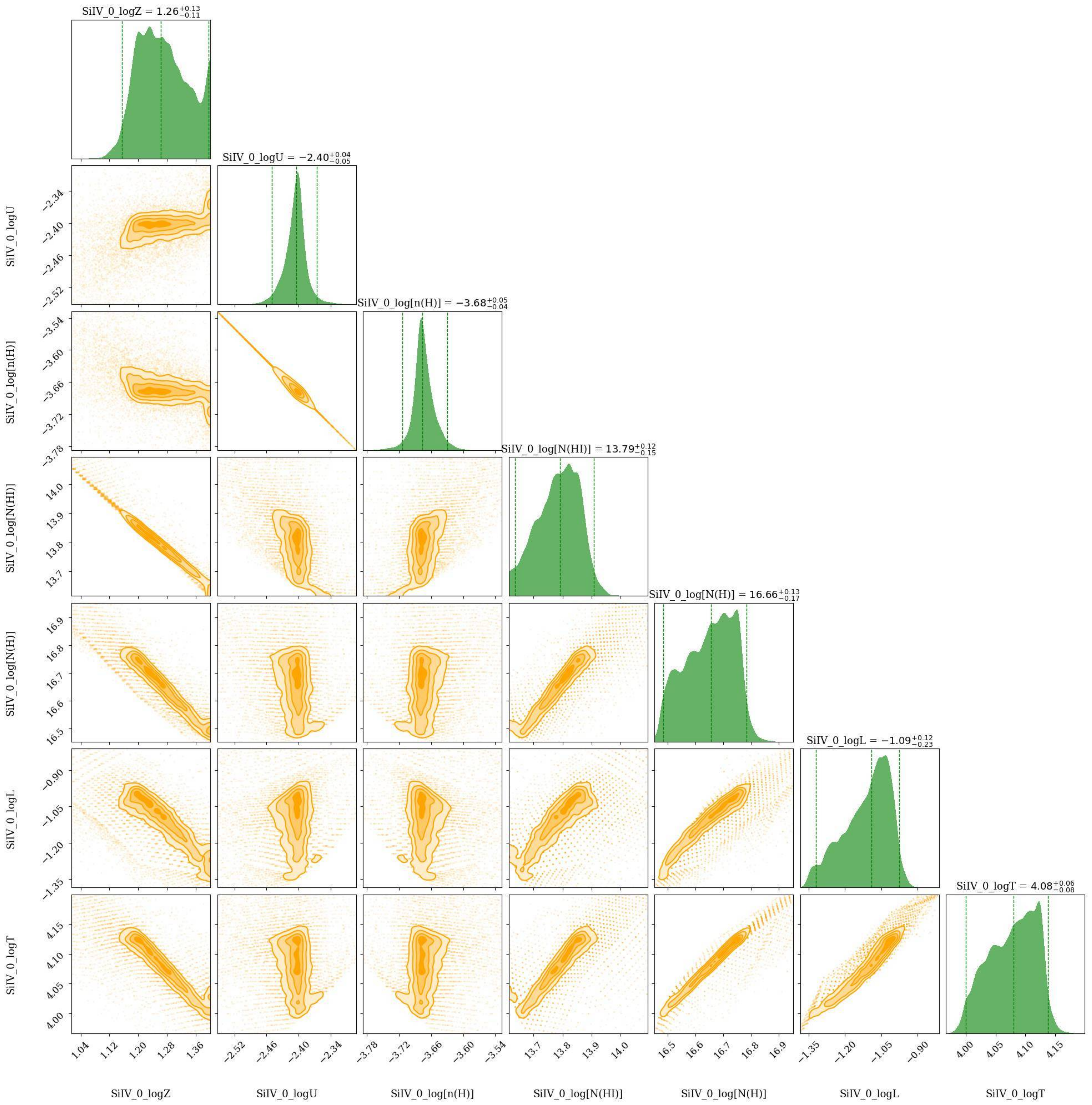}
\caption{The corner plot showing the marginalized posterior distributions for the metallicity ($\log Z$), ionization parameter ($\log U$), and other physical properties of the most blueward intermediate phase traced by {\siiv} of the $z=0.22596$ absorber towards HE0153-4520. The over-plotted vertical lines in the posterior distribution span the 95\% credible interval. The contours indicate 0.5$\sigma$, 1$\sigma$, 1.5$\sigma$, and 2$\sigma$ levels. The model results are summarised in Table~\ref{tab:HE0153modelpar}, and the synthetic profiles based on these models are shown in Figure~\ref{fig:ModelsHE0153sysplot}.}
\label{fig:SiIV0he0153}
\end{center}
\end{figure*}

\begin{figure*}
\begin{center}
\includegraphics[width=\linewidth]{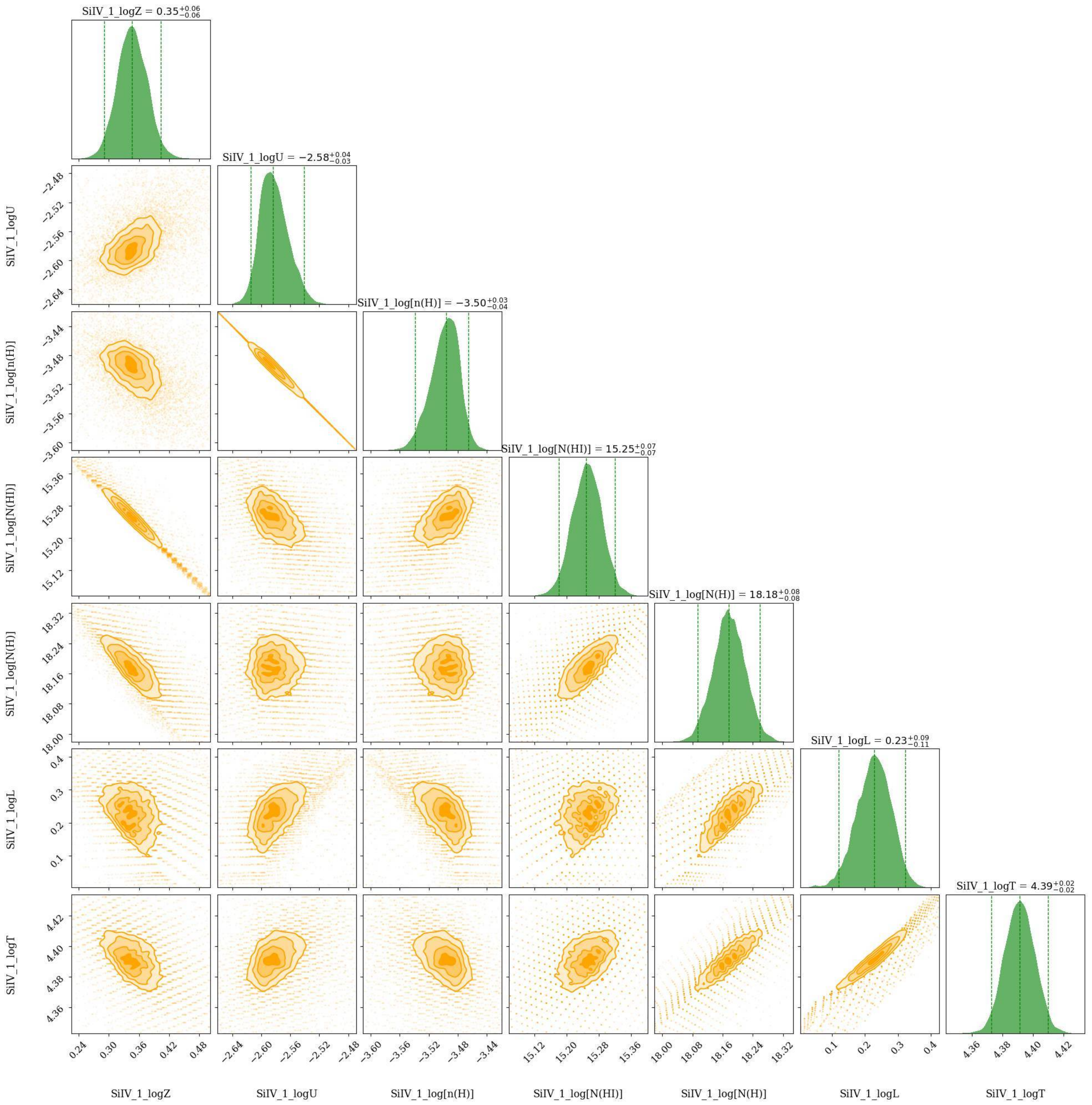}
\caption{The corner plot showing the marginalized posterior distributions for the metallicity ($\log Z$), ionization parameter ($\log U$), and other physical properties of the middle intermediate phase traced by {\siiv} of the $z=0.22596$ absorber towards HE0153-4520. The over-plotted vertical lines in the posterior distribution span the 95\% credible interval. The contours indicate 0.5$\sigma$, 1$\sigma$, 1.5$\sigma$, and 2$\sigma$ levels. The model results are summarised in Table~\ref{tab:HE0153modelpar}, and the synthetic profiles based on these models are shown in Figure~\ref{fig:ModelsHE0153sysplot}.}
\label{fig:SiIV1he0153}
\end{center}
\end{figure*}

\begin{figure*}
\begin{center}
\includegraphics[width=\linewidth]{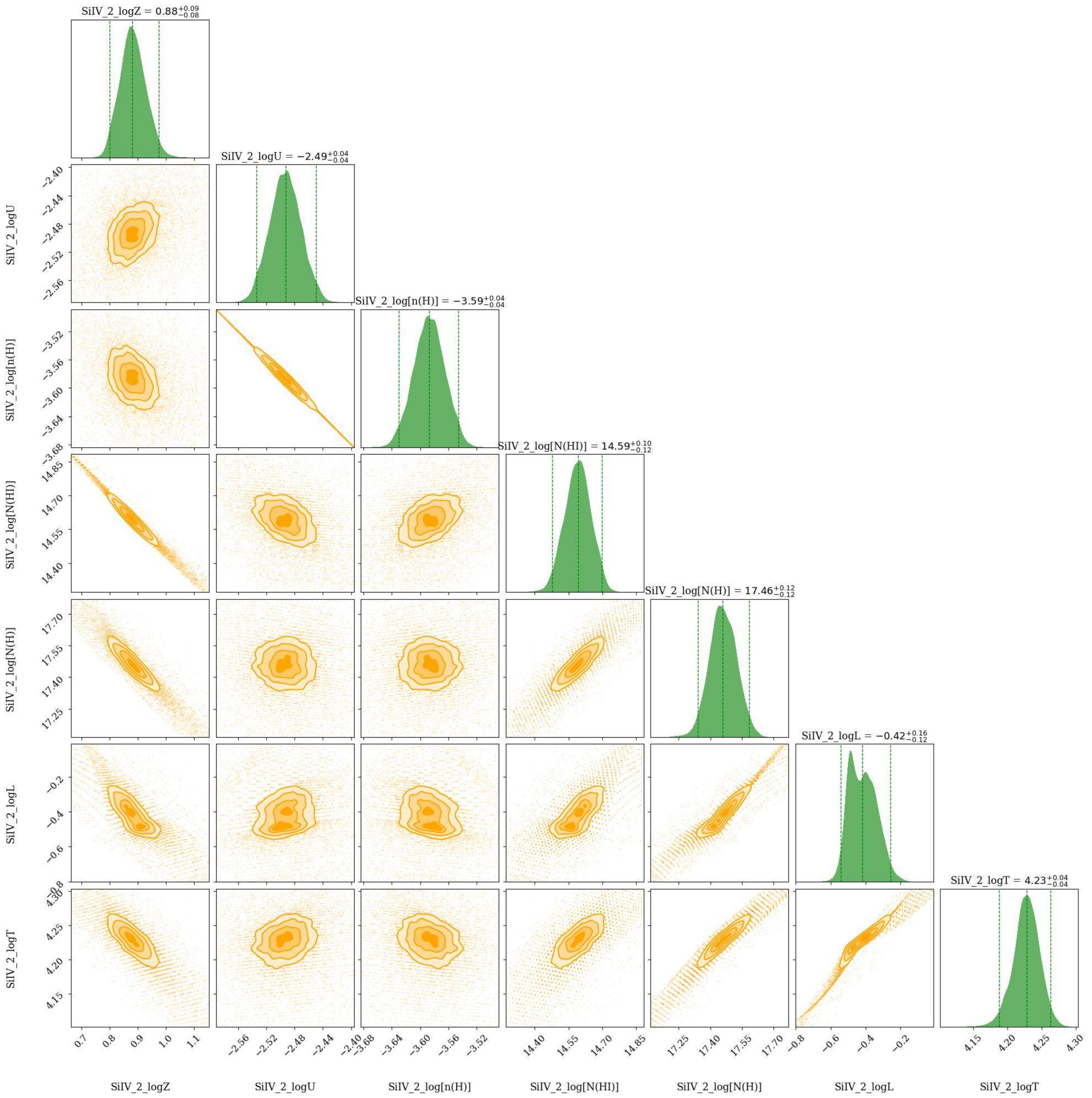}
\caption{The corner plot showing the marginalized posterior distributions for the metallicity ($\log Z$), ionization parameter ($\log U$), and other physical properties of the redward intermediate phase traced by {\siiv} of the $z=0.22596$ absorber towards HE0153-4520. The over-plotted vertical lines in the posterior distribution span the 95\% credible interval. The contours indicate 0.5$\sigma$, 1$\sigma$, 1.5$\sigma$, and 2$\sigma$ levels. The model results are summarised in Table~\ref{tab:HE0153modelpar}, and the synthetic profiles based on these models are shown in Figure~\ref{fig:ModelsHE0153sysplot}.}
\label{fig:SiIV2he0153}
\end{center}
\end{figure*}

\begin{figure*}
\begin{center}
\includegraphics[width=\linewidth]{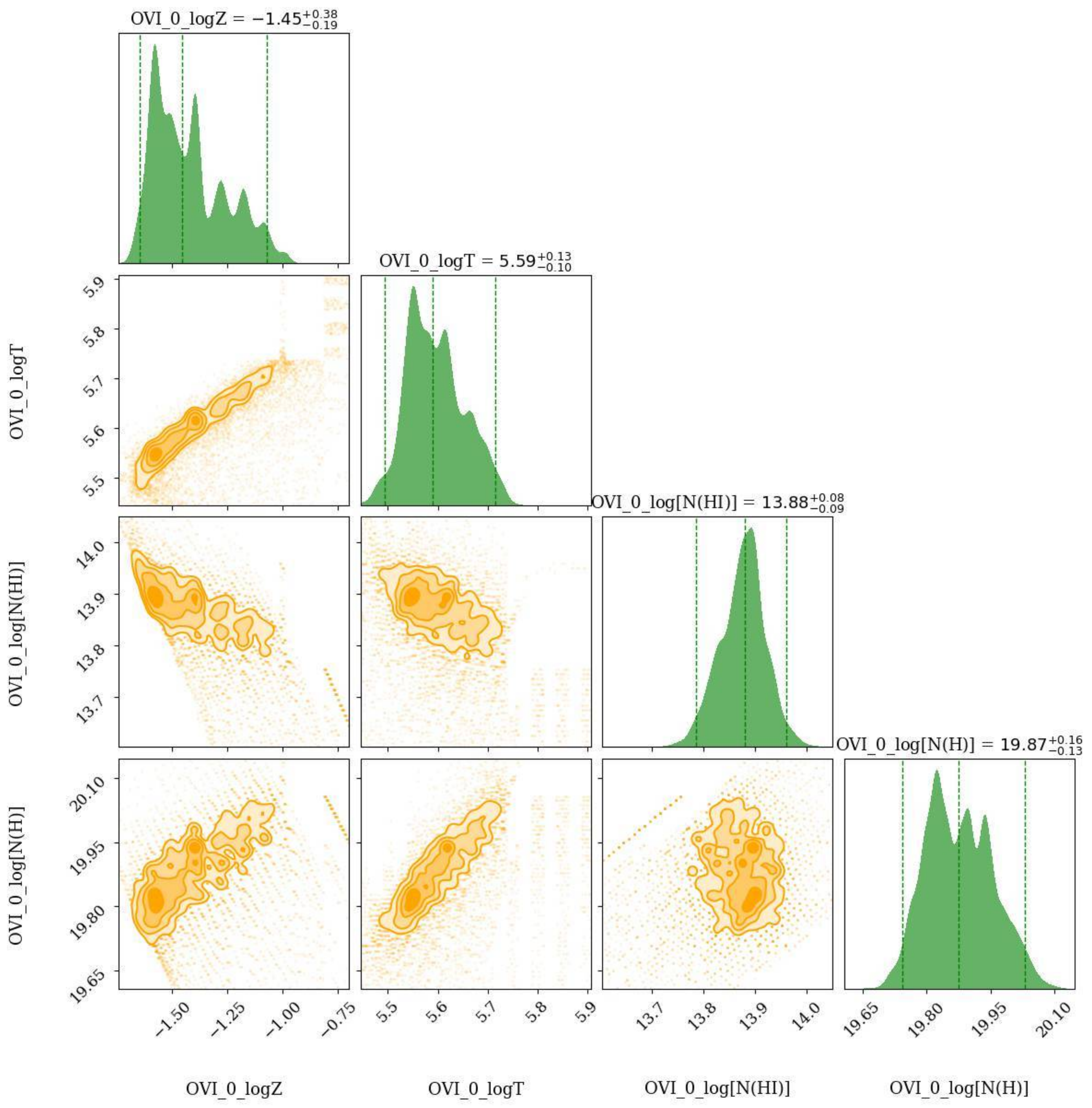}
\caption{The corner plot showing the marginalized posterior distributions for the metallicity ($\log Z$), ionization parameter ($\log T$), and other physical properties of the collisionally ionized high phase traced by {\ovi} of the $z=0.22596$ absorber towards HE0153-4520. The over-plotted vertical lines in the posterior distribution span the 95\% credible interval. The contours indicate 0.5$\sigma$, 1$\sigma$, 1.5$\sigma$, and 2$\sigma$ levels. The model results are summarised in Table~\ref{tab:HE0153modelpar}, and the synthetic profiles based on these models are shown in Figure~\ref{fig:ModelsHE0153sysplot}.}
\label{fig:OVI0he0153}
\end{center}
\end{figure*}

\end{document}